\documentclass[twoside,12pt]{article}
\usepackage{amssymb,amsmath}
\usepackage[dvipdfmx]{graphicx}
\usepackage{color}
\usepackage[breaklinks, colorlinks=true]{hyperref}

\newcommand{\be}{\begin{equation}}
\newcommand{\ee}{\end{equation}}
\newcommand{\bea}{\begin{eqnarray}}
\newcommand{\eea}{\end{eqnarray}}

\newcommand{\LQCD}{\Lambda_{\text{QCD}}}
\newcommand{\Nc}{N_{\text{c}}}
\newcommand{\Nf}{N_{\text{f}}}
\newcommand{\Tc}{T_{\text{c}}}
\newcommand{\diag}{\mathrm{diag}}
\newcommand{\tr}{\mathrm{tr}}
\newcommand{\boldm}{\boldsymbol{m}}
\newcommand{\bn}{\boldsymbol{n}}
\newcommand{\bp}{\boldsymbol{p}}
\newcommand{\br}{\boldsymbol{r}}
\newcommand{\bx}{\boldsymbol{x}}
\newcommand{\bz}{\boldsymbol{z}}
\newcommand{\bS}{\boldsymbol{S}}
\newcommand{\bnabla}{\boldsymbol{\nabla}}
\newcommand{\bsigma}{\boldsymbol{\sigma}}
\newcommand{\calA}{\mathcal{A}}
\newcommand{\calD}{\mathcal{D}}
\newcommand{\calG}{\mathcal{G}}
\newcommand{\calH}{\mathcal{H}}
\newcommand{\calL}{\mathcal{L}}
\newcommand{\calM}{\mathcal{M}}
\newcommand{\calO}{\mathcal{O}}
\newcommand{\calP}{\mathcal{P}}
\newcommand{\calR}{\mathcal{R}}
\newcommand{\imu}{\tilde{\mu}}
\newcommand{\MeV}{\;\text{MeV}}
\newcommand{\feyn}[1]{
  \setbox0=\hbox{\ensuremath{#1}}
  \hbox to\wd0{\hbox to0pt{\hbox to\wd0{\hss/\hss}\hss}\box0}}
\renewcommand{\Re}{\mathrm{Re}}
\renewcommand{\Im}{\mathrm{Im}}
\newcommand{\comment}[1]{}  

\begin{document}
\title{Polyakov loop modeling for hot QCD}
\author{Kenji Fukushima,$^1$ and Vladimir Skokov$^2$\\
\\
{\small $^1$Department of Physics, The University of Tokyo,
7-3-1 Hongo, Bunkyo-ku, Tokyo 113-0033, Japan}\\
{\small $^2$Riken-BNL Research Center, Brookhaven National Laboratory,
Upton, New York 11973, USA}}
\maketitle

\begin{abstract}
  We review theoretical aspects of quantum chromodynamics (QCD) at
  finite temperature.  The most important physical variable to
  characterize hot QCD is the Polyakov loop, which is an approximate
  order parameter for quark deconfinement in a hot gluonic medium.
  Additionally to its role as an order parameter, the Polyakov loop
  has rich physical contents in both perturbative and non-perturbative
  sectors.  This review covers a wide range of subjects associated
  with the Polyakov loop from topological defects in hot QCD to model
  building with coupling to the Polyakov loop.
\end{abstract}
\tableofcontents

\section{Introduction}
\label{sec:intro}

Quantum Chromodynamics, which is commonly abbreviated as QCD, is a
fundamental theory of the strong interaction composed from quarks and
gluons.  Gluons belong to the adjoint representation of the color
SU(3) group, while quarks are in the color fundamental representation.
Conventionally the quark color index runs over \text{red},
\text{green}, and \text{blue}, in analogy to ``three primary colors''
in nature.  Although this is a departure from reality, it is
nevertheless useful to change the number of colors $\Nc$ arbitrarily.
Then the gluonic part in QCD is described by the SU($\Nc$) pure
Yang-Mills theory.  We will refer to the pure Yang-Mills theory as
simply the ``pure gluonic theory'' throughout this review.  We note
that QCD reduces to the pure gluonic theory in the heavy-quark limit,
i.e.\ the limit with all quark masses sent to infinity, which is often
called the quenched limit.

The research area dedicated to reveal microscopic details for the QCD
vacuum structure and QCD phase transitions when the system is
equilibrated at finite temperature, $T$, has been very rich and
active.  The main subject of this review is focused on the physics of
an order parameter of such a hot gluonic system, that is called the
Polyakov loop named after the inventor~\cite{Polyakov:1978vu}.
Regarding the longstanding problem of color confinement in QCD, the
Polyakov loop provides us with a useful view point especially on the
following question:  what causes color confinement?  Precisely
speaking, the problems of ``quark'' confinement and ``gluon''
confinement should be considered separately.  For quark confinement in
the pure gluonic theory, the Wilson loop in the fundamental
representation is an well-defined  measure of confinement,
demonstrating the expected area law in the quark confined phase.  To
realize the area law, the QCD vacuum should accommodate highly
disturbed gauge configurations;  such an extremal case is approached
in the limit of strong gauge coupling, as was first elucidated by
Wilson~\cite{Wilson:1974sk}.  In the strong coupling limit, gauge
fluctuations at each spacetime point become independent of adjacent
ones.  Actually, the strong coupling expansion in gauge theory has
clear resemblance to the high-$T$ expansion in (classical) spin
models.  This analogy is naturally understood in terms of the Polyakov
loop;  the finite-$T$ counterpart of the Wilson loop is the Polyakov
loop correlation function, and the area law of the Wilson loop is
recapitulated as an exponentially decaying behavior of the Polyakov
loop correlation function, which is reminiscent of the spin
correlation function in the \textit{disordered} state.  As a matter of
fact, the Polyakov loop effective model for the hot pure gluonic
theory takes a form of a classical spin model with inverted
temperature.  In this sense the confined phase at low $T$ can be
regarded as a dual of the spin disordered state at high $T$.  Such a
picture of quark confinement from the point of view of condensed
matter physics 
is quite useful for us to deepen our understanding of confinement.

The Polyakov loop provides additional benefits for investigations
of the quark confinement.  The disordered state is by definition a
state with large fluctuations, and the question is what microscopic
quanta can bring such fluctuations into the system.  We know that
magnetic domains and associated domain-walls are microscopic contents
in the ordered and disordered states in spin systems; thus, it is
natural to  consider a gluonic counterpart, namely, the dynamics of
the Z($\Nc$) domain-walls.  Interestingly, the Z($\Nc$) domain-wall is
a solution of the classical equation of motion \textit{including}
thermal fluctuation corrections.  At the same time, we also know that
the pure gluonic theory accommodates other solutions of the classical
equation of motion, i.e.\ instantons.  The finite-$T$ extended
instantons are specifically called the calorons (meaning ``caloric''
instantons at finite $T$) and the Polyakov loop varies on caloron
configurations.  Thus, the calorons are nothing but the Z($\Nc$)
bubbles and the condensation of those bubbles may cause quark
confinement.  Originally, the instantons were supposed to be
responsible for chiral symmetry breaking and quark
confinement~\cite{Callan:1976je,Callan:1977gz} but the
phenomenological model building for confinement based on an instanton
liquid picture was not very successful before the relatively recent
discovery of new calorons that have a non-trivial boundary condition
of the Polyakov loop at large distance.  These intriguing developments
will be closely discussed in Sec.~\ref{sec:polyakov}.  
We note, however, that it is not realistic to cover all relevant
topics and background foundations within this single review.  As a
pragmatic approach, we shall make a decision not to reiterate fairly
well-known parts of the Polyakov loop related physics, for which the
readers can easily find comprehensive reviews.  Let us here recommend
several articles; for center symmetry and the classification of
possible phase transitions, Ref.~\cite{Svetitsky:1985ye} is the most
comprehensive review based on the classic
paper~\cite{Svetitsky:1982gs}.  Some updates for the classification
are found in Ref.~\cite{Fukushima:2013rx}.  For the theoretical
formulation of hot QCD,  including discussion on the Polyakov loop and
the calorons, Ref.~\cite{Gross:1980br} provides a complete
description, and for a modern review on hot QCD, see also
Ref.~\cite{Smilga:1996cm} which contains unique considerations on the
meaning of the Polyakov loop.  Here, in this review, we will not dwell
on the physics of relativistic heavy-ion collision in which a new
state of hot QCD matter, i.e.\ a quark-gluon plasma (QGP) is created
in the laboratory.  Interested readers can consult
Refs.~\cite{Rischke:2003mt,Fukushima:2010bq,Braun-Munzinger:2015hba}
for phenomenological implications of the Polyakov loop to the QGP
physics.  Several established textbooks are also available to study
the heavy-ion collision
physics~\cite{kapusta2006finite,bellac2000thermal,letessier2002hadrons,%
yagi2005quark}.
One might be wondering why we do not touch the lattice-QCD results in
this review.  This is because there are already thorough reviews by
the authors from lattice-QCD groups~\cite{Ding:2015ona} including
recent proceedings~\cite{Ratti:2016jgx,Borsanyi:2016bzg}.  In this
review, we pay more of our attention to semi-analytical sides of the
Polyakov loop physics, particularly about the theoretical formulations
of deconfinement in the pure gluonic theory as addressed in
Sec.~\ref{sec:pure};  we also try to make the present review be
different from others, although we cannot avoid a partial overlap to
keep this review as self-containing as possible.

Now, turning from the pure gluonic theory to QCD with dynamical
quarks, which is the main subject in Sec.~\ref{sec:quarks}, new
possible applications open.  Although the physical meaning of the
Polyakov loop as an order parameter is the most transparent in the
pure gluonic theory, it also captures general screening properties of
any colored excitations, and hence, the thermal excitations of quarks
in QCD should be dictated by the Polyakov loop to a significant
degree.  This is the underlying idea behind the development of the
Polyakov loop augmented chiral effective models, such as the
Nambu--Jona-Lasinio (NJL) model and the quark-meson (QM) model.  This
is an active field of research covering many interesting subjects of
hot QCD matter;  there is also a multitude  of interesting results
from the chiral model studies which cannot be covered in this review
in full.  Therefore, again, we shall take a pragmatic strategy of not
discussing all related works but picking up some selected topics only.
In particular, we will put our emphasis on rather analytical aspects
of the model studies and the implication to the sign problem of the
Dirac determinant at finite density, which also includes recently
developing topics such as the determination of the Polyakov loop
effective potential and the semi-QGP regime.  Because the Polyakov
loop is more sensitive to deconfinement rather than the chiral sector,
though they couple to each other though the Polyakov loop, our
descriptions of the chiral symmetry breaking and restoration will be
minimal in this review.  Interested readers can easily find background
materials;  the classic reviews on the chiral physics can be found
following Refs.~\cite{Klevansky:1992qe,Hatsuda:1994pi} and the
state-of-the-art review including inhomogeneous phases is
Ref.~\cite{Buballa:2014tba}.  Since we already mentioned that the QCD
vacuum has a condensed matter analogy, it is not surprising that
chiral models also have the condensed matter interpretation, which is
known from the very early times traced back to Nambu and
Jona-Lasinio~\cite{Nambu:1961tp}.  In this context readers may want to
consult review articles on condensed matter physics aspects of chiral
symmetry in QCD~\cite{Rajagopal:2000wf,Braun:2011pp}.

\section{Hot QCD, Polyakov Loop, and Confinement}
\label{sec:polyakov}

We explain the quantization procedure for QCD following the 
standard
procedure outlined 
in Ref.~\cite{Gross:1980br} and supplemented with the field
theoretical treatment of the ghost boundary condition.  The Polyakov
loop arises from the free energy in the presence of a test color
charge.  We then proceed to discussions on underlying symmetry that
governs the behavior of the Polyakov loop.  For the rest of this
section, we see various concrete examples of the Polyakov loop
calculations 
using 
perturbative and non-perturbative methods.

\subsection{Quantizing Hot QCD}
\label{sec:hotQCD}

We will evade going into mathematical subtleties of QCD quantization, 
as 
the standard perturbative procedure to quantize hot QCD would
suffice for our present purpose.  The QCD Lagrangian density in
Minkowskian spacetime consists of the pure gluonic part and the quark
part as
\begin{equation}
  \calL(x) = -\frac{1}{4}F_{\mu\nu}^a(x) F^{\mu\nu a}(x)
  + \bar{\psi}(x) (i\feyn{D}-m)\psi(x)\;,
\end{equation}
where the flavor trace is implicitly understood (with $m$ being a
quark mass matrix in flavor space).  Our convention for
the field strength tensor is $F_{\mu\nu}^a
=\partial_\mu A_\nu^a-\partial_\nu A_\mu^a + gf^{abc}A_\mu^b A_\nu^c$
in terms of gluon fields $A_\mu^a(x)$.  We can then read the canonical
momenta from the Lagrangian density as
\begin{align}
  \Pi^{ia}_A(x) &= \frac{\partial \calL}{\partial (\partial_0 A_i^a(x))}
  = -F^{0ia}(x) = -\partial^0 A^{ia}(x)\;,
\label{eq:Pi_A} \\
  \pi_\psi(x) &= \frac{\partial \calL}{\partial (\partial_0 \psi(x))}
  = i\psi^\dag(x)\;.
\end{align}
Several remarks are necessary here.  Because $\pi_\psi(x)$ does not
involve any time derivative, we need to calculate the Dirac brackets
to quantize such a constrained system. It is known, however, that the
correct answer is obtained in a simplified prescription in which
only $\psi(x)$ is treated as a dynamical variable and $\bar{\psi}(x)$
or $\psi^\dag(x)$ is regarded as its canonical momentum.  Another
problem is that there is no canonical momentum for $A_0^a(x)$ because
the antisymmetric tensor 
$F_{\mu\nu}^a$ cannot accommodate 
the term 
$\partial_0 A_0^a(x)$
by definition.  Therefore, we impose the so-called Weyl gauge fixing
condition, $A_0^a=0$, which is used for the second equality in
Eq.~\eqref{eq:Pi_A}.  With these canonical momenta we proceed with 
the Legendre transformation to find the QCD Hamiltonian density as
\begin{equation}
  \calH = -\frac{1}{2}\Pi_{Ai}^a \Pi_A^{ia} + \frac{1}{4}
  F_{ij}^a F^{ija} - i\pi_\psi\gamma^0 \bigl[ -i\gamma^i
  (\partial_i -ig A_i) + m\bigr]\psi\;.
\end{equation}
Now that the Hamiltonian density is explicitly given, we can write
down the QCD ``partition function'' at finite temperature $T$ as
follows;
\begin{equation}
  Z_{\text{QCD}} = \tr\,e^{-\beta H}
  = \int\calD A_i \calD\psi\,\langle A_i-\psi| e^{-\beta H}
    \calP_{\calG} |A_i,\psi\rangle\;.
\end{equation}
Here, the Hamiltonian $H$ represents $H=\int d^3x\,\calH$.  The
anti-periodic boundary condition imposed as
$\int\calD\psi \langle-\psi|\cdots|\psi\rangle$ is 
attributed to our convention for the fermionic complete set;
$1=\int\calD\psi\,|\psi\rangle\langle\psi|$.  It is quite important to
note that $\calP_{\calG}$ is needed to project states out to satisfy
the Gauss law, which eliminates gauge uncertainty in the time
evolution.

The Gauss law operator is a generator for the gauge transformation and
it acts on the field coherent states as
\begin{equation}
  \calG^a(\bx) |A_i,\psi\rangle
  = \biggl( D_i\Pi_A^{ia} - ig \pi_\psi t^a \psi
  \biggr)|A_i,\psi\rangle\;,
\end{equation}
which should be vanishing for physical states selected by
$\calP_{\calG}$.  We note that $t^a$'s are elements of su($\Nc$)
algebra in the fundamental representation.  Then, we can introduce a
Lagrange multipliers $\Theta^a(\bx)$ to express
$\calP_{\calG}$ in the following way,
\begin{equation}
  \calP_{\calG} = \int\calD\Theta\,\mu[\Theta]\,
  \exp\biggl[ -i\int d^3 x\, \Theta^a(\bx)\,\calG^a(\bx)\biggr]\;,
\end{equation}
where $\mu[\Theta]$ represents an appropriate integration measure.  If
we adopt a gauge invariant measure or the Haar measure for the
SU($\Nc$) group integration, as is obvious from discussions above, the
gauge is completely fixed and then $\mu[\Theta]$ turns out to coincide
with the Faddeev-Popov determinant in a certain gauge.  In many cases
it would be more convenient to formulate quantized QCD with more general
Faddeev-Popov determinant rather than a special choice of the gauge
fixing as we have employed above.  For this purpose we extend the
interpretation of $\mu[\Theta]$ to include the gauge fixing constraint
and the Faddeev-Popov determinant.  We then insert the projection
operator at each ``thermal time'' slice and convert the partition
function into a functional representation of
\begin{align}
  Z_{\text{QCD}} &= \lim_{n\to\infty} \int\calD A_i \calD\psi\,
  \langle A_i,-\psi| \Bigl( e^{-\frac{\beta}{n}H}\calP_{\calG}
  \Bigr)^n |A_i,\psi\rangle \notag\\
  &= \int_{\text{(anti-)periodic}}\!\!\!\!\!
  \calD A_\mu \calD\psi \calD\bar{\psi}\,
  \delta(G[A])\,\det\biggl(D_\mu^a
  \frac{\partial G}{\partial A_\mu^a}\biggr)\,
  e^{-S_{\rm G}[A] - S_{\rm F}[A,\bar{\psi},\psi]}\;,
\label{eq:ZQCD}
\end{align}
where the gluon fields are periodic, $A_\mu(x_4=\beta)=A_\mu(x_4=0)$,
while the quark fields are anti-periodic,
$\psi(x_4=\beta)=-\psi(x_4=0)$.  We note that we renamed as
$\Theta\to A_4$ and $i\pi_\psi\gamma^0\to\bar{\psi}$ and replaced
$\mu[\Theta]$ with a conventional set in the Faddeev-Popov
quantization procedure with an arbitrary gauge fixing function $G[A]$.
As a consequence of above manipulations, the theory appears in
Euclidean spacetime and the corresponding actions are
\begin{equation}
  S_{\rm G} = \int^\beta d^4x\,\frac{1}{4}F_{\mu\nu}^a
  F_{\mu\nu}^a\;,\qquad
  S_{\rm F} = \int^\beta d^4x\,\bar{\psi}(i\feyn{D}-m)\psi\;,
\end{equation}
with a short-handed notation; $\int^\beta d^4x\equiv \int_0^\beta dx_4
\int d^3x$.  
For Dirac matrices in the Euclidean space, 
 we use the same convention as in the standard
textbook~\cite{bellac2000thermal} where
$\gamma_4^{\rm E}=i\gamma^0$ and $\gamma_i^{\rm E}=\gamma^i$, so that
$\{\gamma_\mu^{\rm E},\gamma_\nu^{\rm E}\}=-2\delta_{\mu\nu}$, which
is sometimes referred to as the anti-Hermitian convention, for all
$\gamma_\mu^{\rm E}$'s are anti-Hermitian matrices.  Hereafter we use
only the Euclidean notation throughout this review and drop the superscript E. 
We also comment that, according to the
above-mentioned convention, 
the covariant derivative should be changed 
from $\partial_i-igA_i$ in original Minkowskian spacetime to
$\partial^i+igA^i$ in the imaginary-time formalism.  We, however, keep
using the common convention $\partial_i-igA_i$ in both cases by
changing the sign of $g$.  Also at finite  chemical potential
$\mu$, we can simply replace $\partial_4\to\partial_4-\mu$.  This means
that the chemical potential is to be identified as an imaginary
component of the Euclidean gauge field, i.e., $\mu\sim -\Im(g A_4)$.

In frequency space the periodic boundary condition for $A_\mu$ and the
anti-periodic boundary condition for $\psi$ imply that the frequencies
for $A_\mu$ and $\psi$ are discretized as $\omega_n=2\pi n T$ (bosonic
Matsubara frequency) and $\omega_n=2\pi(n+1/2) T$ (fermionic Matsubara
frequency), respectively.  Following the well-known trick with
auxiliary fields we can reformulate Eq.~\eqref{eq:ZQCD} into a more
familiar form with ghost $c$ and anti-ghost $\bar{c}$ fields.
Then, it is quite non-trivial which of the periodic and anti-periodic
boundary conditions the ghost fields should obey.  From an intuitive
argument that the ghost contributions should cancel unphysical
polarizations in $A_\mu$, one would presume that the ghost fields
should be periodic, but the question is how to justify it from the
field theoretical point of view.  To this end, we should have started
with the BRST quantized Hamiltonian in Minkowskian spacetime instead
of choosing the Weyl gauge.  In the BRST quantization procedure,
similarly to $\calP_{\calG}$, the physical states are picked up by the
operator $\calP_{\text{BRST}}$ that projects out states with zero
ghost number.  Then, using the BRST charge
$Q_{\rm B}$ and the ghost number $Q_{\rm C}$, the quantized QCD
partition function takes the following form~\cite{Hata:1980yr}:
\begin{equation}
  Z_{\text{QCD}} = \tr\Bigl[ (1-\{Q_{\rm B},R\})\,
  e^{-\pi Q_{\rm C}}\, e^{-\beta H} \Bigr]\;,
  \label{Eq:ZQ}
\end{equation}
with some operator $R$ that has the ghost number $-1$ (where an exact
form of $R$ is irrelevant here).  Because
$\calP_{\text{BRST}}=1-\{Q_{\rm B},R\}$ already selects only
$Q_{\rm C}=0$ states out, the insertion of $e^{-\pi Q_{\rm C}}=1$ is
formally trivial, but its presence significantly simplifies the final
form.  Then, $\{e^{-\pi Q_{\rm C}}, Q_{\rm B}\} = 0$ follows from
$e^{i\pi}=-1$ and the algebra, $[Q_{\rm C}, Q_{\rm B}]=-iQ_{\rm B}$. 
Hence, owing to the insertion $e^{-\pi Q_{\rm C}}$, 
the $\{Q_{\rm B}, R\}$ part in Eq.~\eqref{Eq:ZQ} does not contribute. 
Thus, the partition function is $Z_{\text{QCD}}=\tr(e^{-\pi Q_{\rm C}-\beta H})$, in
which the remnant of $\calP_{\rm BRST}$ is interpreted as a ghost
chemical potential, namely, $\mu_{\text{ghost}}=i\pi T$.  This
imaginary chemical potential shifts the ``fermionic'' Matsubara
frequency of the ghost fields by $-\pi T$, so that the ghost fields
have the ``bosonic'' Matsubara frequency after all (see discussions in
Ref.~\cite{Hata:1980yr} for more details).

Let us now evaluate the free energy $F_q(\br)$ when a static test
quark is placed at $\bx=\br$.  The Gauss law constraint is modified
as
\begin{equation}
  \calG^a(\bx)|A_i,\psi;q^b(\br)\rangle
  = \biggl[ D_i \pi_A^{ia} - ig\pi_\psi t^a\psi
  + g t^a\delta^{ab}\delta(\bx-\br) \biggr]
  |A_i,\psi;q^b(\br)\rangle\;.
\end{equation}
The color-averaged free energy is then given
by~\cite{McLerran:1980pk,McLerran:1981pb}
\begin{equation}
  e^{-\beta F_q(\br)} = \int\calD A_i \calD\psi \sum_{a=1}^{\Nc}
  \langle A_i,-\psi;q^a(\br)| e^{-\beta H} \calP_{\calG}
  |A_i,\psi;q^a(\br)\rangle
  = \langle \tr_c\, L_3(\br)\rangle\;,
\label{eq:F_q}
\end{equation}
where $\tr_c$ explicitly indicates that this trace is taken in color
space only.  Here, $L(\br)$ is called the Polyakov
loop~\cite{Polyakov:1978vu} defined as
\begin{equation}
  L_3(\br) = \calP \exp\biggl[ ig\int_0^\beta dx_4\, A_4(\br,x_4)
    \biggr]
\label{eq:polyakov}
\end{equation}
in the fundamental representation.  The path ordering $\calP$ appears
naturally from the construction of the path integral.  Replacing
$A_4=A_4^a t^a$ with $A_4^a T^a$, where $T^a$ are elements of
su($\Nc$) algebra in the adjoint representation, we can also define
the adjoint Polyakov loop $L_8$ likewise.

The free energy $F_{q\bar{q}}(\br)$ in the
presence of a static test quark at $\bx=0$ and a static test
anti-quark at $\bx=\br$ can also be computed  by taking the color average separately for
the quark and the anti-quark 
\begin{equation}
  e^{-\beta F_{q\bar{q}}(\br)} = \langle \tr_c\, L_3^\dag(\br)\;
  \tr_c\, L_3(0)\rangle\;.
\end{equation}
Furthermore, it would be convenient to define the traced Polyakov
loops and its expectation values for general representation $\calR$ as
\begin{equation}
  \ell_{\calR}(\bx) = \frac{1}{d_{\calR}}\tr_c\, L_{\calR}(\bx)\;,\qquad
  \Phi_{\calR}(\bx) = \langle \ell_{\calR}(\bx)\rangle\;.
\end{equation}
Here, $d_{\calR}$ denotes the dimension of the representation
$\calR$, so that $\ell_{\calR}$ becomes the unity for vanishing gauge
field (apart from renormalization).  For the quark dynamics we mostly
deal with the fundamental (triplet) representation only, and we often
drop the subscript 3 and use simplified notations in this review such
as $\ell = \frac{1}{\Nc}\tr_c\, L$ and $\Phi=\langle\ell\rangle$ and
so on.  Finally we here make a quick remark that $\Phi_{\calR}$ takes
a real-valued number in general, while $\ell_{\calR}$ can be complex.
We will discuss this point in detail when we consider finite-density
systems.

\subsection{Center Symmetry}
\label{sec:center}

The Polyakov loop is a gauge invariant quantity and nevertheless it is
sensitive to the boundary condition of the gauge transformation.
Under the gauge transformation with a transformation matrix
$V(x)\in$ SU($\Nc$), the gluon fields change as
\begin{equation}
  A_\mu(x) \;\to\; A_\mu'(x) = V(x)\biggl( A_\mu(x)
  -\frac{1}{ig}\partial_\mu \biggr) V^\dag(x)\;,
\end{equation}
which does not affect the action.  However, the boundary condition may
be changed by the boundary property of $V(x)$.  In principle we can
abandon the periodicity of $A_\mu'(x)$, and then the theory is put on
a different manifold than $S^1\times R^3$ (see
Ref.~\cite{James:1990it} for such an example).  If we prefer to keep
the same computational rules on the same manifold, we need to extend
the meaning of symmetry including the manifold structure.  This would
require,
\begin{equation}
  A_\mu'(x_4=\beta) = A_\mu'(x_4=0)\;.
\end{equation}
What is interesting is that this requirement for $A_\mu'(x)$ does not
necessarily impose the periodicity of $V(x)$, that is, a twisted
boundary condition is allowed as
\begin{equation}
  V(x_4=\beta) = z_k\cdot V(x_4=0)\;,\qquad
  z_k = e^{i2\pi k/\Nc}
\label{eq:centertrans}
\end{equation}
for $k=0, 1, \dots, \Nc-1$.  It should be noted that
$\det(z_k V) = (z_k)^{\Nc}\det V=1$.  Because $z_k$'s belong to the
center group $Z_{\Nc}$ of the SU($\Nc$) gauge group, the invariance
under such a gauge transformation with a twist by $z_k$ is called
center symmetry~\cite{Svetitsky:1982gs}.

It would be instructive to take an example of $V(x)$ to deepen some
intuitive understanding on center symmetry.  The simplest example
would be
\begin{equation}
  V(x_4) = \diag\bigl[ e^{i2\pi k x_4/(\beta\Nc)},
  e^{i2\pi k x_4/(\beta\Nc)}, \dots,
  e^{-i2\pi(\Nc-1)k x_4/(\beta\Nc)} \bigr]\;,
\end{equation}
which belongs to SU($\Nc$) for any $x_4$.  Obviously
$V(x_4=0)=1_{\Nc\times\Nc}$ and $V(x_4=\beta)=z_k V(0)$.
Then, $A_4$ is shifted by this center twisted gauge transformation $V$
as
\begin{equation}
  A_4  \;\to\; A_4' = A_4 - \frac{2\pi}{g\beta \Nc}\diag\bigl[ k, k,
  \dots, -(\Nc-1)k \bigr]\;.
\label{eq:period}
\end{equation}
Therefore, the center transformation amounts to a discretized
displacement in $A_4$, and usually such a constant shift or the
so-called large gauge transformation is not relevant for physical
quantities.  However, it is quite easy to see from the
definition~\eqref{eq:polyakov} that the Polyakov loop changes as
\begin{equation}
  L_3 \;\to\; L_3' = z_k\cdot L_3\;.
\end{equation}
As long as center symmetry is not broken, therefore, the Polyakov loop
expectation value, $\Phi_3$, is zero, and this leads to the following
conclusion [see Eq.~\eqref{eq:F_q}];
\begin{equation}
  \text{(center symmetry)} \;\to\; \Phi_3=0 \;\to\;
  F_q = \infty\;,
\end{equation}
which is interpreted as realization of \textit{quark confinement}.  To
summarize the above discussions, the fundamental Polyakov loop
expectation value is an order parameter for center symmetry breaking,
and the confined phase corresponds to the center symmetric vacuum.

It would be an interesting question what happens for gluons in the
adjoint representation.  Using an identity, $\tr L_8 = |\tr L_3|^2 - 1$,
we see that, with a decomposition to the disconnected part as
$\langle|\tr L_3|^2\rangle
= \Nc^2 \Phi_3^2 + \langle |\tr L_3|^2\rangle_{\rm c}$, the adjoint
Polyakov loop is always non-zero, i.e.
\begin{equation}
 \Phi_8 = \frac{\Nc^2}{\Nc^2-1} \Phi_3^2
  + \frac{1}{\Nc^2-1}\bigl( \langle|\tr L_3|^2\rangle_{\rm c}
  - 1\bigr) \;.
\label{eq:Phi8}
\end{equation}
We can give a clear interpretation for the fact that $\Phi_8$ is not
necessarily zero even in the center symmetric vacuum in which
$\Phi_3=0$;  (test) gluons can be always screened by gluons (in a
medium) to form a color singlet with a finite energy, while it was
impossible for a test quark in a gluonic medium.  In other words,
gluons are not sensitive to the center of the gauge group (note that
the adjoint representation is not a faithful representation by the
center of the original group).  The important message is that we
cannot construct an order parameter for \textit{gluon confinement} in
this way.  However, Eq.~\eqref{eq:Phi8} implies that $\Phi_8$ can
behave like an order parameter $\propto \Phi_3^2$ in the large $\Nc$
limit and it may approximately work even for $\Nc=3$.

One might have thought that $\Phi$ looks like a spin variable and the
confined phase may well be characterized by a disordered state in a
corresponding spin system.  The possible connection between hot QCD
and spin systems would be transparent in the original argument by
Polyakov~\cite{Polyakov:1978vu}.  For the rest of this subsection, we
will take a quick look at the original idea, which is actually quite
useful to understand why the QCD \textit{disordered} state appears in
\textit{low} $T$.  To address a possible phase transition of quark
deconfinement at high $T$, it is first indispensable to setup a
theoretical description of confined matter at low $T$.  One way to do
this is to use the strong coupling expansion.  In the leading order of
the strong coupling expansion in the Hamiltonian formalism, only the
chromo-electric fields $\sim g^2 E^2$ contribute to the partition
function and the chromo-magnetic fluctuations $\sim g^{-2}B^2$ are
negligible.  The leading-order QCD partition function at temperature
$T$ reads;
\begin{equation}
  Z_{\text{QCD}} = \sum_{\{n(\bx)\}} \exp\biggl[ - \frac{g^2}{2T}\sum_{\bx}
  \bn(\bx)^2 \biggr]\,\delta(\bnabla\cdot\bn)\;,
\end{equation}
where $\bn(\bx)^2$ denotes an electric flux squared with the trace
over color implicitly taken.  At strong coupling the gauge group is
irrelevant and even the compact U(1) theory realizes confinement via
Dirac monopoles.  In what follows let us consider the simple compact
U(1) gauge theory and only the quantized Dirac strings for $\bn(\bx)$
discarding continuous fluctuations.  Now, for notational simplicity,
we choose the lattice spacing so that $\bn(\bx)$ is quantized to be
integral numbers.  The delta functional constraint represents the
Gauss law, for which we can introduce an auxiliary field $\phi$ as
$\delta(\bnabla\cdot\bn)\to\int \calD\phi\,\mu(\phi)\,
\exp[i\sum \phi(\bnabla\cdot\bn)]$.
Clearly, $\phi$ plays the same role as $\Theta$ in our previous
discussions and it should be interpreted as $A_4$ after all.

Using Poisson's resummation formula,
\begin{equation}
  \sum_{n=-\infty}^\infty f(x+nT) = \frac{1}{T}\sum_{m=-\infty}^\infty
  \tilde{f}(k/T)\,e^{i2\pi kx/T}\;,
\end{equation}
where $\tilde{f}(k)$ is a Fourier transform of $f(x)$, we can rewrite
the partition function, apart from irrelevant overall constants, as
\begin{equation}
  Z_{\text{QCD}} = \int \calD\phi\,\mu(\phi) \sum_{\{m(\bx)\}} \exp\biggl[
  -\frac{T}{2g^2}\sum_{\bx} (\bnabla\phi-2\pi \boldm)^2 \biggr]\;.
\end{equation}
This is a well-known form of the Villain approximated XY spin model.
If the coefficient $T/(2g^2)$ is large enough to suppress large
amplitude fluctuations, the Villain approximation works good, and then
we can make an approximation on the above expression to go
\textit{back} to the XY spin model;
\begin{equation}
  Z_{\text{QCD}} \approx \int\calD\phi\,\mu(\phi)\,
  \exp\biggl( \frac{T}{g^2} \sum_{\text{n.n.}} \bS_i\cdot\bS_j \biggr)\;,
\label{eq:XY}
\end{equation}
where the nearest neighbor interaction arises from discretized
$\bnabla$ on the lattice and $\bS_i=(\cos\phi_i, \sin\phi_i)$.  In
this way we can see that the QCD partition function at strong coupling
is mapped onto the XY spin model with the temperature $\propto 1/T$.
In fact, Poisson's resummation formula connects dual theories with
inverse coupling and temperature, and this clearly explains why the
confined phase in low-$T$ QCD looks like a disordered phase that
usually appears in high-$T$ spin systems.  We can readily identify the order
parameter for spontaneous magnetization in this XY model;
$\langle S_x\rangle=\langle \cos\phi\rangle$, which is nothing but the
traced Polyakov loop expectation value.

\subsection{Strong Coupling Potential for the Polyakov Loop}
\label{sec:strong}

The Hamiltonian formalism is suitable to clarify the physical contents
of the dual spin-like theory, as we saw in the previous subsection,
but is not very convenient practically for more systematic
expansions.  Here, let us see another strong coupling expansion to
calculate the effective potential for the Polyakov
loop~\cite{Polonyi:1982wz,Gross:1983ju,Gocksch:1984yk,Fukushima:2002ew,Fukushima:2003fm}.
The strategy is
straightforward;  first we integrate all spatial gauge fluctuations
and leave only $A_4$ unintegrated, and second, we analyze the phase
transition in the mean-field approximation.

In the lattice gauge theory the partition function is formulated in terms
of the link variables, $U_\mu(\bx)=e^{-igaA_\mu(\bx)}$, and the
plaquettes,
$U_{\mu\nu}(\bx)=U_\mu(\bx)U_\nu(\bx+\hat{\mu})
U_\mu^\dag(\bx+\hat{\nu})U_\nu^\dag(\bx)$.
Then, the $A_4$ or $U_4$ unintegrated partition function is defined as
\begin{equation}
  Z[U_4] = \int\calD U_i\, \exp\biggl[ \frac{1}{g^2} \sum_{\bx, \mu>\nu}
  \tr_c (U_{\mu\nu}+U_{\mu\nu}^\dag)\biggr]\;.
\label{eq:ZU4}
\end{equation}
In the leading order of the strong coupling expansion, the first
nonzero term appears from the contribution of $\Nc$ plaquettes aligned
straightly along the $x_4$ direction, i.e.
\begin{equation}
  Z[U_4] \sim \exp\biggl[ \biggl(\frac{1}{g^2\Nc}
  \biggr)^{N_\tau} \sum_{\text{n.n.}} \tr L_3^\dag(\bx')
  \cdot\tr L_3(\bx)\biggr]\;,
\end{equation}
where $L(\bx)=\prod_{x_4}U_4(\bx,x_4)$ is the Polyakov loop on the
lattice and $N_\tau$ is the number of sites along the periodic $x_4$
direction.  Because $T=1/(N_\tau a)$ and the string tension in the
strong coupling limit is $\sigma=a^{-2}\ln(g^2\Nc)$ (with $a$ being the lattice
spacing), we can write the nearest neighbor interaction strength with
physical quantities, and the Polyakov loop effective theory, that is
exactly the counterpart of the XY model in Eq.~\eqref{eq:XY}, takes
the following form;
\begin{equation}
  Z = \int\calD L\, \exp\biggl[ J
  \sum_{\text{n.n.}} \ell_3^\ast(\bx') \cdot \ell_3(\bx) \biggr]
  \qquad (J=\Nc^2\, e^{-\beta\sigma a})\;,
  \label{eq:Zstrong}
\end{equation}
which defines an effective matrix model.  We will closely discuss the
properties of this model in a later section and here we see a phase
transition in the simplest mean-field (tree-level) approximation.

\begin{figure}
  \includegraphics[width=0.47\textwidth]{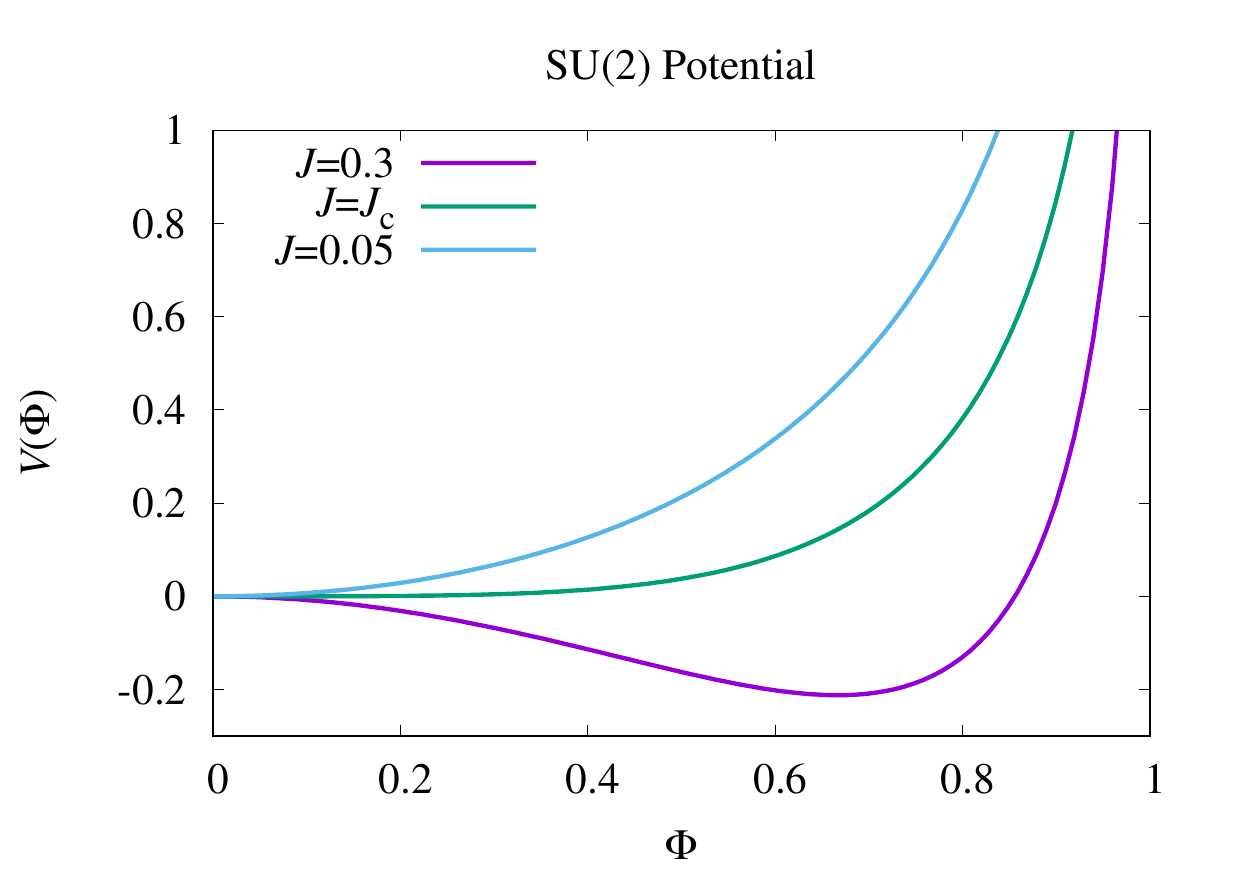}
  \includegraphics[width=0.47\textwidth]{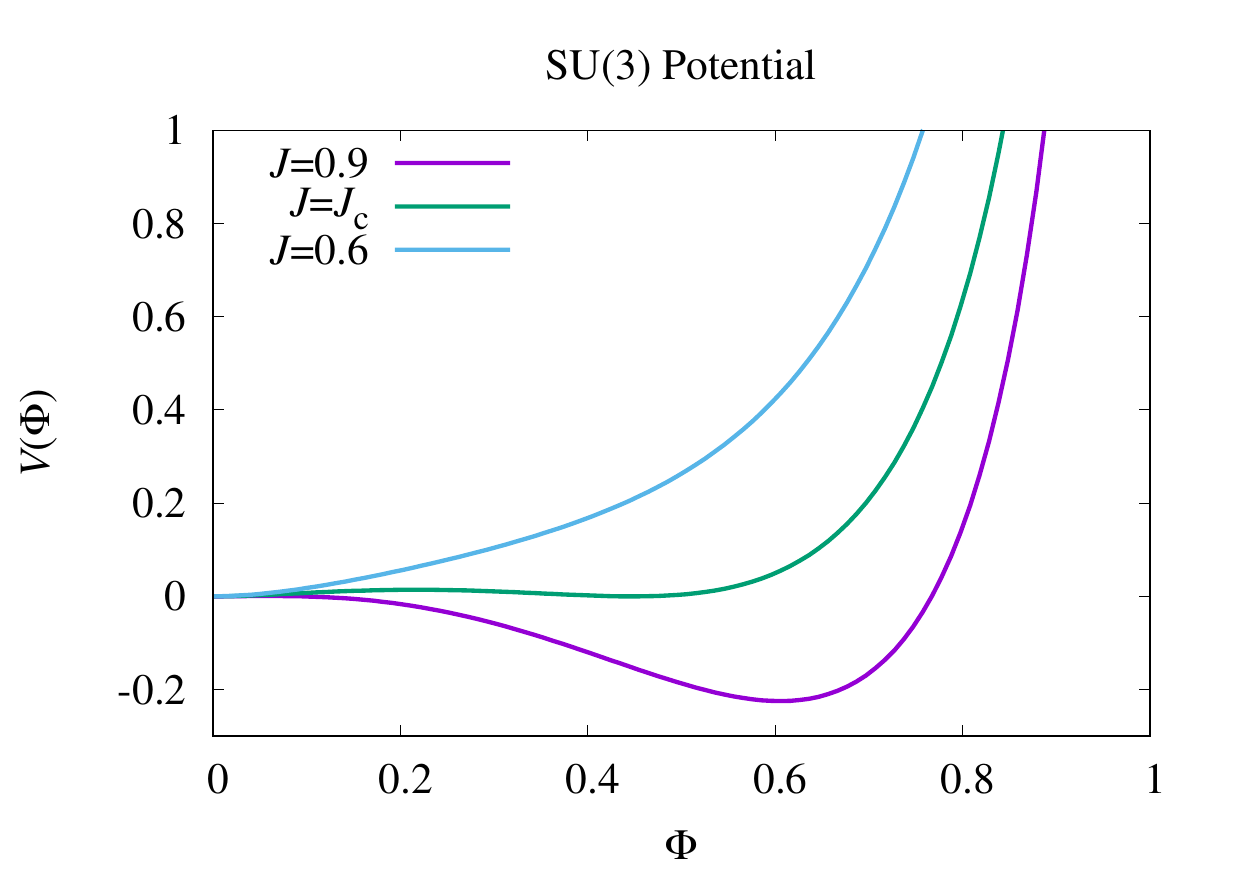}
  \caption{SU(2) (left) and SU(3) (right) effective potentials as a
    function of the Polyakov loop $\Phi$ around the critical value of
    $J$.}
  \label{fig:potential}
\end{figure}

If $J$ is very small, the Polyakov loop expectation value should be
zero, which is caused by the group integration.  Therefore, it is
crucial to include the effect of the group integration that favors the
confined phase.  We can actually take account of this by including the
Haar measure in the effective potential, which is sometimes called the
Vandermonde determinant interaction, for the SU($\Nc$) Haar measure is
given by the Vandermonde determinant.  Then, the effective potential
is,
\begin{equation}
  V[\Phi_3] = -6J |\Phi_3|^2 - \ln H[\Phi_3]\;,
\end{equation}
where $H[\Phi_3]$ represents the Haar measure whose concrete shape
simplifies in the Polyakov gauge in which $A_4$ is diagonal and
static.  In this gauge we can parametrize $A_4$ as
\begin{equation}
  A_4 = \frac{2\pi}{g\beta} \diag(q_1, q_2, \dots, q_{\Nc})
  \qquad \Bigl( \sum_i q_i = 0 \Bigr) \;.
\label{eq:A4}
\end{equation}
Then, the SU($\Nc$) Haar measure associated with the Polyakov loop
integration reads,
\begin{equation}
  H[q] = \prod_{\bx, i<j}
  \Bigl| e^{i 2\pi q_i(\bx)} - e^{i 2\pi q_j(\bx)} \Bigr|^2\;.
\label{eq:Haar}
\end{equation}
For the color SU(2) case we can write the Polyakov loop as
$\Phi_3 = \langle \cos(\pi q)\rangle$ using $q_1=q/2$ and $q_2=-q/2$.
This expression reminds us of $\langle S_x\rangle$ in the XY model
that is dual to hot QCD in the Hamiltonian formalism.  We note that
the Polyakov loop (even before taking the expectation value) always
takes a real value in the SU(2) case, reflecting the pseudo-real
property of the SU(2) group.  In this case, according to
Eq.~\eqref{eq:Haar}, the Haar measure is $H[q]=\sin^2(\pi q)$ and in
the mean-field approximation we have $H[\Phi] = 1-\Phi_3^2$.  The
effective potential including the Haar measure contribution is thus,
\begin{equation}
  V[\Phi_3] = -6J \Phi_3^2 - \ln (1-\Phi_3^2)
  \;\simeq\; (1-6J)\Phi_3^2 + O(\Phi_3^4) \qquad \text{[for SU(2) case]}\;,
\end{equation}
from which we can easily understand that a second-order phase
transition occurs at the critical
coupling, $J_c=1/6$.  The potential behavior near $J_c$ is plotted in
the left of Fig.~\ref{fig:potential}.  It is important to point out
that the perturbative vacuum at $\Phi_3=1$ (or $A_4=0$) is singular
and can never be realized at finite energy.  For the color SU(3) case
$\ell_3$ is generally a complex number, though $\Phi_3$ is real.  This
may in principle induce $\bar{\Phi}_3 \neq \Phi_3$ where
$\bar{\Phi}_3=\langle\ell_3^\ast\rangle$.  Then, again from
Eq.~\eqref{eq:Haar} for the SU(3) Haar measure, the effective
potential in the mean-field approximation reads,
\begin{equation}
  V[\Phi_3] = -6J \bar{\Phi}_3 \Phi_3 - \ln\bigl[ 1-6 \bar{\Phi}_3 \Phi_3
 + 4 (\bar{\Phi}_3^3 + \Phi_3^3) -3 (\bar{\Phi}_3 \Phi_3)^2 \bigr]
  \qquad \text{[for SU(3) case]}\;.
\label{eq:SU3Haar}
\end{equation}
We note that we can safely postulate $\bar{\Phi}_3=\Phi_3$ as long as
the charge parity symmetry holds (that is the case for the pure
gluonic theory).  Because of the presence of the cubic terms,
$\bar{\Phi}_3^3$ and $\Phi_3^3$, the above effective potential has a
first-order phase transition at $J_c\simeq 0.773$.  The potential
behavior near $J_c$ is plotted in the right of
Fig.~\ref{fig:potential}.

So far, we have seen that the perturbative vacuum of $A_4=0$ has an
infinite barrier and its realization is prohibited in the strong
coupling calculation.  Interestingly enough, one can conversely prove
that $A_4=0$ is always the ground state once the Haar measure is
removed by hand from the partition function~\cite{Gocksch:1993iy}, or
in short, one could say; no Haar measure, no confinement!  This
observation strongly suggests that the Haar measure is the
driving-force for quark confinement.

\subsection{Perturbative Potential for the Polyakov Loop}
\label{sec:perturbative}

The perturbative vacuum is by definition empty with $A_\mu=0$, and nevertheless, the
Polyakov loop there is $\Phi=1$, and this means that center symmetry
must be spontaneously broken around $A_4=0$ in the deconfined phase at
high enough temperature.  In other words, if we evaluate the Polyakov
loop effective potential perturbatively, we should find a local
minimum at $A_4=0$.  Let us confirm this in what follows below.

To achieve this goal, let us specify the gauge fixing condition and the
concrete matrix representation of the su($\Nc$) algebra.  The most
convenient choice for systematic higher-order calculations should be
the covariant background gauge for the gauge fixing and the ladder
basis for the representation.  In the background
field method, the gauge fields are split into the quantum fluctuations
$\calA_\mu$ and the background fields $A_{{\rm B}\mu}$, and the
covariant background field gauge is chosen by the condition,
\begin{equation}
  D_{{\rm B}\mu} \calA_\mu = 0\;.
\end{equation}
Then, from the residual symmetry of the effective action
$\Gamma'[\bar{A},A_{\rm B}]$ defined on top of the background
$A_{{\rm B}\mu}^a$, one can prove that
$\Gamma'[\bar{\calA},A_{\rm B}]=\Gamma[\bar{A}=\bar{\calA}+A_{\rm B}]$,
where $\Gamma[\bar{A}]$ is the standard effective action that we want
to know.  From this we see
$\Gamma[A_{\rm B}]=\Gamma'[\bar{\calA}=0,A_{\rm B}]$.  This implies
that we can obtain the effective action by integrating the quantum
fluctuations out around the background fields.  Because we are
interested in the effective potential for the Polyakov loop, we should
take only the temporal component of the background fields and rename
it in the same way as in Eq.~\eqref{eq:A4}, i.e.
\begin{equation}
  A_{{\rm B}4} = \frac{2\pi}{g\beta}\diag(q_1,q_2,\dots,q_{\Nc})
  = \frac{2\pi}{g\beta}\sum_{i=1}^{\Nc} q_i \delta_i \qquad
  \Bigl( \sum_i q_i = 0 \Bigr) \;,
\end{equation}
where, for convenience, we define matrices as
\begin{equation}
  (\delta_i)^{ab} = \begin{cases}
  1 & (a=b=i) \\ 0 & (\text{otherwise})
  \end{cases} \;.
\end{equation}
For the evaluation of the covariant derivative $D_{{\rm B}4}$, the
ladder basis is quite convenient for systematic higher-order
calculations~\cite{KorthalsAltes:1993ca}.  The elements of the Cartan
subalgebra in the ladder basis are defined as
\begin{equation}
  (t_{(n,n)})^{ab} = \frac{\delta^{ab}}{\sqrt{2n(n+1)}}\times
  \begin{cases}
  1 & (a\le n) \\
  -n & (a = n+1)\\
  0 & (n+2\le a \le \Nc)
  \end{cases}
\end{equation}
and off-diagonal ladder elements for $i\neq j$ are
\begin{equation}
  (t_{(i,j)})^{ab} = \frac{1}{\sqrt{2}}\delta^{ai}\delta^{bj}\;.
\end{equation}
and then it is easy to show the following commutation relations,
\begin{equation}
  [\delta_i, t_{(j,k)}] = (\delta_{ij}-\delta_{ik}) t_{(j,k)}\;,\qquad
  [t_{(i,j)}, t_{(k,l)}] = \frac{1}{2}\delta_{il}\delta_{jk}
  (\delta_i - \delta_j)\;.
\end{equation}
Using these commutation relations we can express the covariant
derivative as
\begin{equation}
  D_{{\rm B}4}\calA_\mu = \partial_4 \calA_\mu - ig[A_{{\rm B}4}, \calA_\mu]
  = \partial_4^{(i,j)}\calA_\mu^{(i,j)} t_{(i,j)}\;,
\end{equation}
where $\partial_4^{(i,j)}=\partial_4-2\pi i \delta_{\mu 4} q_{ij}$
with $q_{ij} = q_i - q_j$.  We note that $A_{{\rm B}4}$ or $q_{ij}$
appears like a colored imaginary chemical potential.  Then, the
one-loop integration with respect to $\calA_\mu$ and the ghost fields
leads to the following effective potential,
\begin{equation}
  V_{\rm glue}[q] = \frac{1}{2}\tr\ln\bigl[
    (\partial_4^{(i,j)})^2 + \bnabla^2 \bigr] \cdot
  \tr(\delta_{\mu\nu})
  - \tr\ln\bigl[ (\partial_4^{(i,j)})^2 + \bnabla^2 \bigr]
  = \tr\ln \bigl[ (\partial_4^{(i,j)})^2 + \bnabla^2 \bigr]\;.
\label{eq:glue_oneloop}
\end{equation}
Here the first term multiplied by four polarizations
$\tr(\delta_{\mu\nu})$ results from $\calA_\mu$ fluctuations and the
second term from the ghost fluctuations that eliminate two unphysical
polarizations out from the gluon fluctuations.  From the observation
that $q_{ij}$ is an
imaginary chemical potential, we can immediately conclude that the
above trace in momentum space becomes the grand canonical partition
function with an imaginary chemical potential, that is,
\begin{equation}
  V_{\rm glue}[q] = 2V\int\frac{d^3 p}{(2\pi)^3} \sum_{i>j} \Bigl[
  \ln\bigl( 1-e^{-\beta |\bp| + 2\pi i q_{ij}} \bigr)
  + \ln\bigl( 1-e^{-\beta |\bp| - 2\pi i q_{ij}} \bigr) \Bigr]\;.
\label{eq:weiss_integ}
\end{equation}
We can carry out this momentum integration explicitly, which yields,
\begin{equation}
 V_{\rm glue}^{\rm Weiss}[q]
  = \frac{4\pi^2 V}{3\beta^3} \sum_{i>j} (q_{ij})_{\text{mod1}}^2
  \bigl[ (q_{ij})_{\text{mod1}} - 1 \bigr]^2 \;.
\label{eq:weiss}
\end{equation}
This is often called the (GPY-)Weiss
potential~\cite{Gross:1980br,Weiss:1980rj,Weiss:1981ev} (there are
many derivations and generalizations of the Weiss potential;  for
example, see Refs.~\cite{Megias:2002vr,Megias:2003ui} for the
heat kernel expansion approach to the one-loop effective action).  The modulo
operation is defined as $(q)_{\text{mod1}}=q-\lfloor q\rfloor$, where
$\lfloor\cdots\rfloor$ denotes the floor function.

Surprisingly the recent extension of this result to 2-loop order~\cite{Dumitru:2013xna} 
showed that the shape of the potential remains the same as in Eq.~\eqref{eq:weiss}. 

The Weiss potential has a periodic nature for $q_{ij}$, that is
already obvious in Eq.~\eqref{eq:weiss_integ};  $q_{ij}$ or the
imaginary chemical potential generally appears as an angle variable.
This periodic property is attributed to center symmetry that is a
symmetry associated with a discretized displacement in $A_4$ as we
discussed in Eq.~\eqref{eq:period}.

\begin{figure}
  \includegraphics[width=0.47\textwidth]{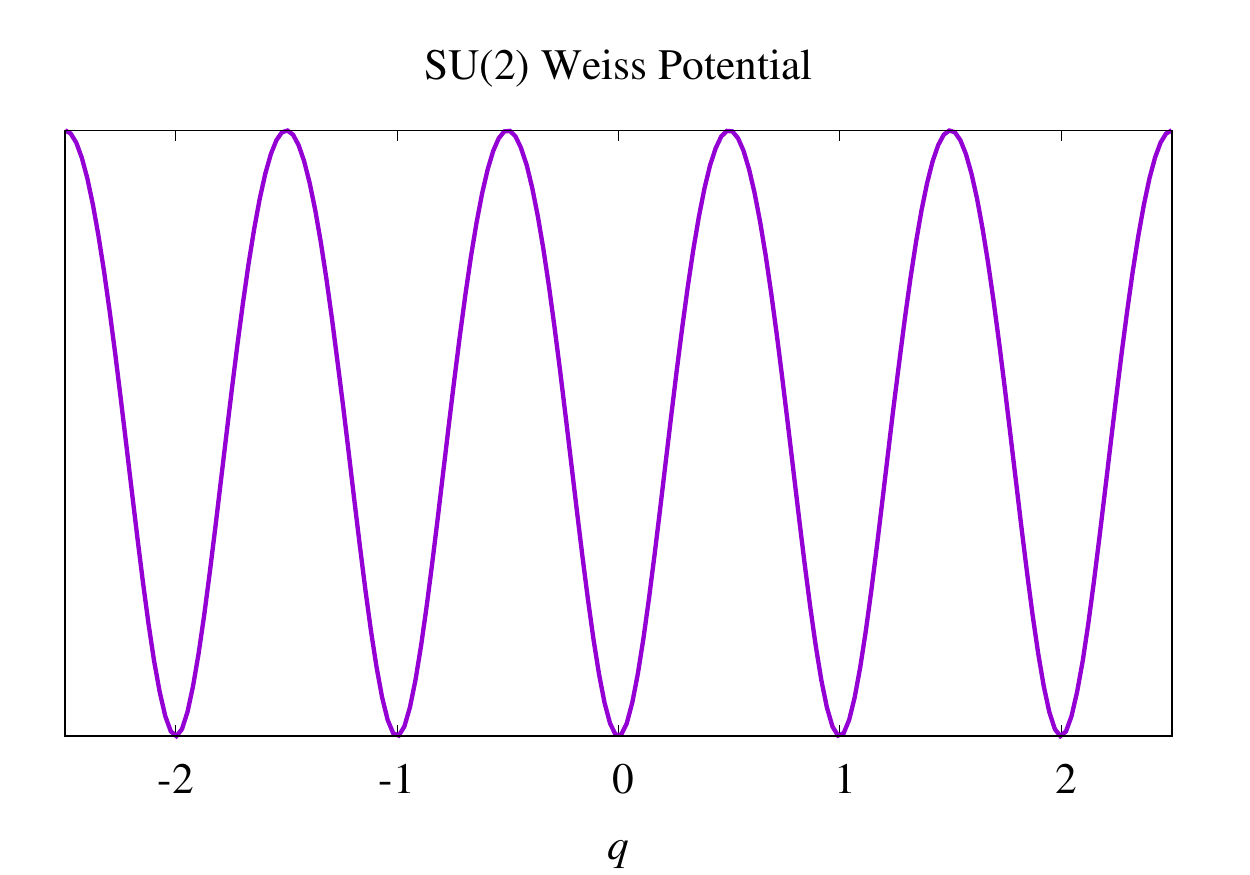}
  \includegraphics[width=0.47\textwidth]{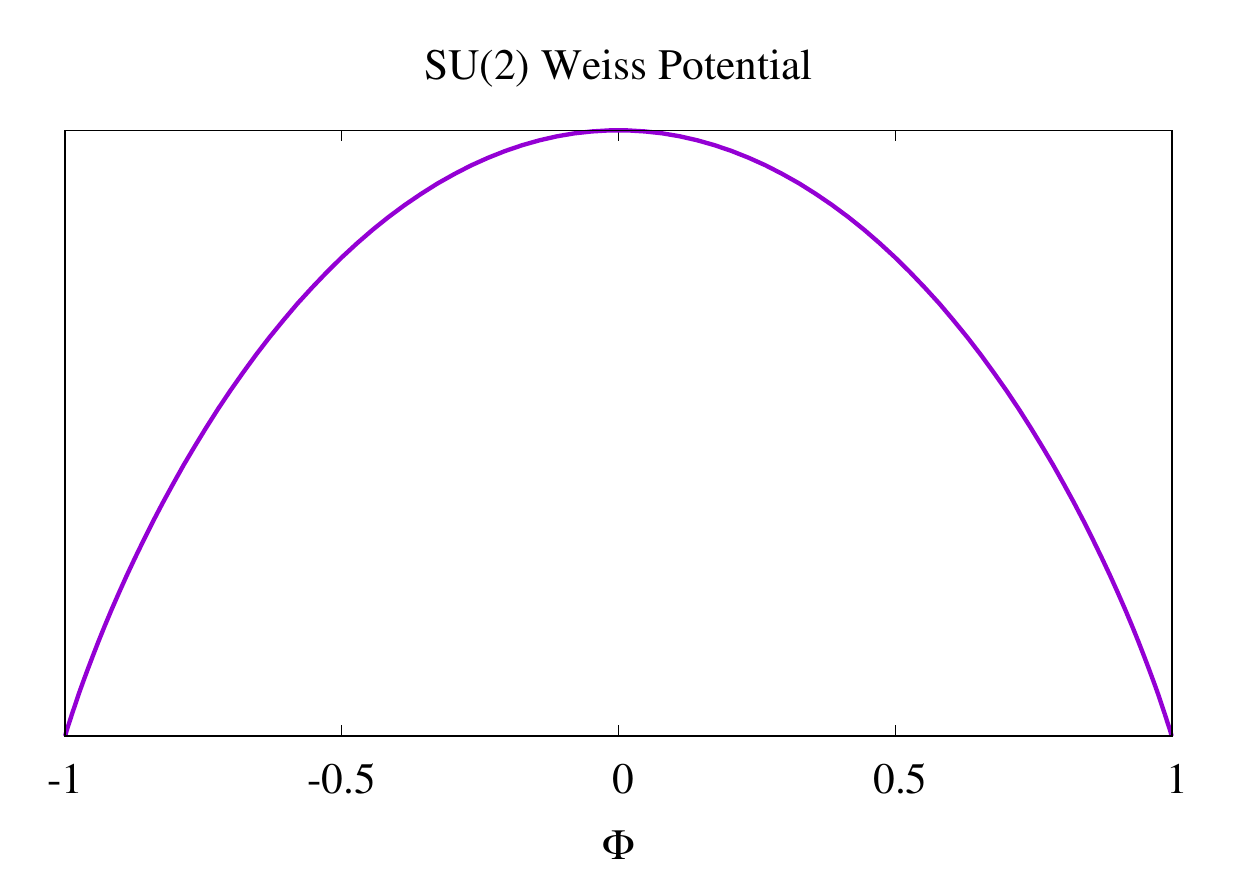}
  \caption{SU(2) Weiss potential as a function of $q$ (left) and
    $\Phi$ (right).}
  \label{fig:su2weiss}
\end{figure}

First let us consider the SU(2) case, for which there is only one
independent variable; $q_1=q/2$ and $q_2=-q/2$.  Then, the SU(2) Weiss
potential has a periodic shape as depicted in the left of
Fig.~\ref{fig:su2weiss}.  We note that the Polyakov loop in this case
is $\Phi=\cos(\pi q)$ (just like the case at strong coupling), and so
a minimum at the perturbative vacuum $q=0$ corresponds to $\Phi=1$,
and center symmetry is spontaneously broken there.  The next minimum
at $q=1$ is a center transformed point with $\Phi=-1$.  One might
think that the perturbation theory may be reformulated around $q=1$
equivalently, but as we see later, the quark
one-loop potential favors $q=0$ and the perturbative vacuum must be
identified at $q=0$ if we assume continuity between the pure gluonic
theory and the massive limit of QCD.\ \ It is straightforward to
change the variable from $q$ to $\Phi$, and the SU(2) Weiss potential
as a function of $\Phi$ is plotted in the right of
Fig.~\ref{fig:su2weiss}.  We can see that the potential minima are
located at $\Phi=\pm 1$, but these points of $\Phi=\pm 1$ look
singular unlike the left of Fig.~\ref{fig:su2weiss}.  Such singular
character originates from the Jacobian from $q$ to $\Phi$, namely,
$dq/d\Phi=-1/[\pi \sin(\pi q)]$, which diverges at $\Phi=\pm 1$.  This
observation of the potential shape is important when we want to
consider the effect of the Polyakov loop fluctuations in physical
observables.

\begin{figure}
  \includegraphics[width=0.47\textwidth]{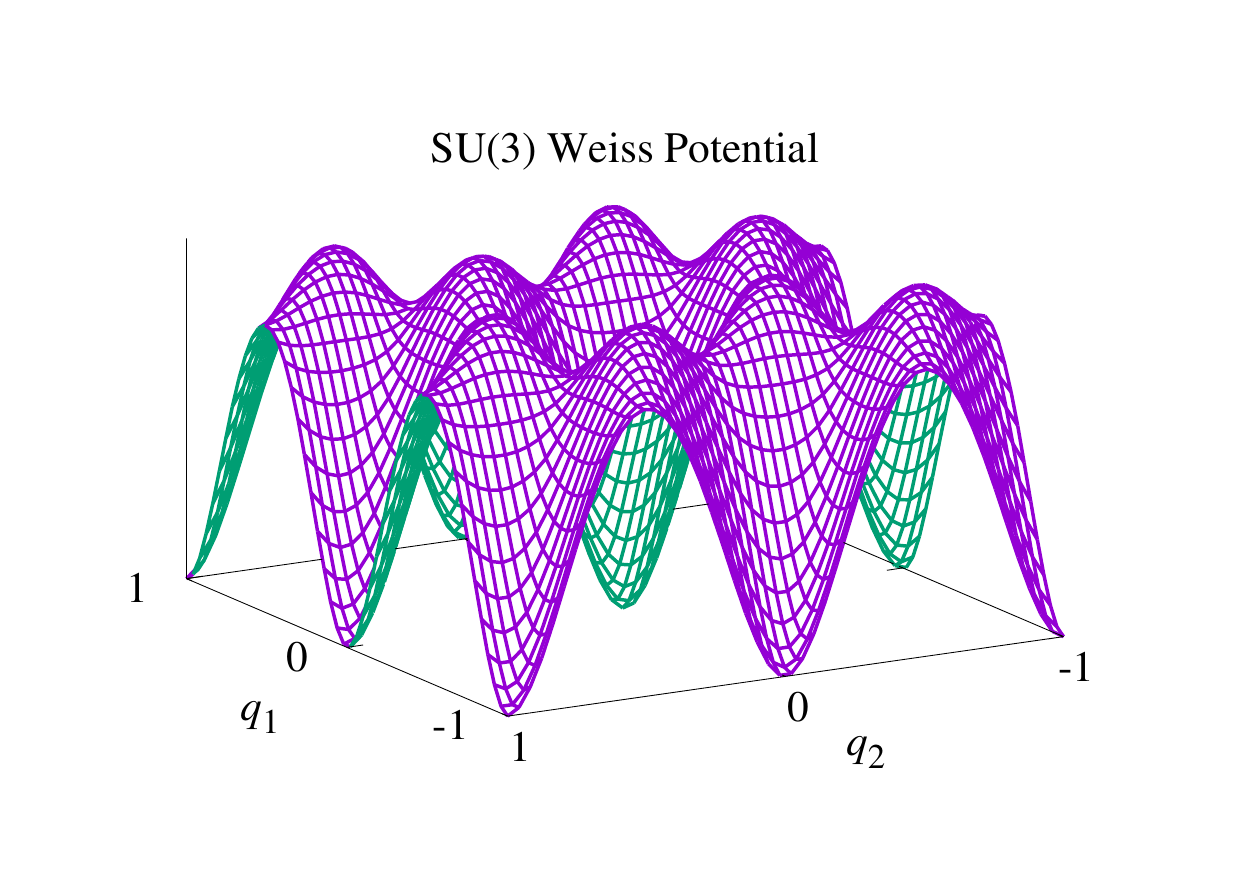}
  \includegraphics[width=0.47\textwidth]{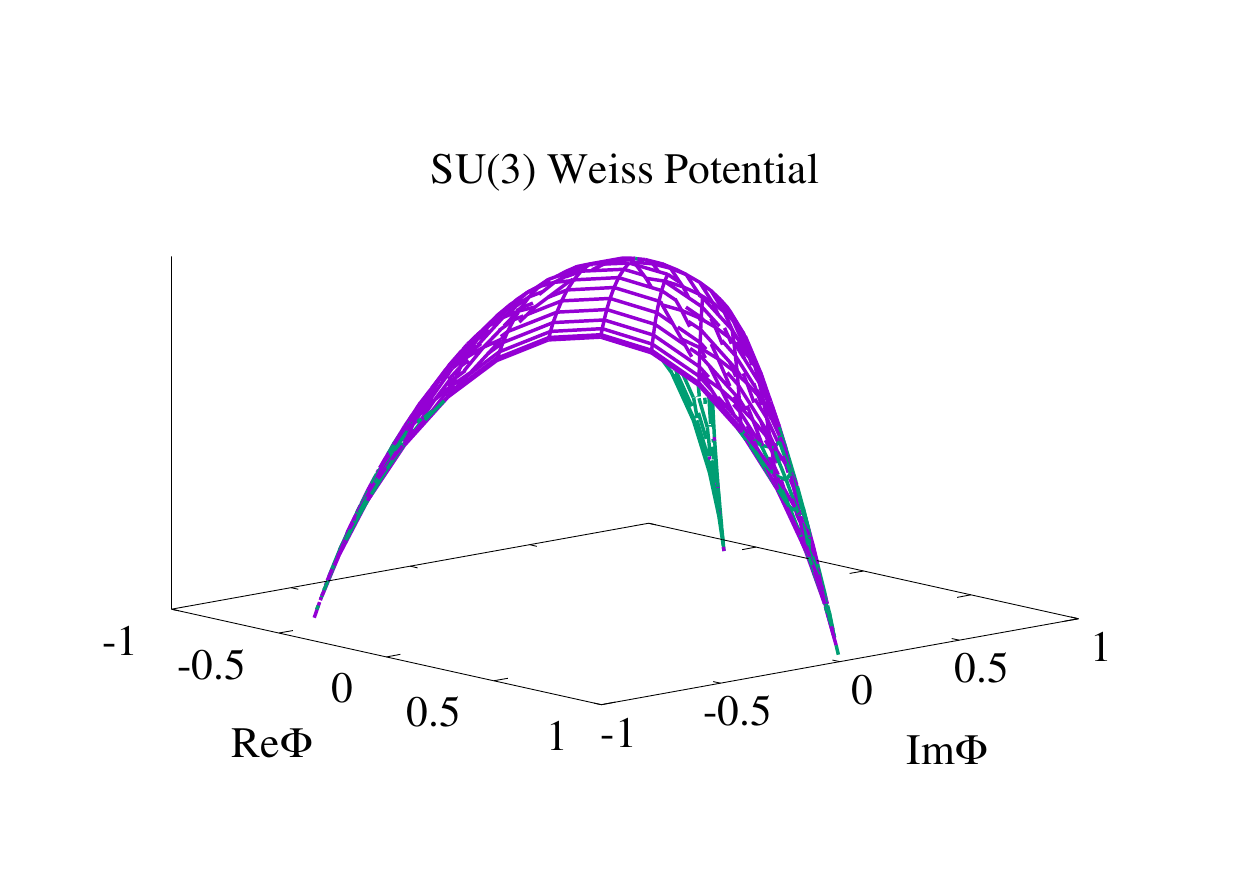}
  \caption{SU(3) Weiss potential as a function of $q_1$ and $q_2$
    (left) and $\Re\Phi$ and $\Im\Phi$ (right).}
  \label{fig:su3weiss}
\end{figure}

Next, the generalization to the SU(3) case is easy to understand.
Now, for the graphical purpose, we choose $q_1$ and $q_2$ as
independent variables and set $q_3=-q_1-q_2$ to draw the SU(3) Weiss
potential in the left of Fig.~\ref{fig:su3weiss}.  We see that one of
the minima is certainly located at the perturbative vacuum $q_1=q_2=0$
and there are degenerate minima at the center transformed points.  It
is not clear which minimum has what value of the Polyakov loop, and so
let us change the variables from $q_1$ and $q_2$ to $\Re\Phi$ and
$\Im\Phi$ as shown in the right of Fig.~\ref{fig:su3weiss}.  In this
case, three points, $\Phi=1$, $e^{2\pi i/3}$, $e^{4\pi i/3}$, are
degenerate and connected by center transformation, among which
$\Phi=1$ is favored by quark loop contributions.

In this way we have confirmed that the perturbative vacuum at $A_4=0$
should be certainly identified as an ordered state with spontaneous
center symmetry breaking.  The potential curvature around the
potential minimum characterizes how strongly symmetry is broken;  in
other words, the Debye screening mass stabilizes the perturbative
vacuum.  From the explicit expression~\eqref{eq:weiss} we can infer
the potential curvature or the Debye screening mass $m_E$ as
\begin{equation}
  \frac{V_{\rm glue}[q]}{\beta V}
  = \frac{4\pi^2}{3\beta^4}\sum_{i>j}(q_{ij})^2 + O(q^3)
  = m_E^2 \tr(A_4^2) = m_E^2 \biggl(\frac{2\pi T}{g}\biggr)^2
    \sum_i q_i^2 \;\Rightarrow\;
  m_E^2 = \frac{\Nc}{3}g^2 T^2\;,
\end{equation}
where we used $\sum_{i>j}(q_{ij})^2 = \Nc\sum_i q_i^2$.  We can
continue such an analysis to read higher-order interaction terms.
The cubic term has an infrared singular origin from infinite sum over
ring diagrams (at zero Matsubara frequency), and we next go to quartic
order;  suppose that the one-loop effective action has quartic terms
such as $\lambda_E(\tr A_4^2)^2 + \bar{\lambda}_E \tr A_4^4$, we can
infer $\lambda_E$ and $\bar{\lambda}_E$ from the Weiss
potential~\eqref{eq:weiss} as
\begin{equation}
  \lambda_E = \frac{g^4}{4\pi^2}\;,\qquad
  \bar{\lambda}_E = \frac{\Nc g^4}{12\pi^2}\;.
\end{equation}
These are exactly the coefficients that appear in the so-called
electrostatic QCD (EQCD).

\subsection{Gauge Configurations and Center Symmetry Restoration}
\label{sec:classical}

In the perturbative regime at high $T$ the Debye screening is the
physical origin of center symmetry breaking and deconfinement.  Then,
one would ask the following question;  what is the physical origin of
confinement at low $T$?  We have seen in Sec.~\ref{sec:strong} that
the Haar measure or the ghost determinant is responsible for
confinement in the strong coupling limit, but confinement should
persist even at weak coupling as long as $T$ is low enough.  We are
now addressing typical gauge configurations as microscopic mechanisms
to restore center symmetry.

\subsubsection{Z($\Nc$) domain walls}
\label{sec:walls}

Slightly above the critical temperature $\Tc$, it is a natural anticipation that
the spatial regions are divided into domains with different center
elements just like the formation of magnetized domains in the spin
systems near $\Tc$.  The possibility of domain-wall formation was
pursued historically in the
cosmological context~\cite{Witten:1984rs} and in
the heavy-ion collision experiment recently~\cite{Asakawa:2012yv},
which is one concrete manifestation of a more general idea of the
semi-QGP that will be argued in Sec.~\ref{sec:semiQGP}.

Now that we have the explicit form of the Weiss potential,
it is easy to find a domain-wall solution of the equation of
motion~\cite{Bhattacharya:1990hk,Bhattacharya:1992qb}.  Let us find
the $Z(2)$ domain wall along the $z$ direction for the color SU(2)
case.  We take the boundary condition as $q(z=-\infty)=0$
(i.e.\ $\Phi=1$) and $q(z\to\infty)=1$ (i.e.\ $\Phi=-1$).  The
effective action with the kinetic term added should be
\begin{equation}
  \Gamma_{\text{glue}}[q] = L_x L_y \int_{-L_z/2}^{L_z/2} dz\,\Biggl\{
  \frac{2\pi^2}{g^2\beta}\biggl[\frac{d q(z)}{dz}\biggr]^2
  + \frac{4\pi^2}{3\beta^3} q(z)^2 \bigl[ 1-q(z) \bigr]^2 \Biggr\}\;.
\label{eq:oneloopG}
\end{equation}
Because we are thinking of a limited range $0\le q\le 1$, we do not
have to put the modulo operator.  The equation of motion leads to the
following solution;
\begin{equation}
  q_c(z) = \frac{1}{1+\exp\Bigl(-\sqrt{\frac{2}{3}}g Tz\Bigr)}\;,
\label{eq:domainwall}
\end{equation}
which is, together with the corresponding Polyakov loop value, plotted
in the left of Fig.~\ref{fig:domain}.  We can evaluate the on-shell
value of the effective action as
\begin{equation}
  \Gamma_{\rm glue}[q_c] = \sigma_t\, (\beta L_xL_y)
  = \frac{4\pi^2 T^2}{3\sqrt{6}g} L_x L_y\;.
\end{equation}
In this way we can estimate the interface surface tension $\sigma_t$
of the $Z(2)$ domain-wall perturbatively, which can be generalized to
SU($\Nc$) as~\cite{Bhattacharya:1990hk,Bhattacharya:1992qb}
\begin{equation}
  \sigma_t = \frac{4(\Nc-1)\pi^2 T^3}{3\sqrt{3\Nc} g}\;.
\end{equation}
It is important to note that the center domain-wall is accompanied by the
magnetic loop or the 't~Hooft loop that is a dual of the Wilson
loop~\cite{'tHooft:1977hy}.

To have an intuitive feeling about how the center domain-wall and the
't~Hooft loop are linked, it would be useful to consider the following
transformation matrix in the Abelian part,
\begin{equation}
  V(z') = \exp\biggl[ -\int d\bx\,\delta(z-z')
  \frac{\delta}{\delta a_z(\bx)}
  \biggr]\;.
\end{equation}
In the same manner as $A_4^3$, we defined $a_z$ from
$A_z^3=\frac{2\pi}{g}a_z$.  Because $V(x)$ is a shift operator, the
unitary transformation (not the gauge transformation) leads to
\begin{equation}
  a_z'(\bx) = a_z(\bx) - \delta(z-z')\;.
\end{equation}
Thus, if we define the Wilson loop as
$W = \exp(ig\int dz\, A_z^3 \frac{\sigma_3}{2})$, it is easy to show,
\begin{equation}
 V(z') W V^\dag(z') = e^{-i\pi\theta(z-z')} W\;.
\end{equation}
Actually, in the canonical quantization, $\delta/\delta a_z$ is
nothing but $-i\frac{2\pi}{g}E_z$ and then the matrix can be rewritten
in a form of the surface integral on $z=z'$ as
$V(z')=\exp[i\frac{2\pi}{g}\int dS_z E_z(z')]$.  To pick up such an
electric flux, we can regard this expression of $V$ as a dual
counterpart of the Wilson loop with magnetic loops on the edges of the
$xy$ plane.  In this sense, $V$ is called the 't~Hooft loop, named
after the pioneer who studies such a \textit{disordered} parameter
first~\cite{'tHooft:1977hy}.

Now let us see that $V(z')$ turns out to be a creation operator of the
center domain-wall at $z=z'$, which can be understood from the
computation of the expectation value, $\langle V(z')\rangle$.  In the
functional integration we should replace $\delta/\delta a_z$ by
$\Pi_z$ which should be integrated out, and after all we reach an
expression, $\langle V(z')\rangle = e^{-\tilde{\Gamma}}$, where
\begin{equation}
  \tilde{\Gamma} = L_x L_y \int_{-L_z/2}^{L_z/2} dz\,\Biggl\{
  \frac{2\pi^2}{g^2\beta}\biggl[\frac{dq(z)}{dz}
  - \delta(z-z') \biggr]^2
  + \frac{4\pi^2}{3\beta^3} q(z)^2 \bigl[ 1-q(z) \bigr]^2 \Biggr\}
  \Biggr|_{\text{minimizing $q$}}\;.
\end{equation}
Here, an additional term by $\delta(z-z')$ imposes the boundary
condition as $q(z\to-\infty)=0$ and $q(z\to+\infty)=1$ for a
configuration that minimizes the above effective action.  We already
know such a configuration that is the center domain-wall, and then the
expectation value of $V$ reads,
\begin{equation}
  \langle V \rangle = e^{-\sigma_t \beta L_x L_y}\;.
\end{equation}
We note that the above expression implies that the 't~Hooft loop shows
the area law in the deconfined phase at high temperature.  So, unlike
the string tension in the Wilson loop, $\sigma_t$ in the 't~Hooft loop
is a perturbatively calculable quantity.  For the
non-perturbative calculation in the lattice-QCD simulation, instead of
using the definition of $V$ in an operator form, one can impose a
domain-wall boundary condition as argued in
Ref.~\cite{deForcrand:2005pb}.

\subsubsection{Z($\Nc$) bubbles and calorons}
\label{sec:calorons}

In the thermodynamic limit an infinitely large domain-wall solution is
not stable because its action or energy diverges 
with $L_x L_y \to \infty$.  
Therefore, in reality, a number of finite-size domain-walls
distribute in space, and its length becomes large only near $\Tc$
where the surface tension is suppressed.

\begin{figure}
  \begin{center}
  \includegraphics[width=0.47\textwidth]{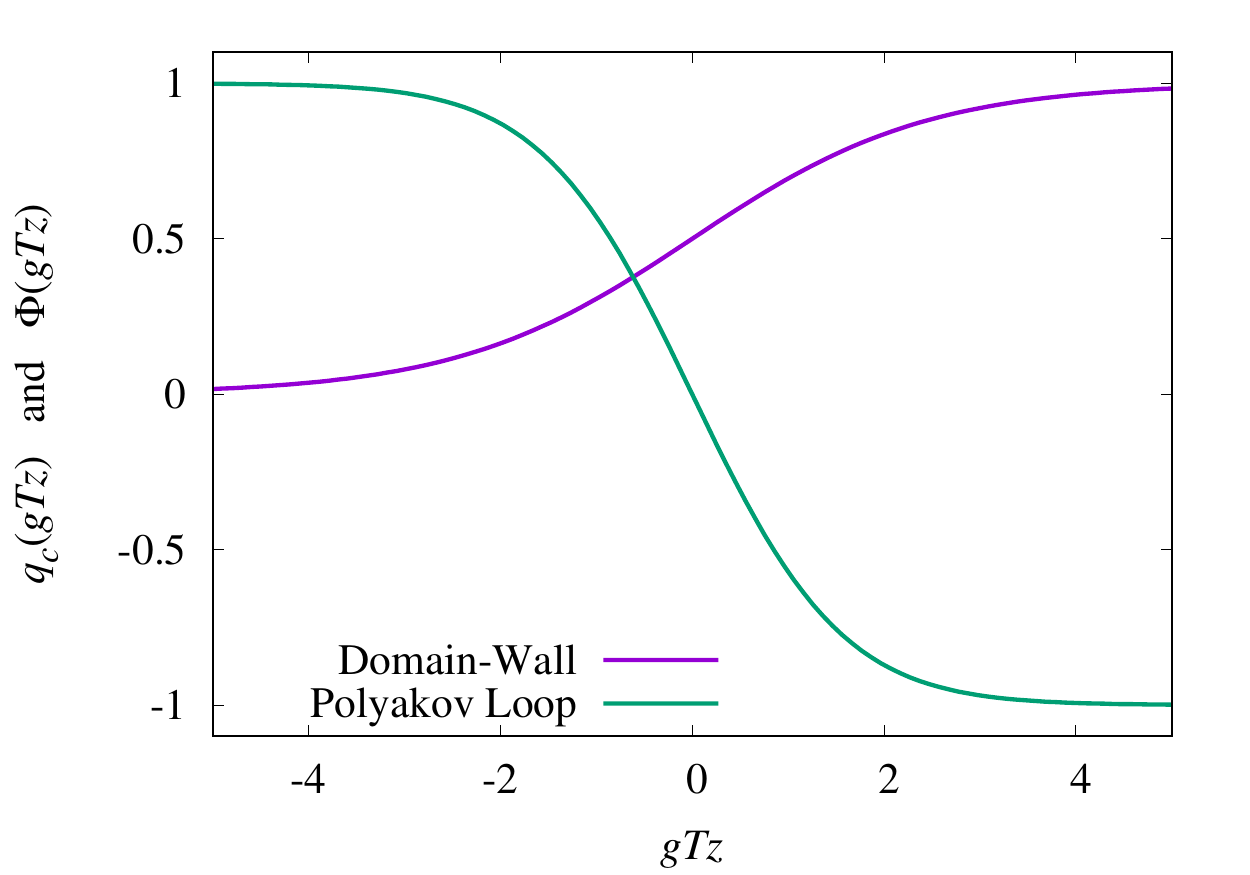}
  \includegraphics[width=0.47\textwidth]{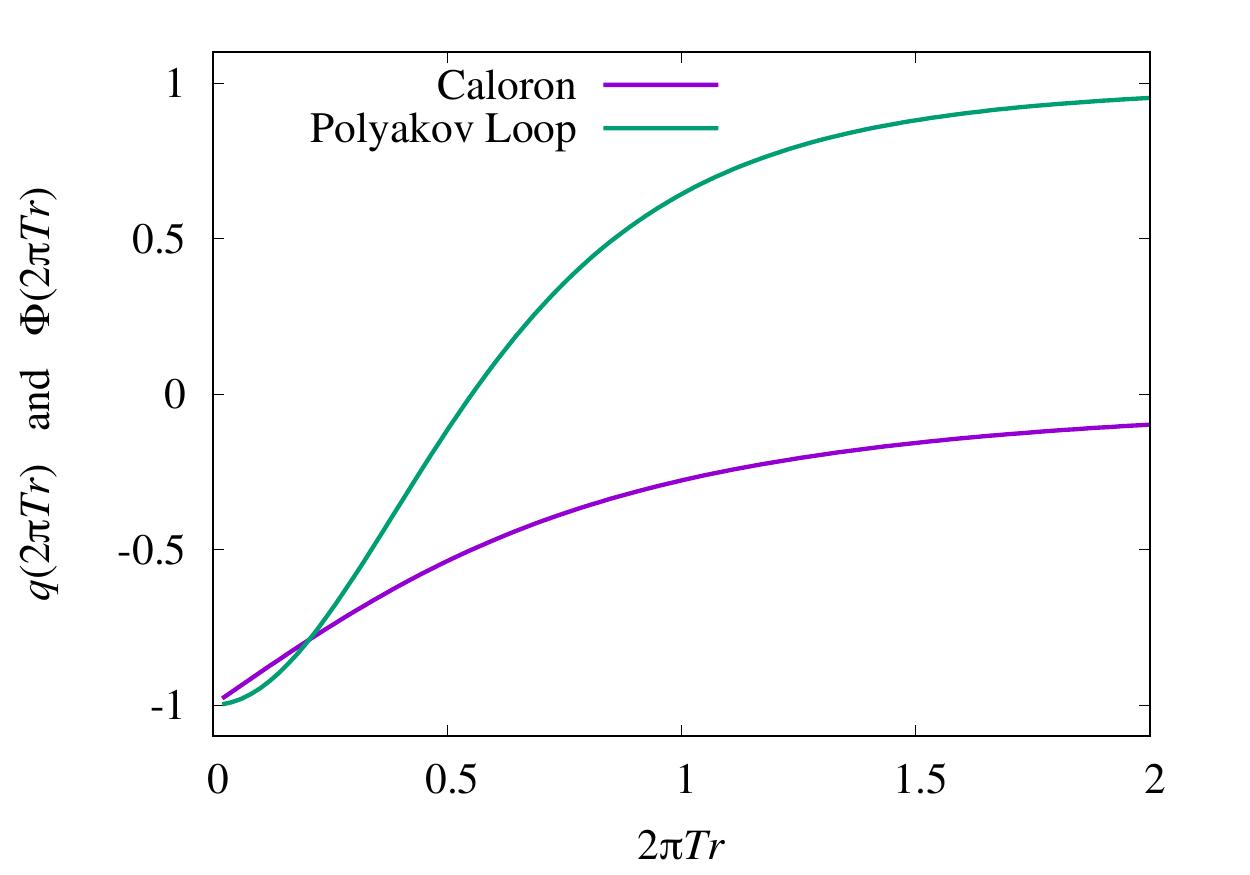}
  \end{center}
  \caption{SU(2) domain-wall solution and associated Polyakov loop
    distribution (left) and SU(2) caloron solution and associated
    Polyakov loop distribution for $\rho'=1$ (i.e.\ $\rho=1/(2\pi T)$)
    (right).}
  \label{fig:domain}
\end{figure}

Away from $\Tc$, rather a picture of instanton distributions should be
a favorable interpretation of disordered state, for the instanton
action stays finite.  At finite temperature the temporal or thermal
direction should be periodically closed, and to construct periodic
instantons, one can place multi-instantons along the $x_4$ axis.  Let
us focus on the SU(2) case only in what follows, but the
generalization to SU($\Nc$) is simple;  we can consider all possible
combinations of SU(2) subgroups in SU($\Nc$) and embed instantons
there.

The single SU(2) instanton centered at $z$ is written as
\begin{equation}
  A_\mu(x) = \frac{1}{2}\bar{\eta}_{\mu\nu}
  \partial_\nu \ln\phi(x)\;,\qquad
  \phi(x) = 1 + \frac{\rho^2}{(x_4-z_4)^2 + |\bx-\bz|^2}
\label{eq:instanton}
\end{equation}
with the t'~Hooft symbols,
$\eta_{\mu\nu}=\eta_{\mu\nu}^a \sigma^a
 = -i(\sigma_\mu\bar{\sigma}_\nu-\delta_{\mu\nu})$ and
$\bar{\eta}_{\mu\nu}=\bar{\eta}_{\mu\nu}^a \sigma^a
 = -i(\bar{\sigma}_\mu \sigma_\nu-\delta_{\mu\nu})$, where
$\sigma_\mu=(1,i\bsigma)$ and $\bar{\sigma}_\mu=(1,-i\bsigma)$ are the
quaternion bases.  For the Polyakov loop calculation the most useful
relation is $\bar{\eta}^a_{4i} = \delta_{ia}$.  By putting infinitely
many instantons at $z_4+n\beta$, the Harrington-Shepard instanton
configuration that satisfies the finite-$T$ periodic boundary
condition is obtained by Eq.~\eqref{eq:instanton} with a modified
$\phi(x)$, that is~\cite{Gross:1980br},
\begin{equation}
  \phi(x) = 1 + \frac{\pi T \rho^2}{|\bx-\bz|}\cdot
  \frac{\sinh(2\pi T|\bx-\bz|)}{\cosh(2\pi T|\bx-\bz|)
  -\cos(2\pi T|x_4-z_4|)}\;.
\label{eq:caloron}
\end{equation}
It is easy to see that Eq.~\eqref{eq:caloron} immediately reduces to
Eq.~\eqref{eq:instanton} in the low-$T$ limit and this $\phi(x)$ has a
period $1/T$ in terms of $x_4$.  This finite-$T$ instanton is
commonly called the ``caloron'' indicating that this special instanton
is manifested in a thermal or caloric environment.  It is then an
interesting question how the Polyakov loop value behaves in the
presence of a caloron.  We can calculate this using the explicit
components of the 't~Hooft symbols;
$\bar{\eta}_{4i}=\bar{\eta}_{4i}^a\sigma^a = \sigma^i$.  Thus, for a
caloron whose center is $z=0$, the $x_4$ integral of the eigenvalue of
$A_4$ simplifies as
\begin{equation}
  \int_0^\beta dx_4\, |A_4|
  = \frac{1}{2}\int_0^\beta dx_4 \frac{d}{dr} \ln\phi(r,x_4)
  = -\pi\Biggl[ 1 - \frac{\rho'^2(r'\cosh r'-\sinh r')
    + 2r'^2\sinh r'}{r' \sqrt{ 2(r'^2+\frac{\rho'^4}
    {4})(\cosh 2r'-1) + 2 \rho'^2 r' \sinh 2r'}} \Biggr]\;,
\end{equation}
where $r'=2\pi T r$ and $\rho'=2\pi T \rho$ are, respectively, the
rescaled dimensionless radial coordinate and instanton size.  The
Polyakov loop in the SU(2) case is simply given cosine of the above
integrated $A_4$ (where $g$ is absorbed in the normalization of $A_4$)
and the numerical profile is shown in the right of
Fig.~\ref{fig:domain}.  From this figure it is clear that $\Phi=-1$ at
the caloron center $r=0$ and $\Phi\to +1$ at infinity
$r\to\infty$.  Therefore, calorons are identified as Z(2) bubbles and
can disturb center elements to realize a disordered and thus center
symmetric ground state.

It is an intriguing attempt to seek for an extension of
calorons with different asymptotic behavior; by introducing an
additional parameter $\omega$ in the SU(2) case, if
$L(r\to\infty)\to e^{i 2\pi \omega\sigma_3}$ is realized,
the asymptotic value of the traced Polyakov loop approaches
$\Phi\to \cos(2\pi\omega)$.  Such a caloron solution has been
discovered and it is called the Kraan,
van~Baal~\cite{Kraan:1998kp,Kraan:1998pm} and Lee,
Lu~\cite{Lee:1998vu,Lee:1998bb} (KvBLL) solution.  Intuitively
speaking, this new caloron is a generalization of the
Harrington-Shepard solution with a gauge rotation by
$e^{i2\pi \omega \sigma_3}$ among temporally
aligned multi-instantons.  Such a construction in the so-called algebraic
gauge does not respect the periodicity along the thermal axis, and a
gauge transformation by $\Omega=e^{-i 2\pi T \omega \sigma_3 x_4}$
turns the solution to the periodic one, which reads after all as
\begin{equation}
  A_\mu(x) = \Omega \biggl\{ \frac{\phi}{2}\,\Re\bigl[
  (\bar{\eta}_{\mu\nu}^1 - i\bar{\eta}_{\mu\nu}^2)
  (\sigma_1 + i\sigma_2) \partial_\nu\chi \bigr]
  + \frac{\sigma_3}{2}\bar{\eta}_{\mu\nu}^3 \partial_\nu \ln\phi
  \biggr\}   \Omega^\dag
  -i \Omega\partial_\mu \Omega^\dag\;.
\end{equation}
Here, one should be careful of $\Re$ being defined by
$\Re W=\frac{1}{2}(W+W^\dag)$.  The $\sigma_3$ direction is intact
from the standard caloron, while the $\sigma_1$ and $\sigma_2$ terms
involve new functions defined as follows;
\begin{equation}
  \phi(x) \equiv \frac{\psi(x)}{\hat{\psi}(x)}\;,\quad
  \chi(x) \equiv \frac{1}{\psi(x)} \biggl[
   e^{-i 4\pi T \bar{\omega}x_4} \frac{\pi T\rho^2}{s}
  \sinh(4\pi T \omega s) +
   e^{i 4\pi T \omega x_4} \frac{\pi T\rho^2}{r}
  \sinh(4\pi T \bar{\omega} r) \biggr]\;,
\end{equation}
where $\bar{\omega}=1/2-\omega$, and for this solution there are two
relative position variables of ``instanton quarks'', i.e.\
$r^2=x^2+y^2+(z+2\pi T\rho^2 \omega)^2$
and $s^2=x^2+y^2+(z-2\pi T\rho^2 \bar{\omega})^2$.  These relative
positions of $r$ and $s$ are aligned on the $z$ axis due to a gauge
choice.  The definitions of dimensionless functions $\psi(x)$ and
$\hat{\psi}(x)$ are,
\begin{equation}
  \hat{\psi}(x) \equiv -\cos(2\pi T x_4) + \cosh(4\pi T\bar{\omega} r)
  \cosh(4\pi T\omega s) + \frac{r^2+s^2-(\pi T \rho^2)^2}{2rs}
  \sinh(4\pi T \bar{\omega}r)\sinh(4\pi T\omega s)\;,
\end{equation}
and
\begin{equation}
 \begin{split}
  \psi(x) &\equiv \hat{\psi}(x)
  + \frac{(\pi T\rho^2)^2}{rs}\sinh(4\pi T\bar{\omega}r)
  \sinh(4\pi T\omega s) \\
  &\qquad + \frac{\pi T\rho^2}{r}
  \sinh(4\pi T\bar{\omega} r)\cosh(4\pi T \omega s)
  + \frac{\pi T\rho^2}{s} \cosh(4\pi T \bar{\omega} r)
  \sinh(4\pi T\omega s) \;.
 \end{split}
\end{equation}
It is easy to make sure that the $\omega\to 0$ limit correctly reduces
these expressions
to Eqs.~\eqref{eq:instanton} and \eqref{eq:caloron} using
$\chi\to 1- \phi^{-1}$ and
$\phi\partial_\nu\phi^{-1}=-\phi^{-1}\partial_\nu\phi$.  The $s$
dependence disappears in this limit.

\begin{figure}
  \begin{center}
  \includegraphics[width=0.45\textwidth]{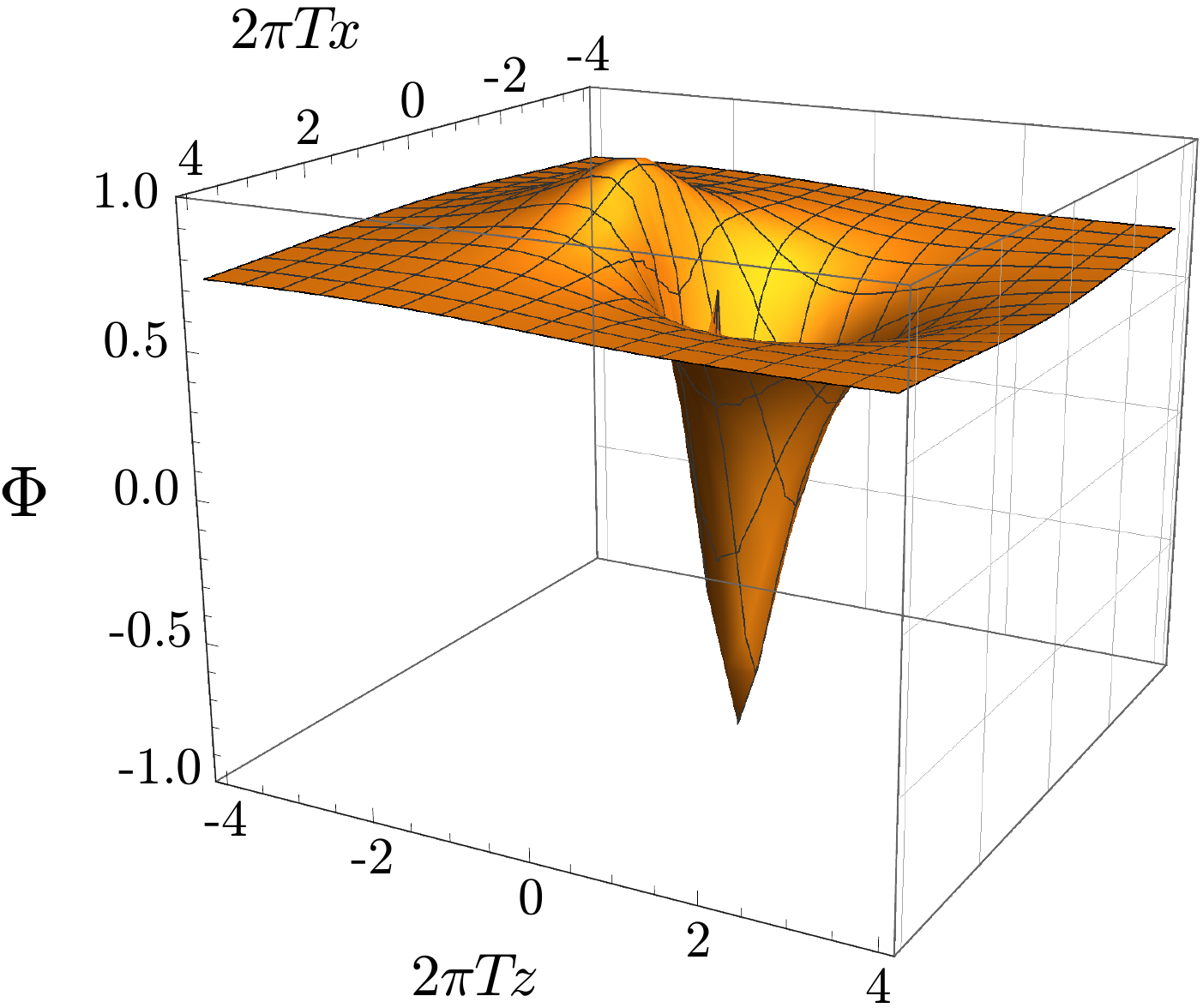}
  \hspace{1em}
  \includegraphics[width=0.45\textwidth]{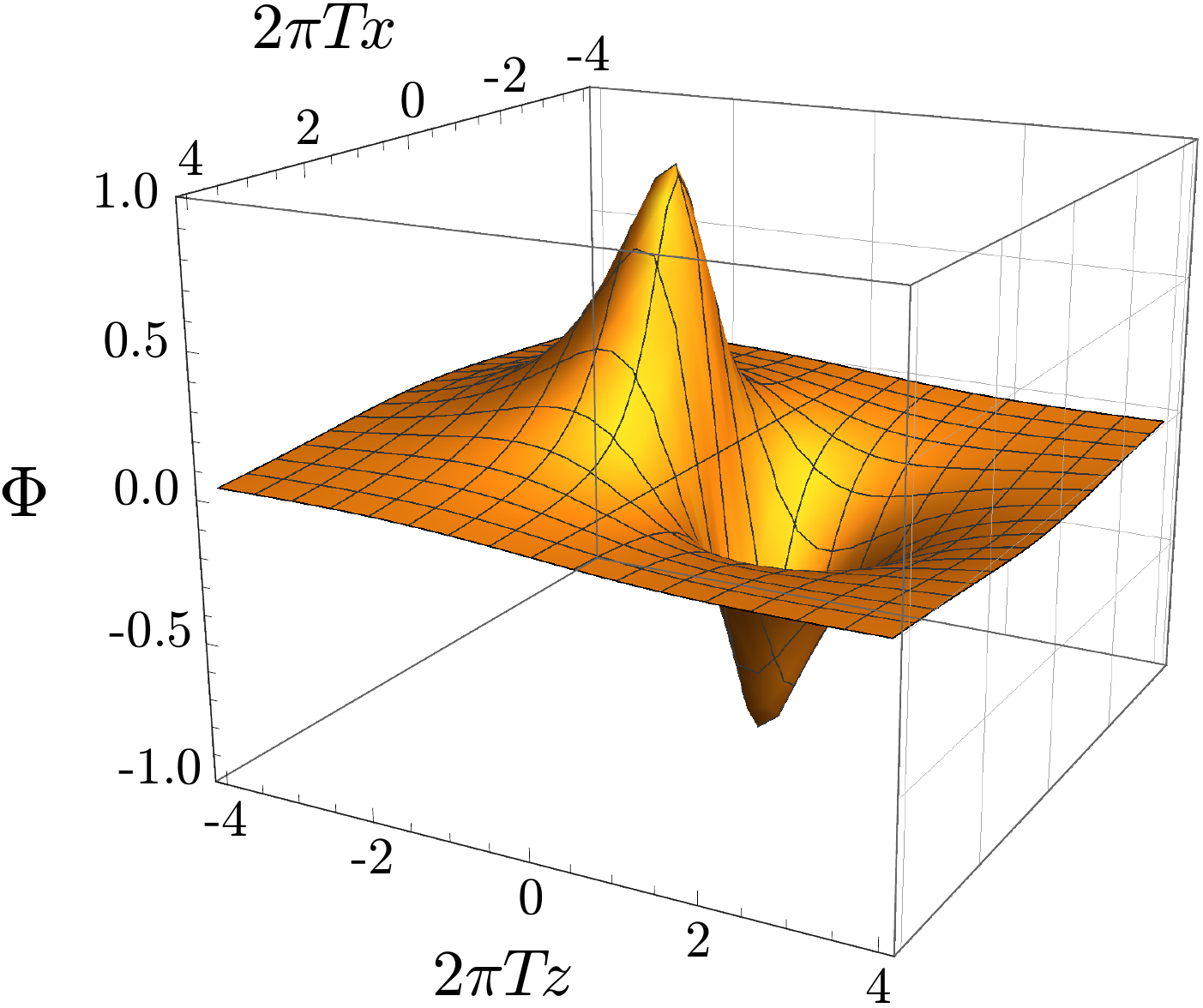}
  \end{center}
  \caption{Polyakov loop distribution in the presence of the KvBLL
    caloron for $\rho=1/(2\pi T)$ and $\omega=1/8$ (left) and the
    maximal holonomy $\omega=1/4$ (right) as a function of $2\pi T x$
    and $2\pi T z$ at $y=0$.}
  \label{fig:KvBLL}
\end{figure}

Shortly after the discovery of the KvBLL caloron, it has been
recognized that monopoles constitute
it~\cite{Lee:1998bb,Kraan:1998sn};  more precisely speaking, for the
winding number $k$ KvBLL caloron, there are $k \Nc$ dyons so that the
total monopole charge is zero and the caloron is a magnetic neutral
object.  Let us now see the Polyakov loop profile in the presence of
one KvBLL caloron for the SU(2) case for simplicity;  some examples
for this case are found in Ref.~\cite{Gerhold:2006sk}.  The 4th
component of the ``algebraic'' gauge potential is,
\begin{equation}
  A_4(x) = \sigma_1 \biggl( \frac{\phi}{2}\partial_1 \Re\chi
  + \frac{\phi}{2}\partial_2 \Im\chi \biggr)
  + \sigma_2 \biggl( -\frac{\phi}{2}\partial_1 \Im\chi
  + \frac{\phi}{2}\partial_2 \Re\chi \biggr)
  + \sigma_3 \biggl( \frac{1}{2}\partial_3 \ln\phi \biggr)\;,
\end{equation}
which is gauge transformed with $\Omega$, and then the Polyakov loop
$L$ should be multiplied by $e^{-i2\pi \omega \sigma_3}$.  Now, for
confinement physics, the most interesting choice of the asymptotic
boundary condition is $\omega=1/4$, which is usually referred to as
the maximal holonomy.  In this case the asymptotic value of the
Polyakov loop away from the caloron center becomes zero, i.e.\
$\Phi\to 0$ at $r\to \infty$.  Then, as shown in the right of
Fig.~\ref{fig:KvBLL}, two peaks in the Polyakov loop distribution
appear corresponding to $\Nc$ dyon constituents.  For a smaller value
of $\omega$, as seen in the left of Fig.~\ref{fig:KvBLL} that depicts
the result for $\omega=1/8$, the shape tends to reduce to that of the
conventional caloron as in the right of Fig.~\ref{fig:domain}.

It has been known analytically and numerically that the dyon contents
become more explicit for $\rho T < 1$, while they come close to be
degenerate for $\rho T > 1$.  This observation implies that
confinement may be caused by dyons in the KvBLL caloron at $T<1/\rho$,
and indeed, it has been found that a gas of the KvBLL calorons with
$\omega=1/4$ leads to confinement~\cite{Gerhold:2006sk}.  The
explanation, however, still awaits to be given for dynamical
determination of the most favorable value of $\omega$ (for an example
of recent calculations of the holonomy potential, see
Refs.~\cite{Larsen:2015vaa,Lopez-Ruiz:2016bjl}).

\section{Phase Transition in the Pure Gluonic Theory}
\label{sec:pure}

We have discussed the Polyakov loop expectation value from various
approaches, and it was so far only the strong coupling expansion that
could describe a phase transition in the pure gluonic theory.  One can
extract the essential feature of such a description to build a more
generic model, that is, the Polyakov loop matrix theory.  Sometimes
even further simplification on the matrix model is useful, which gives
us a parametrized form of the Polyakov loop potential.  The phase
transition in the pure gluonic theory has been studied in detail and
one of the most non-trivial but well-defined questions is the nature
of phase transition for large $\Nc$.  Physically, the deconfinement
phase transition is of second order only for the color SU(2) case,
while it is of first order for the color SU($\Nc\ge 3$) case in
general.  Moreover, the latent heat increases for larger $\Nc$,
meaning that the first-order phase transition becomes stronger if
$\Nc$ gets larger.  Once physical quantities are rescaled with $\Nc$,
however, it is no longer such easy to make a conclusion about the
order of the phase transition, and there is a theoretical suggestion
that a third-order phase transition may occur.  We will take a close
look at this possibility using an effective matrix theory of the
Polyakov loop.

\subsection{Polyakov Loop Models}
\label{sec:pure_models}

We introduce three models here.  The first one is a parametrized
Polyakov loop potential, which is a common ingredient for the Polyakov
loop augmented chiral models as we will elucidate later.  The second
one is an inverted Weiss potential.  This special form of the Polyakov
loop potential is not a frequent choice but the underlying physics
picture provides us with a consistent view of confinement with an
interesting link between the perturbative calculation, the strong
coupling expansion, and the propagator studies.  The last is the
Polyakov loop matrix model, which is quite successful for its simple
appearance.

\subsubsection{Parametrizing the Polyakov loop potential}
\label{sec:potential}

The first strategy to capture the Polyakov loop dynamics is to find a
reasonable parametrization of the Polyakov loop potential instead of
calculating it.  The guiding principle to optimize the parametrization
is the numerical outputs from the lattice
simulation~\cite{Borsanyi:2012ve,Giusti:2014ila,Giusti:2016iqr}.  The Polyakov
loop and the pressure (from which other thermodynamic quantities are
derived) as functions of $T$ have been already well known in the pure
gluonic sector.  Then, we
can make a guess about the minimal number of parameters to reproduce
the lattice numerical outputs;  one needs to adjust at least where the
phase transition takes place, namely, $\Tc$, and how large the
Polyakov loop and the pressure are then, namely, $\Phi(\Tc)$ and
$p(\Tc)$.  Also the large-$T$ behavior of the Polyakov loop
(i.e.\ $\Phi\to 1$) and the pressure (i.e.\ the Stefan-Boltzmann
limit) should be taken into account.  Once we treat these large-$T$
conditions as constraints to eliminate parameters, the minimal number
of parameters should be three corresponding to $\Tc$, $\Phi(\Tc)$, and
$p(\Tc)$.

This program to establish a parametrized Polyakov loop potential was
addressed in Ref.~\cite{Ratti:2005jh} with an Ansatz of the power 
expansion in terms of $t=T/\Tc$.  The SU(3) potential then reads,
\begin{equation}
  T^{-4} V_{\rm glue}(\Phi,\bar{\Phi};T)
  = -\frac{b_2(T)}{2}\bar{\Phi}\Phi - \frac{b_3}{6}(\Phi^3+\bar{\Phi}^3)
  + \frac{b_4}{4}(\bar{\Phi}\Phi)^2\;,
\label{eq:polpotential}
\end{equation}
and the temperature dependence is assumed to appear only in the
quadratic coefficient as
\begin{equation}
  b_2(T) = a_0 + a_1 t^{-1} + a_2 t^{-2} + a_3 t^{-3}\;.
\label{eq:b2}
\end{equation}
We note that $\bar{\Phi}=\Phi$ results from the variation of the
potential.  In this effective potential there are six parameters, one
of which is constrained by the Stefan-Boltzmann limit for the
pressure; $T^{-4}V_{\rm glue}|_{T\to\infty}\to
a_0/2-b_3/3+b_4/4=-16\cdot(\pi^2/90)\simeq -1.75$.  Another constraint
is $\Phi\to 1$ at high $T$.  In the limit of $t^{-1}=0$, we can
calculate the Polyakov loop as
$\Phi\to (b_3+\sqrt{b_3+4a_0 b_4})/(2b_4)=1$, which imposes a
condition; $a_0+b_3-b_4=0$.  The common parameter
set~\cite{Ratti:2005jh} is
\begin{equation}
 a_0 = 6.75\;,\quad
 a_1 = -1.95\;,\quad
 a_2 = 2.625\;,\quad
 a_3 = -7.44\;,\quad
 b_3 = 0.75\;,\quad
 b_4 = 7.5\;.
\end{equation}
We note that we can safely take the high-$T$ limit because the
temperature dependence in the coefficients is only through the power
of $t^{-1}$, and this feature is consistent with the functional form
from the strong coupling expansion; $b_2/2 \propto e^{-\beta \sigma a}$.

The above-mentioned parametrization still has redundancy and ideally
we could have found a better parametrization with two fewer
parameters.  Later a simpler and better parametrization with less
parameters was found as~\cite{Roessner:2006xn}
\begin{equation}
  T^{-4} V_{\rm glue}(\Phi,\bar{\Phi};T) = -\frac{a(T)}{2}
  \bar{\Phi}\Phi + b(T)\ln\bigl[ 1-6\bar{\Phi}\Phi
  + 4(\bar{\Phi}-3+\Phi^3) - 3(\bar{\Phi}\Phi)^2 \bigr]
\label{eq:polpotential2}
\end{equation}
with four parameters in
\begin{equation}
  a(T) = a_0 + a_1 t^{-1} + a_2 t^{-2}\;,\qquad
  b(T) = b_0 t^{-3}\;.
\end{equation}
In this case the Stefan-Boltzmann constraint is quite simple;
$T^{-4} V_{\rm glue}|_{T\to\infty}\to -a_0/2$, which immediately leads
to $a_0=3.51$.  Then, there are only three parameters to be fixed by
the lattice data, and the well tuned values are~\cite{Roessner:2006xn}
\begin{equation}
  a_0 = 3.51\;,\quad
  a_1 = -2.47\;,\quad
  a_2 = 15.2\;,\quad
  b = -1.75\;.
\end{equation}
This latter parametrization~\eqref{eq:polpotential2} has three
advantages over the former~\eqref{eq:polpotential};  the number of
parameters is minimal, $\Phi$ (and $\bar{\Phi}$) never exceeds the
group theoretical upper bound, i.e.\ $\Phi\le 1$, and the role by the
SU(3) Haar measure (see Eqs.~\eqref{eq:Haar} and \eqref{eq:SU3Haar}) is manifest in Eq.~\eqref{eq:polpotential2}.  The
comparison to the lattice simulation data is found for example in the
review of Ref.~\cite{Fukushima:2013rx}.  The agreement is remarkable;
the pure gluonic dynamics for \textit{any} $T>\Tc$ is
almost perfectly characterized by only three free parameters.

We also make a comment on $\Tc$ here.  Strictly speaking, $\Tc$ should
be counted as another model parameter, and $\Tc=270\MeV$ is a common choice
based on the lattice simulation of the pure gluonic theory.  Later,
with dynamical quarks, we will come back to the determination of this
parameter $\Tc$ that should take a backreaction effect from the quark
loops.

It is always a tempting idea to derive such a Polyakov loop potential
from the first-principle approach.  For recent discussions on
improving the Polyakov loop potential, see
Refs.~\cite{Haas:2013qwp,Sasaki:2012bi,Lo:2013hla}, and for lattice
determination, see
Refs.~\cite{Greensite:2013bya,Smith:2013msa,Diakonov:2013lja}.

\subsubsection{Inverted Weiss potential}
\label{sec:inverted}

The ghost or the SU(3) Haar measure causes confinement in the strong
coupling expansion, and the successful parametrization in
Eq.~\eqref{eq:polpotential2} indeed implements the Haar measure form.  Hence,
the modern picture of confinement implies the ghost dominance in the
infrared regime.  This
situation of the ghost dominance makes a sharp contrast to the
perturbative calculation, in which the ghost and the longitudinal
gluon contributions exactly cancel out.

The cancellation no longer holds once we take account of
non-perturbative propagators of the ghost and the gluons.  Such a
procedure to compute the non-perturbative Polyakov
loop potential was first suggested in Ref.~\cite{Braun:2007bx}.
Schematically, the effective potential can be written as [see
Eq.~\eqref{eq:glue_oneloop}]
\begin{equation}
  V_{\rm glue}[q] = \frac{1}{2}\tr\ln D_{\mu\nu}(p) -\tr\ln G(p)\;,
\label{eq:glue_full}
\end{equation}
where $D_{\mu\nu}(p)$ and $G(p)$ represent the gluon and the ghost full
propagators, respectively.  In the Landau gauge these propagators are
parametrized by the dressing functions, apart from the trivial color
index that is simply the unit matrix in color space, as
\begin{align}
  D_{\mu\nu}(p)
  &= Z_A(p^2)(p^2\delta_{\mu\nu} - p_\mu p_\nu)
    + \frac{1}{\xi}Z_L(p^2) p_\mu p_\nu \;,\\
  G(p) &= Z_C(p^2) p^2\;.
\end{align}
As long as $T$ is low, it is natural to anticipate that the infrared
momentum region would be dominant in the integration in
Eq.~\eqref{eq:glue_full}.  Then, to capture the qualitative feature of
the Polyakov loop potential in the confined phase, the so-called
scaling solution should be useful, which is numerically known to
behave as~\cite{Zwanziger:2001kw,Lerche:2002ep,Pawlowski:2003hq}
\begin{equation}
  Z_A(p^2\sim 0) \simeq (p^2)^{\kappa_A}\;,\quad
  Z_L \simeq 1+\xi\;,\quad
  Z_C(p^2\sim 0) \simeq (p^2)^{\kappa_C}\;.
\end{equation}
One might mind about the formal correctness of the scaling solution;
the lattice simulation seems to have falsified the exact scaling
solution in the $p^2\to 0$ limit, but still, for small but finite
$p^2$, the scaling solution makes sense qualitatively and even
quantitatively.  It is not important whether $Z_C(p^2\to\infty)$
strictly diverges or not.  We note that $\kappa_A=-2\kappa_C$ follows
from the non-renormalization property of the ghost-gluon vertex at
vanishing momentum in the Landau gauge.  From the lattice simulation
and the functional method, $\kappa_C\simeq 0.6$ approximately
holds~\cite{Pawlowski:2003hq}.  Then, the power of the momentum
dependence of the propagator is modified, so that the weight for the
gluon part (counting the polarization sum) changes from $4$ to
$3(1+\kappa_A)+1$ and the weight for the ghost part from $1$ to
$1+\kappa_C$.  Then, the non-perturbative effective potential becomes,
\begin{equation}
  V_{\rm glue}^{\rm non-perturbative}[q]
   = (1-4\kappa_C)V_{\rm glue}^{\rm Weiss}[q]
   \;\simeq\; -1.4 V_{\rm glue}^{\rm Weiss}[q]\;.
\end{equation}
Because of the overall minus sign, the perturbative Weiss potential is
\textit{inverted} and the potential minimum is then located at the
confined phase with $\Phi=0$, as is obvious from
Fig.~\ref{fig:su2weiss} for the SU(2) case.  This is one of the
simplest pictures to understand how confinement is realized.
It should be noted that, if the theory has adjoint fermions satisfying
the periodic boundary condition, the simple one-loop calculation leads to the
inverted Weiss potential as above, and the perturbative confinement is
realized~\cite{Kanazawa:2017mgw}.

The phase transition temperature $\Tc$ is characterized by a typical
scale in the ghost and the gluon dressing functions.  For momenta
larger than such a typical scale the propagators are expected to
approach the perturbative ones.  If $T$ is greater than this scale,
contributions from higher momentum regions should dominate
the calculation, and then the Polyakov loop effective potential should
reduce to the Weiss potential that embodies the deconfined phase.
Thus, to quantify $\Tc$, we need more information on the propagators
over wide momentum ranges.

A short-cut approach is a hybrid calculation of the one-loop
integration using the non-perturbative propagators numerically
measured on the lattice.  Such a combination of the ``perturbative''
loop integration and the ``non-perturbative'' propagators might have
sounded weird, but this can be understood as an extension of the
ordinary mean-field type approximation.  Usually, in the mean-field
approximation, all the interaction effects are assumed to be
renormalized into the non-perturbative propagator, and then the
precise form of the propagator is self-consistently determined to
minimize corrections from the interaction effects.  Instead of solving
the self-consistency condition or the gap equation, why shouldn't we
jump to the full propagators already available from the lattice
simulation?  Of course we can, and we in principle need the lattice
propagators at all temperatures in order to locate
$\Tc$ precisely.  Looking at the lattice data, we would recognize that
only the transverse gluons show moderate $T$ dependence.  It is
remarkable that one can hardly see any qualitative difference associated with
the first-order deconfinement transition in the ghost and the gluon
propagators~\cite{Aouane:2011fv}.
Therefore, it is not a bad approximation to utilize even the zero-$T$
propagators for finite-$T$ loop calculations.

Interestingly enough, as demonstrated in Ref.~\cite{Fukushima:2012qa},
the effective potential calculated in such a way successfully captures
even quantitative properties of the deconfinement phase transition,
i.e.\ the phase transition temperature is correctly reproduced.  We
also note that, instead of using the lattice propagator, an
introduction of the gauge mass term can capture the essential features
of the phase transitions as discussed in Ref.~\cite{Fukushima:2013xsa}
and the systematic expansion in general SU($\Nc$) group has been
developed in
Refs.~\cite{Reinosa:2014ooa,Serreau:2013ila,Reinosa:2015gxn}.

\subsubsection{Mean-field approximation of a matrix model}
\label{sec:matrix}

Here let us discuss an extended version of the mean-field
approximation that correctly takes care of the group integration
properties according to Ref.~\cite{Gocksch:1984yk}.
So far, we implicitly assumed a stronger
approximation than the mean-field approximation, that is,
$\Phi=\langle\ell[A_0]\rangle = \ell[\langle A_0\rangle]$.  Such a
treatment would sometimes lead to unphysical results;  the most
obvious example of pathological behavior is found in the Polyakov loop
calculation in color superconductivity, in which a non-zero color
charge emerged even in normal quark matter, which is just an artifact
from improper
approximation~\cite{Roessner:2006xn,GomezDumm:2008sk,Abuki:2009dt}.

To make the following discussions concrete, we adopt a Polyakov loop
effective theory in the strong coupling limit, where the partition
function is given by the group integration in Eq.~\eqref{eq:Zstrong}.
We would emphasize that the idea itself is quite general, as
implemented in Refs.~\cite{Abuki:2009dt,Megias:2004hj}.
Precisely speaking, the treatment in Sec.~\ref{sec:strong} was the
tree-level approximation, and quantum fluctuations should have been taken
into account self-consistently in the genuine mean-field
approximation.  To formulate the mean-field approximation, we define
the mean-field action as
\begin{equation}
  S_{\rm mf}[\ell;z] = -\frac{1}{2}\sum_{\bx} \bigl[
    z^\ast \ell(\bx) + z \ell^\ast(\bx) \bigr]
\end{equation}
with mean-field variables $z$ and $z^\ast$.  We next define the
mean-field average for arbitrary functions of $L$ as
\begin{equation}
  \langle \calO[L]\rangle_{\rm mf} \equiv
  \frac{\displaystyle \int\calD L\, e^{-S_{\rm mf}[\ell]} \calO(L)}
       {\displaystyle \int\calD L\, e^{-S_{\rm mf}[\ell]}}\;.
\end{equation}
For the theoretical foundation of the mean-field approximation the
most important relation is the convexity condition, that is,
$\langle e^{\calO}\rangle_{\rm mf} \ge e^{\langle \calO\rangle_{\rm mf}}$
(see Ref.~\cite{creutz1983quarks} for detailed discussions).  From
this inequality the full partition function is bounded by the
mean-field approximated one as
\begin{equation}
  Z \;\ge\; \exp\bigl\{ \langle -S[\ell]+S_{\rm mf}[\ell;z]
  \rangle_{\rm mf} \bigr\} \cdot \int \calD L\,
  e^{-S_{\rm mf}[\ell;z]}\;,
  \label{eq:mf_inequality}
\end{equation}
where $S[\ell]=-J\sum \ell^\ast(\bx')\cdot \ell(\bx)$ is the original
action that defines the matrix theory.  Our goal is to find an
analytical expression of the right-hand side of the above inequality.
Let us introduce a function of the mean-field variables;
\begin{equation}
  I(z^\ast,z) \equiv \int dL\, e^{-S_{\rm mf}[\ell;z]}\;.
\end{equation}
Then, it is easy to rewrite the mean-field expectation values as
\begin{equation}
  \langle S[\ell]\rangle_{\rm mf} = -V_3\cdot 6J \cdot
  \frac{2\partial \ln I}{\partial z} \cdot
  \frac{2\partial \ln I}{\partial z^\ast}\;,
\end{equation}
and
\begin{equation}
  \langle S_{\rm mf}[\ell]\rangle_{\rm mf} = -\frac{1}{2}V_3
  \biggl( z^\ast\,\frac{2\partial \ln I}{\partial z^\ast}
    + z\,\frac{2\partial \ln I}{\partial z} \biggr)\;,
\end{equation}
where $V_3$ represents the spatial (three dimensional) site number,
and $6$ appears from the number of nearest neighbor sites in three
dimensional space.  By taking the logarithm of the right-hand side of
the inequality~\eqref{eq:mf_inequality}, we can evaluate the
mean-field grand potential $\Omega_{\rm mf}$ that gives the upper
bound of the true grand potential.  In this way, the mean-field grand
potential is found to be
\begin{align}
  \beta\Omega_{\rm mf}(z^\ast,z)/V_3 &= \langle S[\ell]\rangle_{\rm mf}
  - \langle S_{\rm mf}[\ell;z]\rangle_{\rm mf} - \ln \int\calD L\,
  e^{-S_{\rm mf}[\ell;z]} \notag\\
  &= -6J\cdot\frac{2\partial \ln I}{\partial z}\cdot
  \frac{2\partial \ln I}{\partial z^\ast}
  + \frac{1}{2}\biggl(z^\ast\,\frac{2\partial \ln I}{\partial z^\ast}
    + z\,\frac{2\partial \ln I}{\partial z} \biggr) - \ln I\;.
\end{align}
As long as the charge parity symmetry is unbroken (that is broken for
example in a system at finite density), we can set $z=z^\ast=x$, which
simplifies the above mean-field grand potential as
\begin{equation}
  \beta\Omega_{\rm mf}(x)/V_3 = -6J \biggl[ \frac{d}{dx}\ln I(x)
    \biggr]^2 + x^2 \frac{d}{dx} \biggl[ \frac{1}{x} \ln I(x)
    \biggr]\;.
\end{equation}
The stationary condition to minimize the grand potential gives a
relation, $x=12J \langle \ell\rangle$, which manifests the fact that
$x$ is a mean-field variable corresponding to the Polyakov loop
expectation value.  We still need to perform the group integration to
find an explicit form of $I(x)$.  For this purpose it is useful to use
an alternative expression for the Haar measure;
\begin{equation}
  H = \bigl| \epsilon_{i_1\cdots i_{\Nc}} e^{i2\pi q_1 (\Nc-i_1)}
  \cdots e^{i2\pi q_{\Nc}(\Nc-i_{\Nc})}\bigr|^2 \;.
\end{equation}
With this expression it is possible to find the analytical expression
as (see Ref.~\cite{Kogut:1981ez} for detailed derivations)
\begin{equation}
  I(x) = \int dL\,e^{\frac{x}{2}(\ell+\ell^\ast)}
  = \Nc! \sum_m \det_{ij} I_{m-i+j}(x/\Nc)\;.
\end{equation}
Here, $I_n(x)$ is the modified Bessel function of the first kind.  Now
we can proceed to numerics using these expressions;  for
$\Nc=3$ the critical coupling is found to be $J_c\simeq 1.21$, which
shows a sizable difference from $J_c\simeq 0.773$ in the tree-level
approximation seen in Sec.~\ref{sec:strong}.

Because calculations are far more involved as compared to the
Polyakov loop potential approach in Sec.~\ref{sec:potential}, the
above analysis is not very much preferred, but can be useful for
theoretical investigations as demonstrated in
Refs.~\cite{Abuki:2009dt,Fukushima:2006uv}.

\subsection{Large $\Nc$ Limit}
\label{sec:largeNc}

It is generally hard to find an analytical answer in the
non-perturbative sector of QCD, and then it is useful to consider a
modified version of QCD that would have desirable properties for
analytical purposes.  Some examples include two-dimensional QCD,
two-color QCD, QCD in the strong coupling limit, many-flavor QCD, and
so on.  The most successful idea along these lines is to increase the
number of colors, $\Nc$, to infinity.  In particular two-dimensional
QCD at $\Nc\to\infty$ is called 't~Hooft model~\cite{THOOFT1974461}
that resembles QCD.\ \
It is a widely accepted folklore that the $\Nc=3$ real world
may share most of non-perturbative aspects with the $\Nc\to\infty$
simplified world.  In this subsection we address what is known so far
about the pure gluonic dynamics and the phase transition in the large
$\Nc$ limit.

\subsubsection{Gross-Witten phase transition}
\label{sec:GW}

In two spacetime dimensions, the dynamics of the SU($\Nc$) pure
gluonic theory is significantly simpler than that in four dimensions
owing to the absence of the transverse dimensions and associated gluon
excitations.  In a lattice formulation of the theory, this
simplification is reflected by an independence of the plaquette
variables.  We can understand this in the Weyl gauge in which the
independence is obvious;  all temporal link variables are trivial and
consequently the spatial links can be represented as a time-ordered
product of independent plaquette variables.  
Thus the partition function~\eqref{eq:ZU4} reduces to
\begin{equation}
  Z  = \Biggl\{ \int dU \exp\biggl[ \frac{1}{g^2} \tr_c (U+U^\dagger)
    \biggr] \Biggr\}^{N_p}\;,
  \label{Eq:ZGW}
\end{equation}
where $N_p = V_2/a^2$ is the number of plaquettes with $a$ and $V_2$
denoting the lattice spacing and the two-dimensional spacetime
volume, respectively.  Here, the integration is performed over a
single unitary matrix, $U \in$ U($\Nc$)~\footnote{In the large $\Nc$
  limit, the difference between U$(\Nc)$ and SU$(\Nc)$ can be
  ignored.}.  Hence the problems is reduced to the computation of the
single-site partition function;
\begin{equation}
  z = \int dU \exp\biggl[ \frac{1}{g^2} \tr_c (U+U^\dagger) \biggr]\;. 
  \label{Eq:zGW}
\end{equation}
The integrand depends on the eigenvalues of $U$, which we denote by
$q_i$ with $i=1,\dots,\Nc$.  By a unitary transformation $T$, the
matrix $U$ can be rewritten as $T D T^\dag$ with a diagonal matrix, 
$D = \diag(e^{i 2\pi q_1}, \dots, e^{i 2\pi q_{\Nc}})$.  The
integration measure is thus expressed in the form of the Vandermonde
determinant as follows;
\begin{equation}
  dU \propto dT \prod_i d q_i \prod_{i<j} \bigl| e^{i 2\pi q_i }
    - e^{i 2\pi q_j } \bigr|^2\;,
    \label{Eq:dU_GW}
\end{equation}
modulo an irrelevant normalization factor.  Therefore the partition
function is proportional to
\begin{equation}
  z \propto \int d^{\Nc}q \, \exp \Biggl\{ \Nc^2 \biggl[
    \frac{2}{\lambda}\cdot \frac{1}{\Nc} 
    \sum_{i=1}^{\Nc} \cos (2\pi q_i)  + 
    \frac{1}{\Nc^2} \sum_{i \ne j} \ln \bigl| e^{i 2\pi q_i}
    - e^{i 2\pi q_j} \bigr| \biggr] \Biggr\}\; ,
  \label{Eq:z_Sums}
\end{equation}
where we introduced the 't~Hooft coupling as
$\lambda = g^2 \Nc$.  We take the large $\Nc$ limit keeping $\lambda$
constant, and then we can replace the discrete sums with the
continuous integrals as
\begin{equation}
  \frac{1}{\Nc} \sum_i \;\to\; \int_0^1 d x\;, \qquad
  \frac{1}{\Nc^2} \sum_{i\ne j} \;\to\; \mathrm{P}\int_0^1 dx dy\;,
  \label{Eq:SumToInt}
\end{equation}
where P denotes the Cauchy principal value.  In what follows below, to
simplify the notation, we denote the large $\Nc$ limit of $q_i$ by
$q_x$.  Then, the steepest descent method yields the energy as a
function of the coupling $\lambda$ as
\begin{equation}
  - E(\lambda)  =  \lim_{\Nc\to\infty} \frac{\ln z}{\Nc^2} = 
  \frac{2}{\lambda} \int dx  \cos (2\pi q_x)
  + \mathrm{P}\int dx dy  \ln \bigl| \sin [\pi(q_x-q_y)] \bigr|
  \label{Eq:Sq}
\end{equation}
with the most probable $q_x$ given by the stationarity condition,
\begin{equation}
  \frac{2}{\lambda} \sin (2 \pi q_x) = \mathrm{P} \int dy\,
  \cot [\pi(q_x-q_y)] \;.
  \label{Eq:EoM_x}
\end{equation}
We can solve this equation by introducing a positive-definite density
of eigenvalues that is defined by
\begin{equation}
  \rho(q) = \frac{dx}{dq}\;,
  \label{Eq:rho_definition}
\end{equation}
satisfying a proper normalization condition 
\begin{equation}
  \int_{-q_0}^{q_0} dq\, \rho(q) = \int_0^1 dx = 1\;,
  \label{Eq:RhoNormalCond}
\end{equation}
where we assumed that the eigenvalues may be separated by a single
gap, i.e.\ $q^2\le q_0^2 \le 1/4$.  Then, Eq.~\eqref{Eq:EoM_x} can be
rewritten into the following equivalent form,
\begin{equation}
  \frac{2}{\lambda} \sin(2\pi q) =  \mathrm{P} \int 
  dq'\,\rho(q')	\cot[ \pi(q-q') ] \;.
  \label{Eq:EoM_rho}
\end{equation}
It is instructive to rewrite this in terms of the complex variables;
$z = e^{i 2\pi q}$ and $z' = e^{i 2\pi q'}$, that is,
\begin{equation}
  \frac{2 i}{\lambda}  \biggl( \frac{z - z^{-1}}{2}\biggr) + i
  = \frac{1}{\pi} \mathrm{P} \int_C 
  dz'\, \frac{\rho(z')}{z'-z} \;.
  \label{Eq:EoM_rho_z}
\end{equation}
For a given contour $C$, the inversion of this equation, also known
under the rubric of the Hilbert transform, is a very well-known
mathematical problem (see Ref.~\cite{muskhelishvili2012singular}).  

In the strong coupling regime, $\lambda \ge 2$, the eigenvalues are
expected to be distributed over the entire circle, that implies
$q_0=1/2$.  In this case $C$ is a closed contour on $|z'|=1$, so that
the inversion of Eq.~\eqref{Eq:EoM_rho_z} trivially results in
$\rho(q) = 1 + \frac{2}{\lambda} \cos(2\pi q)$.  We note that the
distribution of the eigenvalues is properly normalized and is positive
for any $\lambda \ge 2$.

In the weak coupling regime, $\lambda < 2$, one has to solve the
problem for $q_0<1/2$.  Hence, the contour $C$ represents an arc 
and the solution is given by
$\rho(q) = \frac{4}{\lambda} \cos (\pi q)\cdot
\sqrt{\frac{\lambda}{2} - \sin^2 (\pi q)  }$.
At $\lambda = \lambda_c \equiv 2$, two solutions in the strong and the
weak coupling regimes coincide to yield the distribution;
$\rho(q)=1+\cos(2\pi q)$.

\begin{figure}
  \includegraphics[width=0.45\linewidth]{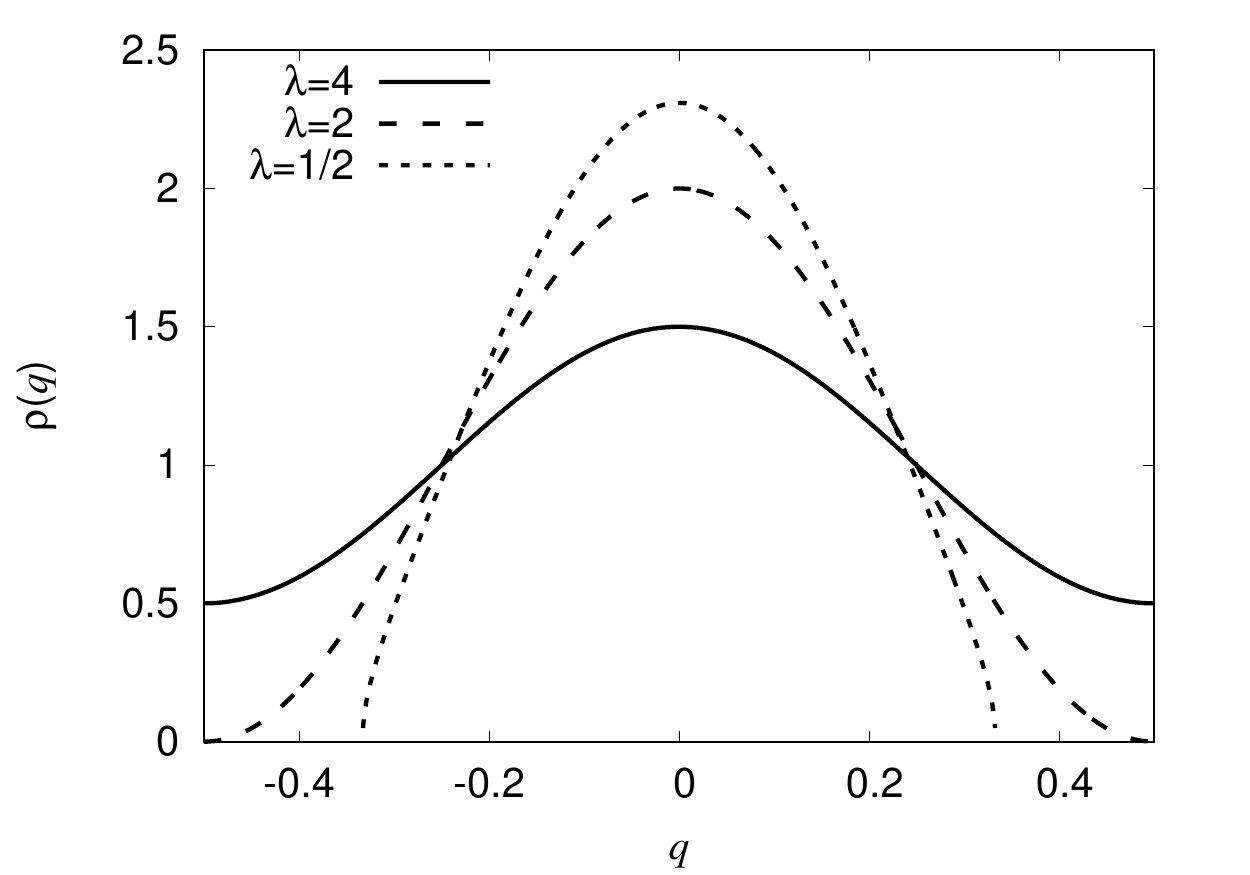}
  \includegraphics[width=0.45\linewidth]{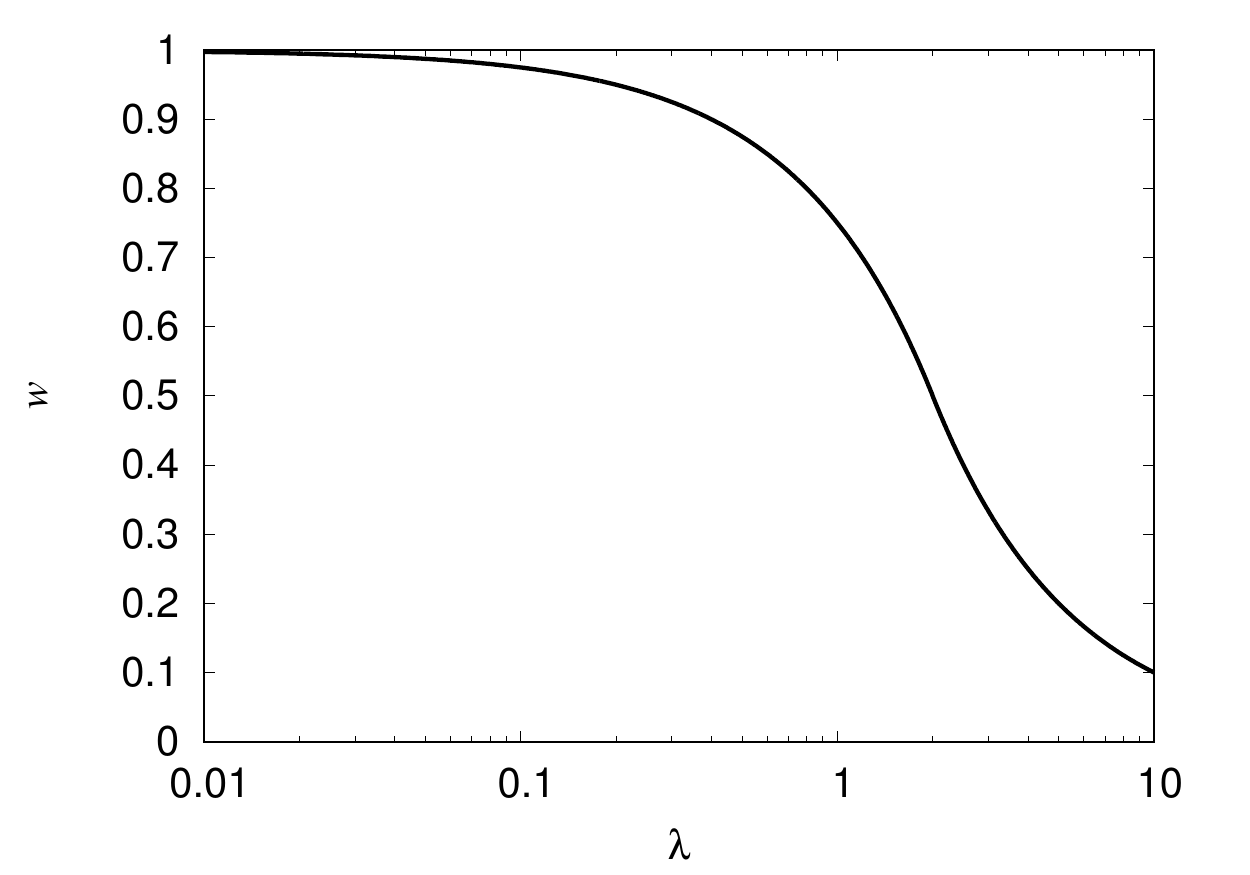}
  \caption{Eigenvalue distribution for different values of the
    coupling constant (left) and the expectation value of the Wilson
    loop as a function of the coupling constant (right).}
  \label{Fig:GW_rho}
\end{figure}

Hence, to summarize the above, it has been established that the
distribution of the eigenvalues is given by two distinct analytic
functions;
\begin{equation}
  \rho(q) = 
  \begin{cases}
    \displaystyle \frac{4}{\lambda} \cos(\pi q)\cdot
    \sqrt{\frac{\lambda}{2} - \sin^2(\pi q) }
    \quad\text{with}\quad \sin^2(\pi q_0) = \frac{2}{\lambda}
    & \quad\text{for}\quad \text{if}~  \lambda < 2\;, \\
    \displaystyle 1 + \frac{2}{\lambda} \cos(2\pi q)
    \quad\text{with}\quad q_0 = \frac{1}{2}
    & \quad\text{for}\quad \text{if}~  \lambda \ge 2\;.
  \end{cases}
  \label{Eq:rho_combined}
\end{equation}
We can see the behavior of $\rho(q)$ for different values of $\lambda$
in the left panel of Fig.~\ref{Fig:GW_rho}.  The corresponding
expectation value of the traced Wilson loop, that is given by $\omega
= \int dq \rho(q) \cos(2\pi q)$ in the large $\Nc$ limit, takes the
following expressions,
\begin{equation}
  \omega = 
  \begin{cases}
    \displaystyle 1 - \frac{\lambda}{4}
    & \quad\text{if}\quad \lambda < 2\;,\\
    \displaystyle \frac{1}{\lambda}
    & \quad\text{if}\quad \lambda \ge 2\;,
  \end{cases}
  \label{Eq:w_combined}
\end{equation}
whose behavior is seen in the right panel of Fig.~\ref{Fig:GW_rho}.
Here, we refer to Ref.~\cite{Buividovich:2015oju} for the interpretation
of the phase transition in terms of ``saddle'' condensation.

Now we can proceed to the calculation of the energy by solving the
stationary condition.  Indeed, we can immediately integrate
Eq.~\eqref{Eq:EoM_rho} with respect to $q$ in the range from 0 to $q$.
Then, we multiply this integrated function of $q$ by $\rho(q)$ and
finally perform the integration over $q$ in the range from
$-q_0$ to $q_0$ to reach,
\begin{equation}
  \mathrm{P} \int dq dq'\, \rho(q) \rho(q')
  \ln \bigl|\sin[\pi(q-q')]\bigr| =
  \mathrm{P} \int dq\,\rho(q) \ln |\sin(\pi q)|
  - \frac{1}{\lambda} \biggl[ \int dq \rho(q)\,
    \cos(2\pi q) - 1 \biggr]\;,
  \label{Eq:2to1}
\end{equation}
and, we can therefore rewrite the energy into the following form,
\begin{equation}
  - E(\lambda) = \frac{1}{\lambda} \biggl[ \int dq \rho(q)\,
    \cos(2\pi q) + 1 \biggr]
  + \mathrm{P} \int dq\, \rho(q) \ln |\sin(\pi q)|\;.
  \label{Eq:S_extr}
\end{equation}
Computing the elementary integrals in the above expression we finally
get a simple expression for the energy as
\begin{equation}
  -E(\lambda) =
  \begin{cases}
    \displaystyle \frac{2}{\lambda}
    + \frac{1}{2} \ln \frac{\lambda}{8} - \frac{3}{4}
    & \quad\text{if}\quad \lambda < 2\;,\\[6pt]
    \displaystyle \frac{1}{\lambda^2} - \ln 2
    & \quad\text{if}\quad \lambda \ge 2\;.
  \end{cases}
  \label{Eq:GW_E}
\end{equation}
This energy expression has an interesting property.  As a function of
$\lambda$, the energy, its first, and second derivatives are all
continuous functions.  The third derivative, $E'''(\lambda)$, turns
out to be discontinuous.  Hence, the critical coupling $\lambda_c=2$
is actually a critical point for a \textit{third-order} phase
transition.  So far, the above results are relevant for the large
$\Nc$ pure gluonic theory in two spacetime dimensions, but it can be
generalized to four dimensional cases under certain conditions;
namely, if the large $\Nc$ limit could be interchangeable with the
strong coupling expansion, the pure gluonic theory in four dimensions
may exhibit the same third-order phase transition in the large $\Nc$
limit~\cite{Gross:1980he}.

\subsubsection{Effective matrix theory}
\label{sec:matrix_GW}

Section~\ref{sec:GW} was a review for the analytical results known for
the lattice Wilson action in two dimensions.  Here, we consider a
matrix model for the Polyakov loops in four dimensions and explain how
the theory may exhibit novel behavior in the large $\Nc$ limit.

The Polyakov loop potential can be computed analytically in several
limits including the high-temperature limit as we already saw in
Sec.~\ref{sec:perturbative} and the small-sphere limit as we will see
in Sec.~\ref{sec:sphere} later.  Here, let us adopt a simple and
general setup to perform an effective potential analysis.  We shall
ignore fluctuations and kinetic terms constituting $1/\Nc$
corrections, and then we can express the partition function in a form
of the group integration as follows,
\begin{equation}
  Z = \int \calD L\, e^{ - \Nc^2 V(L)}\;,
  \label{eq:Z_effGW}
\end{equation}
where 
the Polyakov loop
potential is postulated to have a power-series form in $|\ell_3|^2$,
i.e.\ 
\begin{equation}
  V(L) = -m^2  |\ell_3|^2 + \kappa_4 |\ell_3|^4
  + \kappa_6 |\ell_3|^6 + \dots\;.
  \label{eq:vL_effGW}
\end{equation}
Our goal is to carry out the group integration with respect to $L$ in
Eq~\eqref{eq:Z_effGW} except for $\ell_3$.  In order to achieve this,
we use a common trick to construct the so-called constrained effective
potential~\cite{KorthalsAltes:1993ca} introducing a delta function as
\begin{equation}
  Z = \int \calD L \int d\gamma\, \delta(\gamma - \ell_3)\,
  e^{ -\Nc^2 V(\gamma)}\;.
  \label{eq:z_constrained}
\end{equation}
Replacing the order of integration and using the Fourier
representation for the delta function we obtain,
\begin{equation}
  Z = \int \frac{\omega}{2\pi} \int d\gamma\,
  e^{ - i \Nc^2 \omega \gamma - \Nc^2 V(\gamma)}
  \int \calD L\, e^{ i \Nc^2  \omega \ell_3}\;.
  \label{eq:z_constrained_Zw}
\end{equation}
Now the group integration simplifies with the Haar measure or the
Vandermonde determinant taken into account explicitly, leading to
 \begin{equation}
   \int \calD L\, e^{ i \Nc^2  \omega \ell_3}
   = \int d\ell\,
   \exp \bigl[ i \Nc^2  \omega \ell - \Nc^2 V_{\rm Vdm} (\ell) \bigr]\;,
   \label{eq:vdm}
 \end{equation} 
and then the partition function after the $\omega$ and the $\gamma$
integrations should read,
\begin{equation}
  Z = \int d \ell\, e^{ -\Nc^2 V_{\rm eff}(\ell)}
\end{equation}
with the effective potential defined by 
\begin{equation}
  V_{\rm eff} (\ell) = V(\ell) + V_{\rm Vdm}(\ell)\;.
  \label{eq:veff_gw}
\end{equation}
From this expression it is clear that $V_{\rm Vdm}(\ell)$, the
explicit form of which we will discuss soon later, corresponds to
$-\ln H$ we have encountered in the strong coupling calculation in Sec.~\ref{sec:strong}.  It
is noted that the determination of $H$ for the SU(2) and the SU(3)
cases was straightforward directly from the Vandermonde
determinant~\eqref{eq:Haar}, but what we should consider here is to
find it for the general SU($\Nc\to \infty$) case.  To this
end, we compare the left-hand and the right-hand sides of
Eq.~\eqref{eq:vdm}.  Because $\Nc$ is infinitely large, the stationary
point approximation should work for the $\ell$ integration, and the
stationary condition with respect to $\ell$ leads to
\begin{equation}
  \tilde{\omega} \equiv i\omega
  = \frac{\partial V_{\rm Vdm}}{\partial \ell}\;.
  \label{eq:stationarity_effGW}
\end{equation}
This is a condition to fix $\ell(\omega)$, but we can conversely solve
$\omega$ for a given $\ell$.  Then, we shall consider the left-hand side of
Eq.~\eqref{eq:vdm} for a real-valued $\tilde{\omega}$.  Such a
replacement of $\omega\to\tilde{\omega}$ enables us to make use of the
results in the previous subsection, particularly the
solution~\eqref{Eq:rho_combined} and \eqref{Eq:w_combined} with the
mapping $2/\lambda \to \tilde \omega$.  Thus, the stationary point
solution reads,
\begin{equation}
  \ell_3(\tilde \omega)  =
  \begin{cases}
    \displaystyle 1 - \frac{1}{2 \tilde{\omega}}
    & \quad\text{if}\quad \tilde{\omega} > 1\;,\\
    \displaystyle \frac{\tilde{\omega}}{2}
    & \quad\text{if}\quad \tilde{\omega} \le 1\;.
  \end{cases}
  \label{Eq:w_combined_effGW}
\end{equation}
We can eliminate $\tilde{\omega}$ by combining
Eqs.~\eqref{eq:stationarity_effGW} and \eqref{Eq:w_combined_effGW}
to arrive at an equation to be integrated for $V_{\rm Vdm}$.  In this
way we can fix the explicit form of $V_{\rm Vdm}$ and the final
results are
\begin{equation}
  V_{\rm Vdm} (\ell ) =
  \begin{cases}
    \displaystyle \ell^2
    & \quad\text{if}\quad \ell < 1/2\;,\\
    \displaystyle -\frac{1}{2} \ln \bigl[ 2(1-\ell) \bigr]
    + \frac{1}{4}
    & \quad\text{if}\quad \ell \ge 1/2\;.
  \end{cases}
  \label{Eq:V_combined_effGW}
\end{equation}

The Vandermonde potential $V_{\rm Vdm}(\ell)$ can also be interpreted
as the Legendre transform of the left-hand side of Eq.~\eqref{eq:vdm}
with $i \omega$ playing a role of external field.
The original eigenvalue repulsion inherent in the Vandermonde
determinant is translated into a logarithmic singularity at
$\ell=1$, which guarantees that the Polyakov loop stays below the
unity for any $V(L)$ or any coefficients in the
parametrization~\eqref{eq:vL_effGW} in the total effective
potential~\eqref{eq:veff_gw}.

The most remarkable property of the Vandermonde potential
is that $V_{\rm Vdm}(\ell)$, $V_{\rm Vdm}'(\ell)$, and
$V_{\rm Vdm}''(\ell)$ are continuous functions, while the third
derivative, $V_{\rm Vdm}'''(\ell)$ is discontinuous at $\ell=1/2$.

\begin{figure}
  \includegraphics[width=0.48\linewidth]{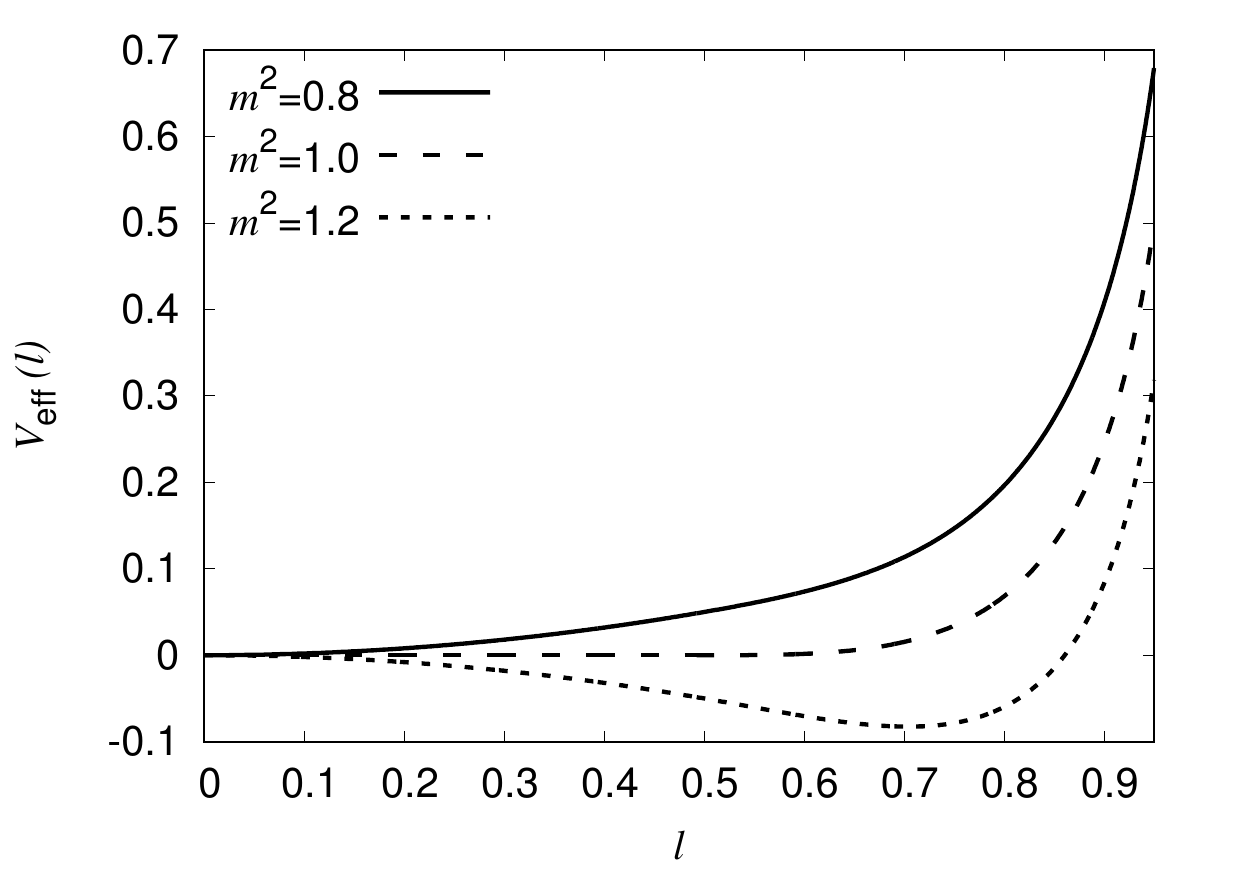}
  \includegraphics[width=0.48\linewidth]{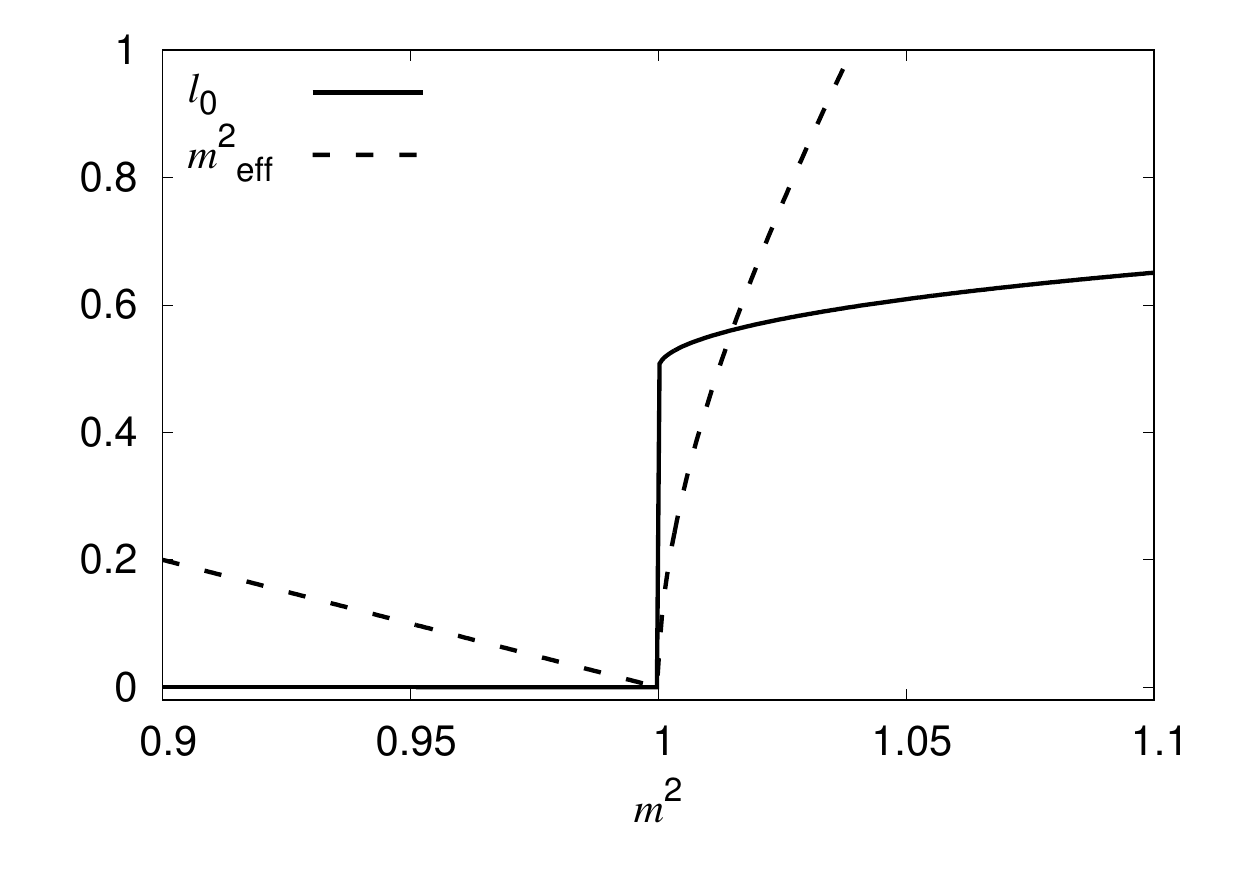}
  \caption{Effective potential at the transition as a function of the
    Polyakov loop (left) and the expectation value of the Polyakov
    loop and the effective mass squared as a function of $m^2$ (right).}
\label{fig:gweff_potential_loom_mass}
\end{figure}

The analysis of the phase diagram for the effective
potential~\eqref{eq:veff_gw} can be readily done.  Let us consider
only non-zero quadratic term first, that is, $m^2\neq 0$ and
$\kappa_n=0$ in Eq.~\eqref{eq:vL_effGW}.  In the large $\Nc$ limit the
stationary point approximation or the mean-field approximation would
be exact, and we can simply identify the minimum of the effective
potential as the expectation value $\Phi=\langle\ell\rangle$.  The
confining vacuum with a stable minimum at $\Phi=0$ is favored for any
$m^2<1$.  The potential around this minimum has only quadratic term.
For $m^2>1$ the sign of the quadratic term changes leading to the
deconfined phase with the expectation value given by
\begin{equation}
  \Phi = \frac{1}{2} \biggl(1 + \sqrt{1-\frac{1}{m^2}}\biggr)\;.
  \label{eq:l0_gw}
\end{equation}
The behavior of the effective potential around the critical point is
depicted in the left of Fig.~\ref{fig:gweff_potential_loom_mass}.
Also, we see that the Polyakov loop jumps from 0 to $1/2$ at the phase
transition at $m^2=1$, which is clearly seen in the right panel of
Fig.~\ref{fig:gweff_potential_loom_mass}.

As shown by the dashed line in the left panel of
Fig.~\ref{fig:gweff_potential_loom_mass} the potential behavior at
$m^2=1$ is quite peculiar;  it is flat and constantly vanishing for
$\ell<1/2$.  Near the onset, for $\ell \to 1/2^+$, the potential
arises from zero as $\sim 4/3(\ell-1/2)^3$.  The effective mass
(i.e.\ the potential curvature mass of the Polyakov loop) inferred
from $m_{\rm eff}^2 = \partial^2 V_{\rm eff} / \partial \ell^2|_{\ell=\Phi}$
is continuous and goes to zero regardless of whether the critical
point is approached by $m^2\to1^+$ or $m^2\to1^-$ as illustrated in
the right panel of Fig.~\ref{fig:gweff_potential_loom_mass}.
Therefore, we can say that the $m^2=1$ point exhibits aspects of both
first- and second-order phase transitions.  Additionally, when the
transition is approached by $m^2\to 1^+$, the expectation value of the
Polyakov loop demonstrates critical behavior associated with the
second-order transition;
$\Phi-1/2 \propto (\delta t)^\beta$ with a critical exponent
$\beta = 1/2$, where we introduce the dimensionless temperature as
$m^2-1\propto \delta t \equiv T/T_c - 1$.  Moreover, the specific heat
diverges at $m^2\to1^+$ as
$c_v \propto (\delta t)^{-\alpha}$ with $\alpha = 1/2$, while it is
vanishing for $m^2<1$.  By adding an external field~\footnote{For any
  non-zero external field the transition is of third order.  This is
  due to flatness of the potential, i.e.\ the absence of the potential
  barrier at the transition for zero external
  field~\cite{Green:1983sd,Schnitzer:2004qt,Dumitru:2004gd}.},
we can extract another critical exponent as $\delta = 2$.  It is
remarkable that the critical exponents satisfy ordinary Griffith's
scaling relation; $2-\alpha = \beta (1+\delta)$.  For this reason this
phase transition is often referred to as a
\textit{critical first-order} phase transition.

\begin{figure}
  \centering
  \includegraphics[width=0.48\linewidth]{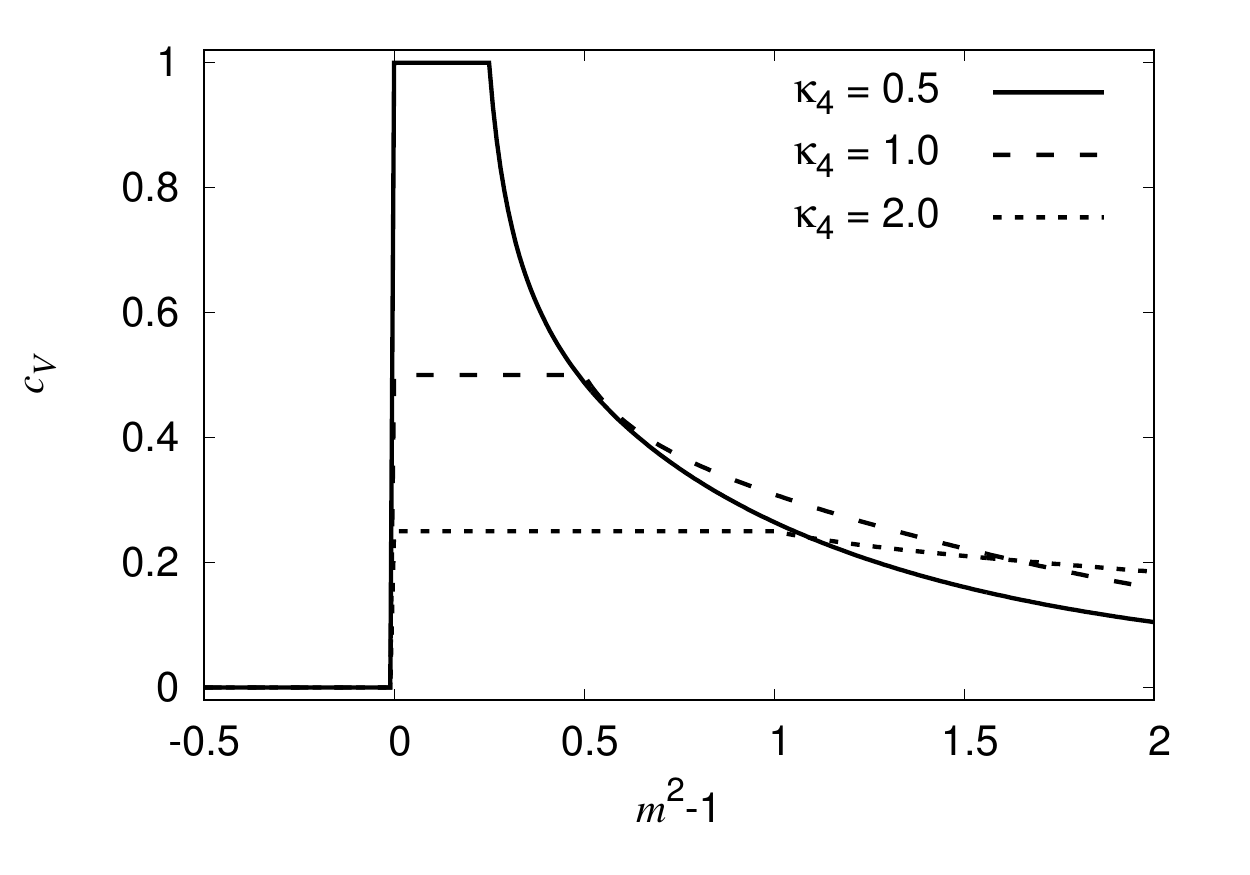}
  \caption{The specific heat, $c_V$, as a function of
    $m^2-1$ for positive quartic couplings.}
\label{fig:cv_kappa4}
\end{figure}

Before closing this subsection, let us discuss effects of non-zero
$\kappa_4$ in the potential~\eqref{eq:vL_effGW}.  For $\kappa_4>0$
some simple computations show that there is an ordinary second-order
phase transition at $m^2 = 1$; this conclusion is not affected by the
non-polynomial part of the Vandermonde potential because the
expectation value of the Polyakov loop develops continuously from
zero.  What is special about the system is that there is one more
phase transition at $m^2 = 1 + \kappa_4/2$.  At this latter critical
value of $m^2$, the expectation value of the Polyakov loop is $1/2$
and the third derivative of the effective potential with respect to
$m^2$ is discontinues.  Thus, this transition at $m^2=1+\kappa_4/2$ is
of third order.  A more detailed analysis presented in
Ref.~\cite{Aharony:2003sx} shows that this transition is associated
with non-zero expectation values of the higher order Polyakov loops,
defined as ${\rm tr} (L^n) $ with $n>1$. 
Indeed, for $m^2<1 + \kappa_4/2$, the higher order Polyakov loops
vanish; their values are nonzero only for $m^2> 1 + \kappa_4/2$.

We illustrate the transition behavior by plotting $c_V$ as a function
of $m^2$ (which is physically interpreted as the temperature) in
Fig.~\ref{fig:cv_kappa4}.  As is common in the mean-field
approximation, the specific heat is discontinuous at the second-order
phase transition, $m^2=1$, due to vanishing critical exponent for the
specific heat.  The third-order phase transition at
$m^2 = 1 + \kappa_4/2$ leads to a kink structure in the specific heat.
The maximum of $c_V$ is proportional to $1/\kappa_4$; thus when
$\kappa_4$ approaches zero, two transitions meet at $m^2=1$ and $c_V$
diverges there.

\begin{figure}
  \centering
  \includegraphics[width=0.8\linewidth]{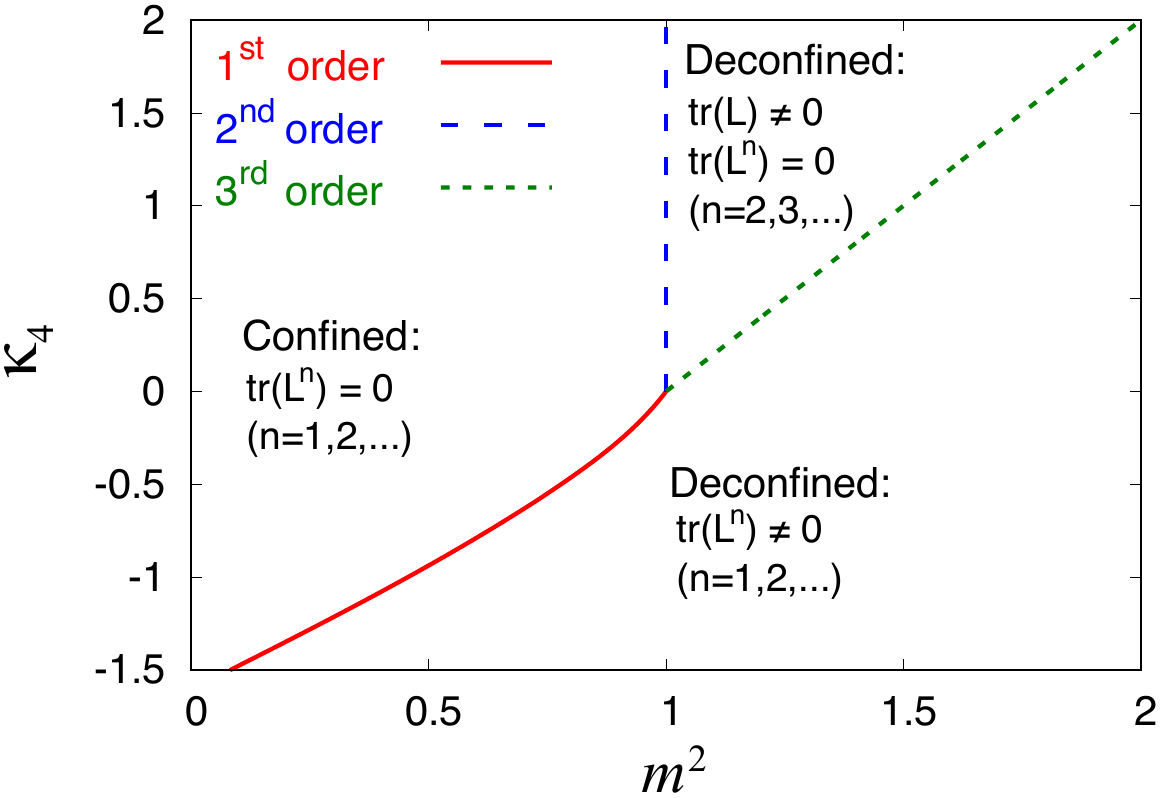}
  \caption{The phase diagram for the effective potential with non-zero
    quartic coupling.}
\label{fig:phasediagramGW}
\end{figure}

For $\kappa_4<0$, there is a single transition of first order at
$m^2<1$.  For the first-order phase transition in general, the
transition line can be found from the energy comparison of two minima
of the effective potential.  This results in a parametric form,
\begin{align}
  m^2 & = \frac{e^{2 v}}{(2 e^v -1)^2} ( 3 + 4 v - 2 e^v)\;, \\
  \kappa_4 & =  \frac{8 e^{4 v}}{(2 e^v -1)^4} ( 1 +  v - e^v)
\end{align}
with some $v$ ranging from $0$ to $\infty$.  In this case, at the
phase transition, the Polyakov loop jumps from $0$ to a value greater
than $1/2$.  We summarize these observations in the phase diagram
illustrated in Fig.~\ref{fig:phasediagramGW}.  All three transitions
meet at a supercritical point, where the system exhibits properties of
both first- and second-order phase transitions, as we already stated.

In the next subsection, we consider a weakly coupled theory on a small
3-sphere, i.e.\ $S^1\times S^3$; although the derivation is more
cumbersome, the main features are essentially the same as those in the
simple matrix model discussed above, including the structure of the
phase diagram presented in Fig.~\ref{fig:phasediagramGW}.  We note
that a similar effective theory with a quadratic deformation of the
perturbative action was considered in Ref.~\cite{Pisarski:2012bj}.
Although some quantitative aspects of the phase transition are
different (e.g.\ the critical exponents significantly differ), the
qualitative pattern of the phase transition and the phase diagram
share the same features with what we have seen in this subsection.

\subsubsection{Weakly coupled theory on a small 3-sphere}
\label{sec:sphere}

As was discussed in Sec.~\ref{sec:perturbative} the pure gluonic
thermodynamics allows for an analytical treatment only at very high
temperature.  At temperatures in the vicinity of the deconfinement 
temperature, the numerical methods on lattice discretized spacetime
are most useful~\cite{Boyd:1996bx,Lucini:2012gg,Borsanyi:2012ve}, for
the coupling constant is not necessarily small enough there.
Nevertheless, there is a theoretical possibility to study the
thermodynamics and the deconfinement phenomenon in the weak coupling
limit using perturbative approaches, if the theory is formulated on a
compact space; more specifically a three-dimensional sphere $S^3$ with
finite radius $R$.  This provides us with an additional parameter $R$
or a dimensionless combination, $R\LQCD$, which interpolates between
the strong coupling regime for large $R$ and the weak coupling regime
for small $R$.  In particular, we can anticipate that the perturbation
theory should work when the compactification radius is much smaller
than the strong interaction scale, i.e.\ $R\LQCD \ll 1$.

Phase transitions for a theory on a compact manifold should be always
continuous and thermodynamic quantities are non-singular due to a
limited number of degrees of freedom in a finite volume.
There is a notable exception of $\Nc\to\infty$ that makes the number
of degrees of freedom infinite regardless of the volume of the system;
in what follows below we consider the phase transition of a weakly
coupled theory on a small 3-sphere in the
``thermodynamic limit'' in a sense of the $\Nc\to\infty$ limit.

At weak 't~Hooft coupling $\lambda = g^2 \Nc$, the partition function
can be reduced to an integral with respect to a single unitary matrix
by leaving the zero mode only after integrating out heavier modes with
a gap of order of $1/R$.  Alternatively, the leading order results of
the perturbative expansion in $\lambda$ (i.e.\ free theory) can also
be obtained by counting the gauge invariant states in free Yang-Mills
theory, as was done in Refs.~\cite{Aharony:2003sx,Sundborg:1999ue}.
Here we follow the former approach along the lines outlined in
Sec.~\ref{sec:perturbative}.

The pure gluonic or Yang-Mills partition function can be expressed as
\begin{equation}
  Z = \int d\alpha \int \calD A_i\, \Delta_A \Delta_\alpha\,
  e^{-S_{\rm YM}(A, \alpha)}\;,
  \label{Eq:S3_Z}
\end{equation}
where $\Delta_A$ is the Faddeev-Popov determinant associated with the
partial gauge fixing, $\partial_i A^i = 0$, while $\Delta_\alpha$
corresponds to the residual gauge fixing,
$\partial_0 \alpha = 0$, where the zero-mode gauge potential is
defined as
$\alpha(t) \equiv \frac{1}{V_3} \int_{S^3} A_0$.  An explicit
evaluation of $\Delta_\alpha$ gives $d\alpha \Delta_\alpha = d U$,
where $d U$ represents a group integration for the unitary matrix 
$U = e^{i \beta \alpha }$ with $\beta$ being a period of temporal
$S^1$.  Therefore, the partition function can be rewritten as follows;
\begin{equation}
  Z = \int dU\, e^{-S_{\rm eff}(\alpha)} \quad\text{with}\quad
  e^{-S_{\rm eff}(\alpha)} = \int \calD A_i\, \Delta_A\,
  e^{-S_{\rm YM}(A, \alpha)}\;.
  \label{Eq:S3_Seff}
\end{equation}
The effective action $S_{\rm eff}(\alpha)$ can be evaluated
diagrammatically as a power-series of the coupling constant.  The
lowest order calculation is rather straightforward and repeats the
one-loop computation with only one minor difference;  the eigenvalues
of the Laplacian operator should be replaced with those for vector
spherical harmonics, which is denoted by $\Delta^2$ here.  On a sphere
$S^3$, the eigenvalues are degenerate, and for an integer
$\Delta \ge 1$, the degeneracy factor is given by
$n_\Delta = 2 (\Delta^2 - 1)$.  From now on, we rescale all quantities
to absorb the mass dimension by the appropriate power of $R$; in other
words, we choose a special unit with which $S^3$ is a unit-radius
sphere.  One-loop calculation immediately leads us to
\begin{equation}
  S_{\rm eff}(\alpha) = \frac{1}{2} \tr_c \sum_\Delta n_\Delta \bigl[
    \beta \Delta + \ln(1 - e^{-\beta \Delta + i \beta \alpha})
    + \ln(1 - e^{-\beta \Delta - i \beta \alpha}) \bigr]\;,
  \label{Eq:S3_Seff_prefinal}
\end{equation}
where $\tr_c$ is a trace over the color adjoint representation that
$\alpha$ belongs to.  A more transparent representation of the
effective action emerges once the logarithms are expanded.  We can
then take the summation with respect to the Laplacian eigenvalues
$\Delta$, and we also switch to the fundamental representation.  Then,
we find,
\begin{equation}
  S_{\rm eff}(\alpha) = \frac{1}{2} \beta\Nc^2 \sum_\Delta \Delta n_\Delta 
  - \sum_{n=1}^\infty \frac{z(e^{-\beta n})}{n} \tr(U^n)\,
  \tr({U^\dagger}^n)\;,
  \label{Eq:S3_Seff_final}
\end{equation}
where
\begin{equation}
  z(e^{-\beta n}) = \sum_{\Delta=1}^\infty n_\Delta e^{-\beta\Delta n}
  = 2\,\frac{3 e^{-2 \beta n} - e^{-3 \beta n}}{(1-e^{-\beta n})^3}\;.
  \label{Eq:S3_z}
\end{equation}
Using these results, the partition function can be readily evaluated.
In terms of the eigenvalues of the unitary matrix
[see Eq.~\eqref{Eq:dU_GW}], the partition function takes the following
form;
\begin{equation}
  Z = \int dU\, e^{-S_{\rm eff}} = \int d q_i \exp\biggl\{
  - \sum_{i\ne j} \ln \bigl| \sin[\pi(q_i-q_j)] \bigr|
  - \sum_{i,j} \sum_n \frac{z(e^{-\beta n}) }{n}
  \cos[2 n\pi (q_i-q_j)] \biggr\}\;.
  \label{Eq:S3_Z_fin}
\end{equation}
We note that both terms can be combined into a common infinite sum by
virtue of
\begin{equation}
  \ln \bigl|\sin(\pi x)\bigr| = -\ln 2
  - \sum_{n=1}^\infty \frac1n \cos (2 n \pi x)\;.
\end{equation}

In the large $\Nc$ limit, the saddle point approximation becomes exact
and the partition function and the distribution of the eigenvalues can
be found analytically; see Refs.~\cite{Aharony:2003sx,Sundborg:1999ue}
for details.  Numerical analysis for finite $\Nc$ was performed in
Ref.~\cite{Hollowood:2012nr} (see also Ref.~\cite{Christensen:2013xea}
for higher order extensions).

Close to the deconfinement transition $z(e^{-\beta_c}) = 1$, where the
sign of $|{\rm tr} L|^2$ changes rendering an instability of the
trivial solution, the partition function can be approximated by the
effective matrix model described in the previous subsection with a
non-trivial quadratic coupling.  Actually, by taking into account the
first perturbative correction to the free limit, we get the effective
matrix theory with negative $\kappa_4$~\cite{Aharony:2005bq}.  The
analysis of the phase structure thus repeats the previous subsection.

\section{Coupling to Quarks}
\label{sec:quarks}

QCD has not only gluons but quarks also, and the inclusion of quarks
would drastically change confinement argument based on center symmetry.
Quarks explicitly break center symmetry since the center twisted gauge
transformation changes the boundary condition for fermions;  after the
transformation they no longer satisfy the anti-periodic boundary condition
but are multiplied by a center element;  the transformed field,
$V(x)\psi(x)$,  with the boundary condition~\eqref{eq:centertrans}
satisfies,
\begin{equation}
  V(x_4=\beta)\psi(x_4=\beta) = -z_k\cdot V(x_4=0)\psi(x_4=0)\;,
\end{equation}
so that $\psi(x)$ is sensitive to the center. This also means that
quark excitations are significantly affected by the realization of
center symmetry in a gluonic medium.

In this section, we will first see how quark contributions would
change the perturbative Polyakov loop potential, and next we will turn
our view point over to discuss how the Polyakov loop background can
in effect capture color screening effects on quarks.  This latter
observation underlies the Polyakov loop augmented building of chiral
effective models.  The idea can be easily generalized to not only
color fundamental quarks but also color adjoint gluons.

\subsection{Polyakov Loop Potential from Quarks}
\label{sec:PLquarks}

Even without concrete calculations, it is straightforward to rewrite
Eq.~\eqref{eq:weiss_integ} to infer the quark one-loop contribution to
the Polyakov loop potential~\cite{Weiss:1981ev}.  In
Eq.~\eqref{eq:weiss_integ} gluons belong to the adjoint
representation, and this is why $q_{ij}$ appears there.  For quarks
$q_{ij}$ should be replaced with $q_i$ in the fundamental
representation.  For $\Nf$ massless quarks including also a finite
chemical potential $\mu$, the potential contribution immediately reads
from such a mapping as
\begin{align}
  V_{\rm quark}[q] &= -2\Nf T V\int\frac{d^3 p}{(2\pi)^3} \sum_{i=1}^{\Nc}
  \Bigl[ \ln\bigl( 1 + e^{-\beta(|\bp|-\mu)+2\pi i q_i} \bigr)
  + \ln\bigl( 1 + e^{-\beta(|\bp|+\mu)-2\pi i q_i} \bigr) \Bigr]
  \notag\\
  &= -\Nf V\frac{4\pi^2}{3\beta^4}\sum_{i=1}^{\Nc} \biggl( q_i+\frac{1}{2}
  -i\frac{\beta\mu}{2\pi}\biggr)_{\text{mod1}}^2 \biggl[ \biggl(
  q_i + \frac{1}{2} - i\frac{\beta\mu}{2\pi}\biggr)_{\text{mod1}}
  - 1 \biggr]^2\;.
\label{eq:weiss_quark}
\end{align}
This additional contribution to the Weiss potential manifestly breaks
translational symmetry and thus center symmetry.

\begin{figure}
  \centering
  \includegraphics[width=0.47\textwidth]{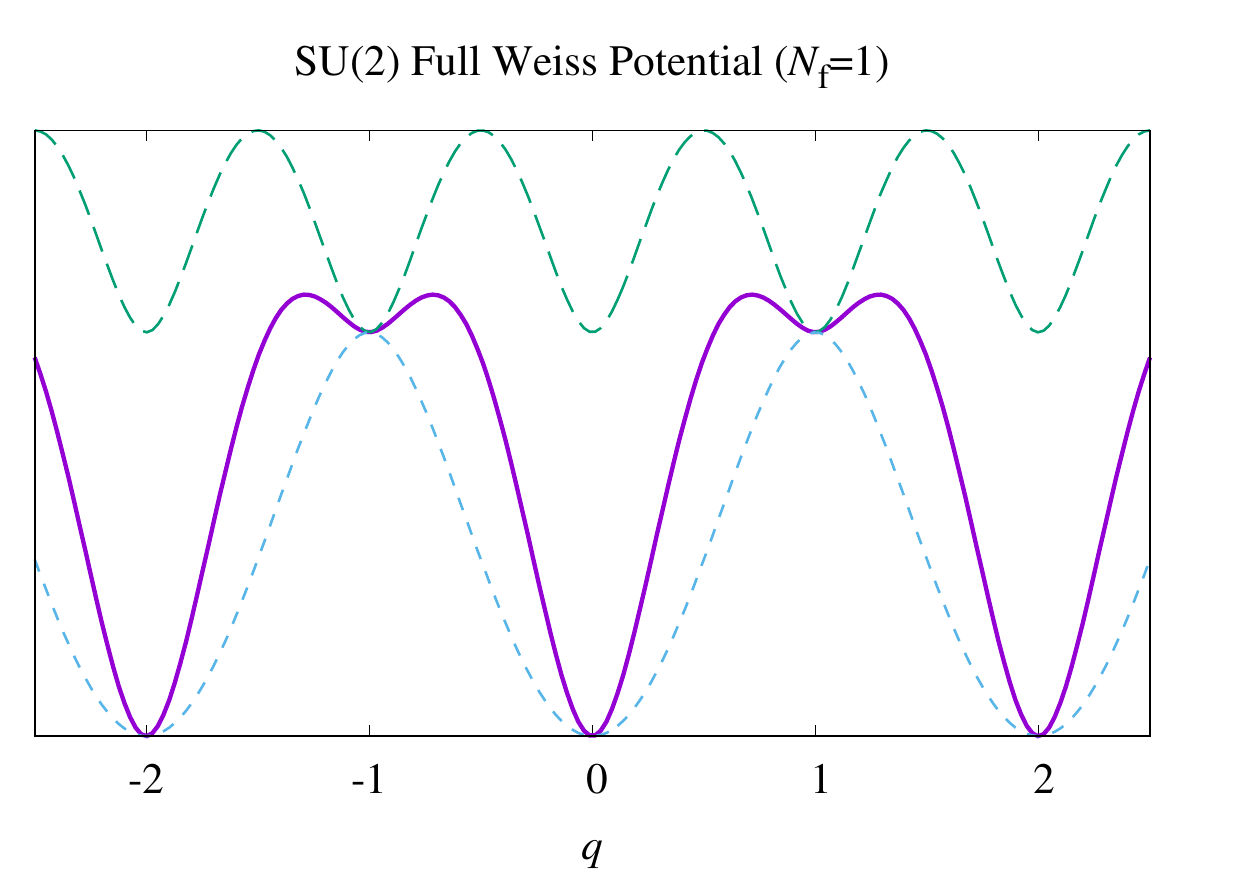}
  \includegraphics[width=0.47\textwidth]{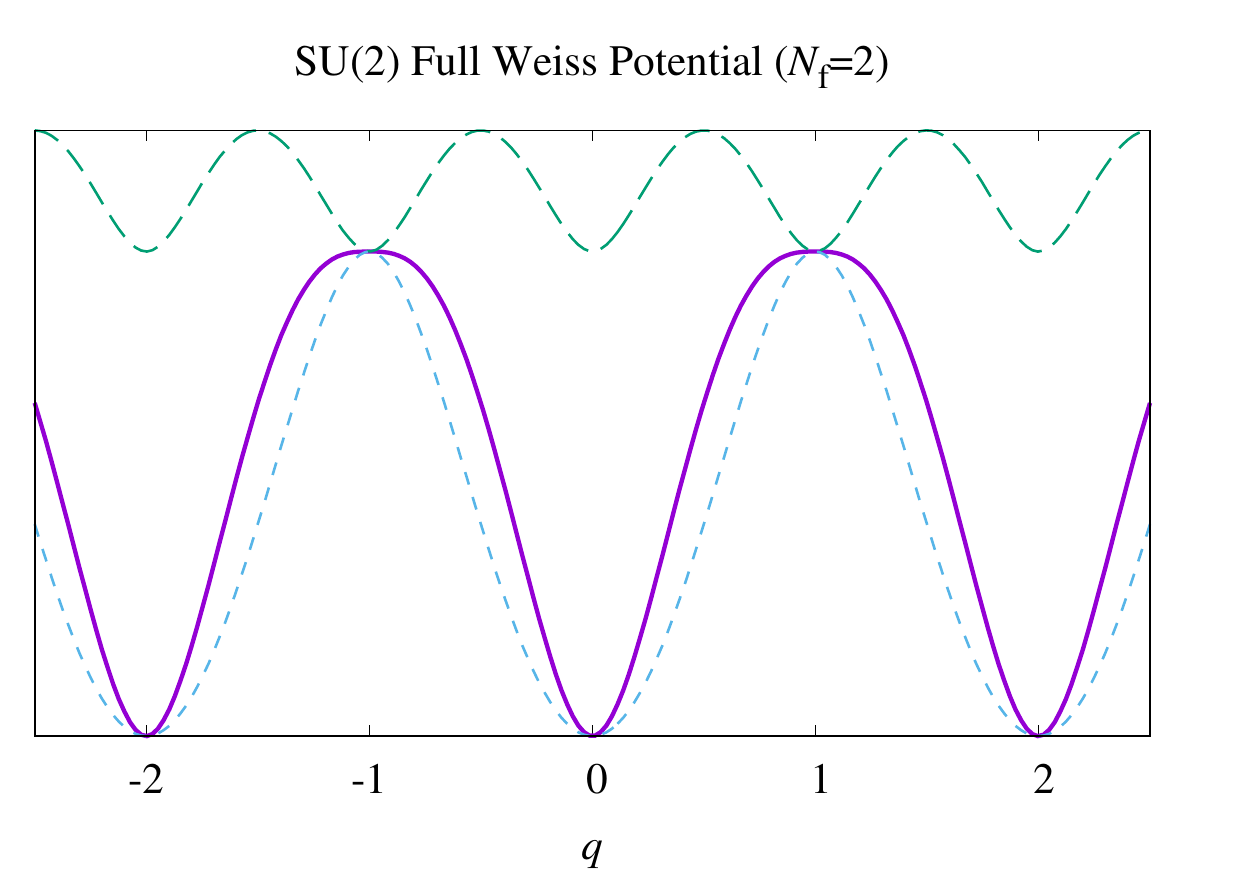}
  \caption{SU(2) full Weiss potential with the gluonic contribution
    (shown by the dashed lines) together with the quark contribution
    (shown by the dotted lines) for $\Nf=1$ (left) and $\Nf=2$
    (right) at $\mu=0$.}
  \label{fig:su2fullweiss}
\end{figure}

To visualize the center symmetry breaking more transparently, we make
a plot for the SU(2) full (i.e.\ gluonic $+$ quark) Weiss potential
for $\Nf=1$ and $\Nf=2$ in Fig.~\ref{fig:su2fullweiss} in the $\mu=0$
case.  As is clearly noticed in Fig.~\ref{fig:su2fullweiss} the
perturbative vacuum at $q=0$ is still a global minimum of the
effective potential, while the center transformed point at $q=1$ is at
best only a local potential minimum due to the center symmetry breaking.
Interestingly, for $\Nf=1$ as in the left panel of
Fig.~\ref{fig:su2fullweiss}, there still remains a meta-stable state
corresponding to a local minimum at $q=1$.  With $\Nf=2$ flavors the
local minimum disappears and only the perturbative vacuum is
energetically allowed.

From the potential curvature around $q=0$ we can deduce the Debye mass
correction from quarks (including $\mu$).  The perturbative vacuum
becomes more stabilized by an increase in the Debye mass by
\begin{equation}
  \delta m_E^2 = \Nf \biggl( \frac{g^2 T^2}{6}
  + \frac{g^2 \mu^2}{2\pi^2}\biggr)\;,
\end{equation}
which correctly reproduces the expression for the Debye
screening mass arising from quark one-loop contributions
at finite $T$ and $\mu$.  Now, one may be naturally
tempted to look more closely at global structures of the finite-$\mu$
Weiss potential away from the perturbative vacuum.  However,
Eq.~\eqref{eq:weiss_quark} is complex for non-zero $\mu$, and then we
can no longer interpret it as a thermodynamic potential.  In other
words, for complex $V_{\rm quark}[q]$, the most favored $q$ is not
uniquely determined energetically from thermodynamic principles.

The origin of theoretical difficulties from complex $V_{\rm quark}[q]$
would be more understandable if we express Eq.~\eqref{eq:weiss_quark}
in a different but equivalent way.  To write explicit expressions
down, let us consider the QCD ($\Nc=3$) case specifically below (note
that the Polyakov loop is always real in the $\Nc=2$ case and the sign
problem becomes serious for $\Nc\ge 3$).  In this case with
$\Nc=3$, it is easy to take the sum over $i=1,\dots\Nc=3$ explicitly
before the momentum integration in Eq.~\eqref{eq:weiss_quark},
so that we finally arrive at
\begin{equation}
  \begin{split}
    V_{\rm quark}[q]
    &= -2\Nf T V \int\frac{d^3 p}{(2\pi)^3}
    \tr\Bigl[ \ln\bigl[ 1+L\,e^{-\beta(\varepsilon_p - \mu)} \bigr]
      + \ln\bigl[ 1+L^\dag e^{-\beta(\varepsilon_p + \mu)} \bigr] \Bigr]\\
    &= -2\Nf T V \int\frac{d^3 p}{(2\pi)^3}
    \Bigl[ \ln\bigl(1+3\ell\,e^{-\beta(\varepsilon_p - \mu)}
    + 3\ell^\ast e^{-2\beta(\varepsilon_p - \mu)}
    + e^{-3\beta(\varepsilon_p - \mu)} \bigr) \\
    & \qquad\qquad\qquad\qquad\quad
    + \ln\bigl(1+3\ell^\ast e^{-\beta(\varepsilon_p + \mu)}
    + 3\ell\, e^{-2\beta(\varepsilon_p + \mu)}
    + e^{-3\beta(\varepsilon_p + \mu)} \bigr) \Bigr]\;,
  \end{split}
  \label{eq:Vquark}
\end{equation}
where a finite mass is introduced through
$\varepsilon_p\equiv\sqrt{p^2+m^2}$.  We can then precisely spot what
makes $V_{\rm quark}[q]$ complex.  In general $\ell$ is a complex
number, but as long as $\mu=0$, the first and the second lines of
Eq.~\eqref{eq:Vquark} are complex conjugate to each other and the sum
of them takes a real number.  A finite $\mu$ would destroy this
balance and the complex nature of $V_{\rm quark}[q]$ is attributed to
complex $\ell$ and $\ell^\ast$ with unbalanced weights.
Such an observation of complex $V_{\rm quark}[q]$ is
a very concrete realization of the sign problem, which we will closely
discuss later in Sec.~\ref{sec:density}.

\subsection{Polyakov Loop in Chiral Models}
\label{sec:chiral_models}

It is worth noting that the general form of $V_{\rm quark}[q]$ in
Eq.~\eqref{eq:Vquark} is a result from the one-loop quark integration,
and nevertheless, it is a full expression in the quark sector for
given gauge (Polyakov loop) configurations.
Therefore, the validity region of
Eq.~\eqref{eq:Vquark} is not necessarily restricted to the
perturbative regime only but can be extended to more general regimes
where the quasi-particle picture makes sense.

This opens a wider range of applications of Eq.~\eqref{eq:Vquark}
beyond the perturbative Weiss potential.  Here, we will see a
successful example of utilizing Eq.~\eqref{eq:Vquark} in quark
models to consider the Polyakov loop effects on chiral symmetry.

\subsubsection{PNJL model}
\label{sec:PNJL}

The effective potential~\eqref{eq:Vquark} (together with the gluonic
contributions) is supposed to be minimized
to determine the expectation value of $q$, and at the same time, we
can make another interpretation for Eq.~\eqref{eq:Vquark};  for a given
$q$, this expression represents how the quark excitations are affected
in the presence of the Polyakov loop background.  Actually, the
physical meaning of Eq.~\eqref{eq:Vquark} would become even clearer in
such an interpretation;  in fact, $\ell$ and $\ell^\ast$ emerge in
Eq.~\eqref{eq:Vquark} as colored chemical potentials.

If $\ell$ and $\ell^\ast$ are just the unity, the logarithmic terms
simplify as
$\ln[1+3e^{-\beta(\varepsilon_p\mp\mu)}+3e^{-2\beta(\varepsilon_p\mp\mu)}
+e^{-3\beta(\varepsilon_p\mp\mu)}]=3\ln[1+e^{-\beta(\varepsilon_p\mp\mu)}]$.
Thus, $V_{\rm quark}[q=1]$ is nothing but the grand canonical
partition function for free quarks.  On the other hand, in the limit
of vanishing $\ell$ and $\ell^\ast$, $V_{\rm quark}[0]$ is reduced to
the grand canonical partition function for free particles with an
effective temperature lowered by factor $3$.  In this case of
$\ell=\ell^\ast=0$, all single-quark-type excitations
($\propto e^{-\beta(\varepsilon_p\mp\mu)}$) and diquark-type excitations
($\propto e^{-2\beta(\varepsilon_p\mp\mu)}$) are diminished by color
screening and only color-singlet combinations of three quarks
survive.

It is obvious from the above arguments that, if some sort of phase
transition occurs for $\ell=1$ at $T=\Tc$ by quasi-quark excitations,
the phase transition under $\ell=0$ would be delayed to $T>3\Tc$.  (The
  underlying mechanism for the increase in the transition temperature
  parallels the canonical ensemble approach; see
  Refs.~\cite{Oleszczuk:1993yf,Fukushima:2002bk}.)
One direct implication from this observation would be the finite-$T$ chiral
phase transition affected by external $\ell$ and $\ell^\ast$.  This
possibility was pioneered in Ref.~\cite{Meisinger:1995ih} by means of
the Nambu--Jona-Lasinio (NJL) model to describe the chiral sector.
It was then Ref.~\cite{Fukushima:2003fw} that first demonstrated a
successful treatment of both the chiral order parameter and the
Polyakov loop as dynamical degrees of freedom, which was later named
as the PNJL (i.e.\ Polyakov loop augmented Nambu--Jona-Lasinio) model
in Ref.~\cite{Ratti:2005jh}.
For chiral symmetry in QCD and the
details about the NJL model,
Refs.~\cite{Klevansky:1992qe,Hatsuda:1994pi,Buballa:2014tba} are
comprehensive reviews.
For more modern view points including recent
highlights from the lattice-QCD simulations, see
Ref.~\cite{Fukushima:2013rx}, and
for a related attempt of the first successful treatment of
both the Polyakov loop and the glueball states.
see Ref.~\cite{Sannino:2002wb}.  Here, since the main focus in this
review is the Polyakov loop physics, we shall give a minimal
explanation about chiral symmetry and its breaking.  If there are
$\Nf$ massless Dirac fermions, they are decomposed into $\Nf$
right-handed Weyl fermions and $\Nf$ left-handed Weyl fermions.  The
Dirac operator is invariant under unitary rotations in $\Nf$ flavor
space.  Thus, an idealized version of QCD with $\Nf$ massless flavors
would accommodate the following global symmetry:
\begin{equation}
  {\rm U}(\Nf)_{\rm L} \times {\rm U}(\Nf)_{\rm R}
  = {\rm SU}(\Nf)_{\rm L} \times {\rm SU}(\Nf)_{\rm R}
    \times {\rm U}(1)_{\rm A} \times {\rm U}(1)_{\rm V}\;.
\end{equation}
Chiral symmetry refers to
${\rm SU}(\Nf)_{\rm L}\times {\rm SU}(\Nf)_{\rm R}$, which is
spontaneously broken down to a vectorial subgroup,
${\rm SU}(\Nf)_{\rm V}$, that is a symmetry under simultaneous
rotations of right-handed and left-handed fermions.  We note that
${\rm U}(1)_{\rm A}$ is explicitly broken by the axial anomaly.  These
symmetry breaking patterns constrain how the low-energy limit of QCD
should look like.  In this way, chiral symmetry and its breaking
provide us with an important guiding principle to build low-energy
effective models of QCD.\ \ A condensate of quark and anti-quark, commonly
called the chiral condensate, is an order parameter for the
spontaneous breaking of chiral symmetry, i.e.\ chiral order parameter.
We will denote it as
$\langle\bar{q}q\rangle$ in the general context, or sometimes use
$\langle\bar{u}u\rangle$, $\langle\bar{d}d\rangle$, and
$\langle\bar{s}s\rangle$ when different flavors need to be distinguished.

The first physics motivation to build such a model with both the chiral
order parameter and the Polyakov loop was a challenge to understand
why two (approximate) critical temperatures are located so closely to
each other as observed in the lattice-QCD simulation.  The chiral
phase transition is associated with an in-medium change in terms of
dressed quark masses including interaction clouds, while colored
excitations are liberated at the deconfinement phase transition.
These are apparently different physics phenomena, and there is no
\textit{a priori} reason why they are strongly entangled.  One might
think that QCD has a unique scale, $\LQCD$, which would explain a
single $\Tc$.  However, such an argument would not be strong enough to
constrain anything beyond order estimates.  As explained above, chiral
symmetry is exact
only when all quark masses are zero, i.e.\ $m=0$, and center symmetry
exists only in a pure gluonic medium, i.e.\ $m=\infty$.  Hence, two
phenomena belong to two opposite limits from QCD, and the critical
temperatures may well be different by a factor.  Summarizing this makes the
following table;
\begin{center}
  \begin{tabular}{cp{2em}cp{2em}c}
    \fbox{$m=\infty$ limit} && \fbox{$0<m<\infty$} && \fbox{$m=0$ limit} \\
    Center Symmetry & $\Longleftarrow$ & No Exact Symmetry
    & $\Longrightarrow$ & Chiral Symmetry \\
    broken at $T_D(m=\infty)$ && $T_D(m) \simeq T_\chi(m)$
    && broken at $T_\chi(m=0)$
  \end{tabular}
\end{center}
It is very important to keep in mind that the critical temperatures,
$T_D$ for deconfinement and $T_\chi$ for chiral restoration, are
well-defined only in the quenched ($m=\infty$) and the chiral ($m=0$)
limits, respectively.  A finite (non-zero and non-diverging) mass
induces explicit symmetry breaking as if it were like a magnetic field
in spin systems.  There is then no clear way to define the critical
temperature for smooth crossover, unless a first-order phase
transition persists.  Usually some ``prescription'' to defined the
so-called pseudo-critical temperature is employed to define $T_D$ and
$T_\chi$ for $0<m<\infty$.  One of the most reasonable choices is the
peak temperature of the order parameter susceptibility $\chi(T)$;  the
pseudo-critical temperature $\Tc$ maximizes $\chi(T)$.  Another choice
would be the inflection temperature of the order parameter;  $\Tc$
maximizes the $T$-derivative of the order parameter.  The lattice-QCD
simulation for various quark masses strongly suggests that
$T_D(m)\simeq T_\chi(m)$ for any $m$.

\begin{figure}
  \centering
  \includegraphics[width=0.35\textwidth]{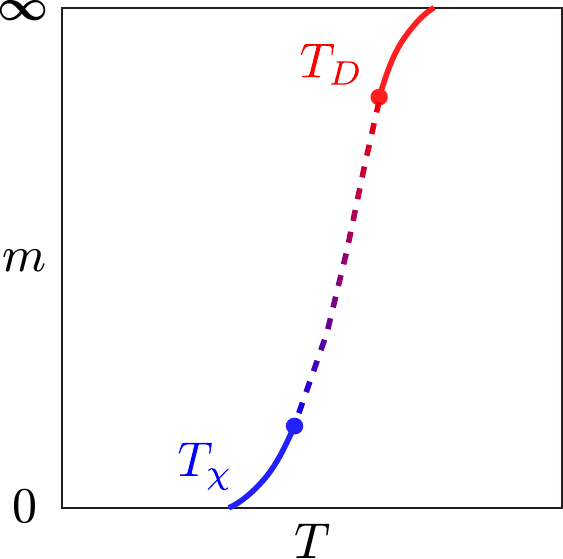}
  \caption{Schematic picture of a scenario on the possible connection
    between the chiral and the deconfinement critical lines.}
  \label{fig:link}
\end{figure}

Such a tight connection between two crossovers is schematically illustrated in
Fig.~\ref{fig:link}.  The deconfinement point at $m=\infty$ starts
with a first-order phase transition for $\Nc \ge 3$ and the
first-order line ends at a critical point (which will be more
explained later in Sec.~\ref{sec:critical}).  On the other hand, at $m=0$
the chiral phase transition is in general of first order again for
$\Nf \ge 3$ and the first-order line again ends at a critical point.
It is likely that these two critical points of deconfinement and
chiral restoration are connected by a \textit{single} line of
crossover, which is a scenario proposed in Ref.~\cite{Hatta:2003ga}.
In physics language the scenario in Fig.~\ref{fig:link} implies that
the sigma meson and the glueball are mixed together leading to a
single critical field out of the chiral condensate and the Polyakov
loop.  According to this scenario, it is expected that the low-lying
scalar glueball should become light near the critical
point, as partially confirmed in the lattice QCD simulation~\cite{Ishii:2001zq}.

One appreciable feature of the PNJL model is that it can provide us
with a quantitative tool to investigate simultaneous phase transitions
of chiral symmetry restoration and quark deconfinement, which goes far
beyond a qualitative conjecture in Ref.~\cite{Hatta:2003ga} and
a parametrization in Ref.~\cite{Mocsy:2003qw}.  To see the pragmatic strength of
the PNJL model, let us elaborate the model details below according to
Ref.~\cite{Fukushima:2008wg}.

In the conventional NJL model the QCD interaction is approximated by a
point-like four-quark vertex.  The simplest chiral symmetric combination of
four-quark interaction is,
\begin{equation}
  \calL_S = \frac{g_S}{2} \sum_{a=0}^{\Nf^2-1}
  \Bigl[ (\bar{\psi}\lambda_a\psi)^2
  + (\bar{\psi}i\gamma_5\lambda_a\psi)^2 \Bigr]\;.
\label{eq:Ls}
\end{equation}
For the $\Nf=3$ case $\lambda_a$ are Gell-Mann matrices in flavor
space.  This interaction is invariant under flavor rotations having
${\rm U}(\Nf)_{\rm L}\times{\rm U}(\Nf)_{\rm R}$ symmetry.  Among this
flavor symmetry, ${\rm U}(1)_{\rm A}$ must be broken by the axial
anomaly as mentioned above, and the explicit breaking can be
incorporated by the Kobayashi-Maskawa-'t~Hooft (KMT)
interaction~\cite{Kobayashi:1970ji,tHooft:1976up} whose explicit form is,
\begin{equation}
  \calL_A = g_D \bigl[ \det\bar{\psi} (1-\gamma_5)\psi + \text{h.c.}
  \bigr]\;,
\label{eq:KMT}
\end{equation}
where the determinant is taken with respect to flavor indices.  Let us
concretely address the $\Nf=3$ case specifically including strange
quarks for later convenience.

For $\Nf=3$ there are three independent chiral condensates, namely,
$\langle\bar{u}u\rangle$, $\langle\bar{d}d\rangle$, and
$\langle\bar{s}s\rangle$.  Non-zero values of these condensates break
chiral symmetry ${\rm SU}(3)_{\rm L}\times{\rm SU}(3)_{\rm R}$
(if $m_u=m_d=m_s=0$) down to ${\rm SU}(3)_{\rm V}$ leading to 8
Nambu-Goldstone (NG) bosons.  In the mean-field approximation,
four-fermionic interactions are approximated as
\begin{equation}
  (\bar{u}u)^2 = [(\bar{u}u-\langle\bar{u}u\rangle)
    + \langle\bar{u}u\rangle]^2
  \;\simeq\; -\langle\bar{u}u\rangle^2 + 2\langle\bar{u}u\rangle\bar{u}u
\label{eq:meanfield}
\end{equation}
in the $u$-quark sector of Eq.~\eqref{eq:Ls} and similar terms appear
in the $d$-quark and the $s$-quark sectors.  The first term in
Eq.~\eqref{eq:meanfield} represents a $u$-quark contribution to the
condensation energy, $V_{\rm cond}$, and the sum over three flavors
yields,
\begin{equation}
  V_{\rm cond}[\langle\bar{q}q\rangle]
  = g_S\bigl( \langle\bar{u}u\rangle^2
  + \langle\bar{d}d\rangle^2 + \langle\bar{s}s\rangle^2 \bigr)
  + 4g_D\langle\bar{u}u\rangle\langle\bar{d}d\rangle
  \langle\bar{s}s\rangle\;,
  \label{eq:Vcond}
\end{equation}
where the last term arises from the KMT interaction~\eqref{eq:KMT}.
The second term in Eq.~\eqref{eq:meanfield} represents a mass
correction by the mean-field interaction.  Together with additional
mass corrections from the KMT interaction~\eqref{eq:KMT}, the
mean-field masses can be expressed as
\begin{equation}
  M_i = m_i - 2g_S\langle \bar{q}_i q_i\rangle - g_D
  \epsilon^{ijk} \langle \bar{q}_j q_j\rangle
  \langle \bar{q}_k q_k\rangle\;,
\end{equation}
where $i$, $j$, $k$ run in flavor space, $u$, $d$, $s$.  Once the
mean-field masses are fixed, the vacuum energy from the zero-point
oscillation can be written down in the following form,
\begin{equation}
  V_{\rm zero}[\langle\bar{q}q\rangle]
  = -2\Nc\sum_{i=u,d,s} \int \frac{d^3 p}{(2\pi)^3}\,
  \varepsilon_i(p)
\end{equation}
with $\varepsilon_i(p)=\sqrt{p^2+M_i^2}$.  The above momentum
integration does not converge and we need to introduce a
regularization for the momentum integration.  The choice of the
regularization is a part of the model definition, and the simplest
choice would be a sharp momentum cutoff at $p=\Lambda$.  We note that,
when the gauge symmetry is concerned, a better regularization such as
the proper-time regularization and the Pauli-Villars regularization
would be an indispensable choice.  With a sharp cutoff, we can carry out the
analytical integration, which after all reads,
\begin{equation}
  V_{\rm zero} = -\frac{\Nc\Lambda^4}{4\pi^2}\sum_{i=u,d,s}
  \biggl[ \Bigl(1+\frac{\xi_i^2}{2}\Bigr)\sqrt{1+\xi_i^2}
    -\frac{\xi_i^4}{2}\sinh^{-1}\xi_i \biggr]\;,
\end{equation}
where we defined a dimensionless mass parameter, $\xi_i\equiv M_i/\Lambda$.  The vacuum energy, $V_{\rm zero}$,
tends to make $\xi$ or $M$ larger, while the condensation energy,
$V_{\rm cond}$, tends to make $M$ smaller, and the energetically most
favored value of the dynamical mass $M$ is determined as a result of
the competition between these two energies.  The analysis would be
quite easy if $g_D=0$ or $\Nf=2$ (for which the KMT interaction is
also a four-fermionic term) and $m_i=0$ (i.e.\ chiral limit).  Under
such simplification the total energy can be expanded for small $\xi$
as
\begin{equation}
  V_{\rm zero}+V_{\rm cond} \;\simeq\;
  -\frac{9\Lambda^4}{4\pi^2} \biggl[ 1
    + \Bigl(1-\frac{\pi^2}{3 g_S\Lambda^2}\Bigr)\xi^2 \biggr]\;.
\end{equation}
From this expression it is clear that $\xi\neq 0$ is energetically
favored for strong coupling, $g_S\Lambda^2 > \pi^2/3$.  This is how
chiral symmetry is spontaneously broken by a BCS-like mechanism in the QCD vacuum.  So far, we
implicitly assumed a second-order phase transition in coupling space,
but once the condensation energy from the KMT interaction is taken
into account, it generates a cubic term with respect to the chiral condensate
as seen in Eq.~\eqref{eq:Vcond}, and the order of the phase transition
must be first~\cite{Pisarski:1983ms}.  For more discussions on the order of
the chiral phase transition and the possible QCD critical point (that
are not covered by the present review), see
Refs.~\cite{Fukushima:2013rx,Fukushima:2010bq}.

In the standard NJL model, the medium effect is implemented by the
grand canonical partition function in the quasi-particle
approximation, and in the PNJL model, this part is augmented with the
Polyakov loop coupling.  Therefore, the finite-$T$ and the
finite-$\mu$ terms acquire the Polyakov loop dependence via the
following replacement;
\begin{equation}
  \begin{split}
  V_{\rm medium} &= -2\Nc T\sum_{i=u,d,s} \int\frac{d^3 p}{(2\pi)^3}\,
  \Bigl\{ \ln\bigl[ 1 + e^{-\beta(\varepsilon_i-\mu)} \bigr]
  + \ln\bigl[ 1 + e^{-\beta(\varepsilon_i+\mu)} \bigr] \Bigr\} \\
  &\to -2T\sum_{i=u,d,s} \int\frac{d^3 p}{(2\pi)^3}\,
  \tr\Bigl\{ \ln\bigl[ 1+ L\,e^{-\beta(\varepsilon_i-\mu)} \bigr]
  + \ln\bigl[ 1+ L^\dag e^{-\beta(\varepsilon_i+\mu)} \bigr] \Bigr\}\;,
  \end{split}
\end{equation}
which is inspired from Eq.~\eqref{eq:Vquark}.  We can explicitly take the color
trace as in Eq.~\eqref{eq:Vquark}.  In many cases the mean-field
approximation is assumed, in which $\ell$ and $\ell^\dag$ are
replaced, respectively, with $\Phi$ and $\bar{\Phi}$.
Then, the medium part of the energy finally reads,
\begin{equation}
  \begin{split}
    V_{\rm medium}[\langle\bar{q}q\rangle,\Phi,\bar{\Phi}]
    &= -2T \sum_{i=u,d,s} \int\frac{d^3 p}{(2\pi)^3}
    \Bigl\{ \ln\bigl[ 1+3\Phi\,e^{-\beta(\varepsilon_i - \mu)}
    + 3\bar{\Phi}\, e^{-2\beta(\varepsilon_i - \mu)}
    + e^{-3\beta(\varepsilon_i - \mu)} \bigr] \\
    & \qquad\qquad\qquad\qquad\qquad
    + \ln\bigl[ 1+3\bar{\Phi}\, e^{-\beta(\varepsilon_i + \mu)}
    + 3\Phi\, e^{-2\beta(\varepsilon_i + \mu)}
    + e^{-3\beta(\varepsilon_i + \mu)} \bigr] \Bigr\}\;.
  \end{split}
  \label{eq:medium}
\end{equation}
This expression may look equivalent to Eq.~\eqref{eq:Vquark}, but we
must use caution here.
In the above it is assumed that the Polyakov loop
fluctuations are negligible.  If the logarithmic terms are expanded,
there are higher order terms involving $\ell^n$ ($n>1$), and in
general, $\langle \ell^n\rangle \neq \Phi^n$.  Once $\Phi$ develops a
substantial value in the deconfined phase,
$\langle \ell^n\rangle \simeq \Phi^n$ can hold in a good approximation,
but such a treatment may badly break down when $\Phi\simeq 0$ in the
confined phase.  Then, a better formulation of the mean-field
approximation involving the group integration,
as discussed in Sec.~\ref{sec:matrix}, would be much more
desirable.  Such sophisticated treatments of the Polyakov loop can be
found in Refs.~\cite{Abuki:2009dt,Megias:2004hj}.

To make use of the PNJL model to study quantitative estimates in the
physical energy unit, we should fix the NJL model parameters;
\begin{equation}
  m_u\;,~~ m_d\;,~~ m_s\;,~~ g_S\;,~~ g_D\;,~~ \Lambda
\end{equation}
to reproduce the hadronic properties in the QCD vacuum, namely,
$m_\pi$, $m_\sigma$, $m_K$, $m_\eta$, $f_\pi$ and an empirical value
of the constituent quark mass.  The standard choice of parameters can
be found in Refs.~\cite{Fukushima:2008wg,Fu:2007xc}.  With a favorite
choice of the gluonic potential, $V_{\rm glue}[\Phi,\bar{\Phi}]$, the
total energy to be minimized is,
\begin{equation}
  V_{\rm PNJL} = V_{\rm glue}[\Phi,\bar{\Phi}]
  + V_{\rm zero}[\langle\bar{q}q\rangle]
  + V_{\rm cond}[\langle\bar{q}q\rangle]
  + V_{\rm medium}[\langle\bar{q}q\rangle,\Phi,\bar{\Phi}]\;.
  \label{eq:VPNJL}
\end{equation}
In the NJL setup the sharp momentum cutoff in $V_{\rm zero}$ restricts
the validity region up to energies below $\Lambda$, but this can be
relaxed by replacing the cutoff with the momentum-dependent form
factor.  Such an extended version of the model is referred to as the
non-local (P)NJL model;  see
Refs.~\cite{Hell:2009by,Radzhabov:2010dd,Pagura:2011rt,Kashiwa:2011td,Contrera:2012wj}.

In this review we would not address QCD phase diagram studies using
the PNJL model, for the results cannot avoid model-dependent
assumptions.  Instead, we focus on rather general features of the model,
especially on the relation between chiral restoration and deconfinement.
(See Refs.~\cite{Kovacs:2009zj,Bruckmann:2011cc,Cossu:2016scb} for
analytical and lattice numerical studies on the Dirac eigenvalue distributions
affected by the Polyakov loop.)
We note that the medium part~\eqref{eq:medium} does not require any
ultraviolet regularization.  We can of course define a model with a
regularization imposed on the medium part too, but then the
Stefan-Boltzmann limit would not be correctly satisfied.  In a special
case of $\Phi=1$ (i.e.\ $L$ is a unit matrix) and $\mu=0$, the above
(without $\Lambda$) can be expanded for small $\xi$ as
\begin{equation}
  V_{\rm zero}+V_{\rm cond}+V_{\rm medium}[\Phi=1] \;\simeq\;
  -\frac{9\Lambda^4}{4\pi^2}\biggl[ 1
    + \Bigl( 1 - \frac{\pi^2}{3g_S\Lambda^2}
    - \frac{\pi^2 T^2}{3\Lambda^2}\Bigr)\xi^2 \biggr]\;.
\end{equation}
This immediately leads to the second-order phase transition
temperature given by
\begin{equation}
  \Tc(\Phi=1) = \Lambda \sqrt{\frac{3}{\pi^2} - \frac{1}{g_S\Lambda^2}}\;,
\end{equation}
at which the chiral condensate goes zero.  In the same way for
$\Phi=0$, the transition temperature is reduced as
\begin{equation}
  \Tc(\Phi=0) = 3\sqrt{3}\Tc(\Phi=1)\;.
\end{equation}
Here, $3$ appears from the effectively reduced temperature by factor
$3$ as is obvious in Eq.~\eqref{eq:Vquark}, and $\sqrt{3}$ is from the
overall coefficient of $V_{\rm medium}$.

\begin{figure}
  \centering
  \includegraphics[width=0.8\textwidth]{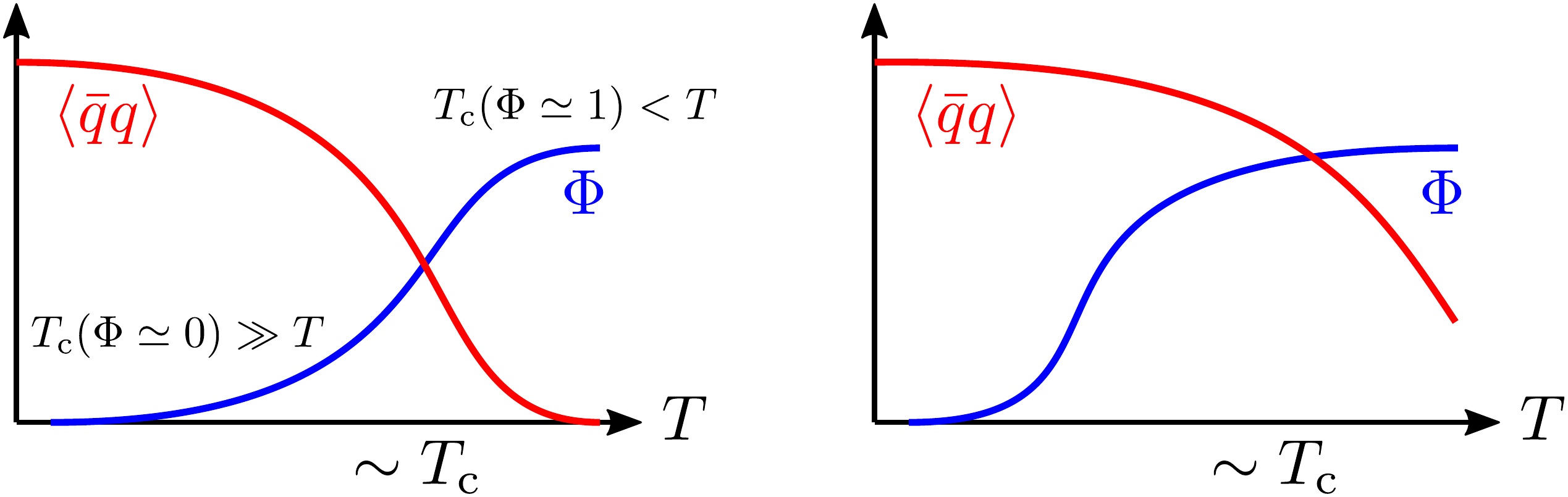}
  \caption{Schematic picture for simultaneous changes of two order
    parameters, i.e.\ the chiral condensate $\langle\bar{q}q\rangle$
    and the Polyakov loop expectation value $\Phi$.  If $\Phi$ arises
    at temperatures larger than $\Tc(\Phi\simeq 1)$,
    $\langle\bar{q}q\rangle$ decreases simultaneously (left).  It is
    also possible in the model that $\langle\bar{q}q\rangle$ starts
    decreasing at temperatures greater than the onset of $\Phi$
    (right).}
  \label{fig:schematic}
\end{figure}

The fact that the small Polyakov loop would push the chiral phase
transition toward higher temperature can give us an intuitive argument
to explain how the chiral restoration occurs almost simultaneously
as deconfinement.  As schematically illustrated in the left panel of
Fig.~\ref{fig:schematic}, the traced Polyakov loop $\ell$ or its
expectation value $\Phi$ is small at low temperature, and as long as
$\Phi\simeq 0$ the critical temperature is $3\sqrt{3}$ times larger, so
that $T$ cannot reach the critical temperature.  Therefore, the
chiral condensate, $\langle\bar{q}q\rangle$, is hardly affected by the
temperature and the chiral restoration is hindered by small $\Phi$ in
this way.  At high temperature, center symmetry is broken in the
deconfined phase and $\Phi\simeq 1$ is approached.  Then, the critical
temperature becomes the standard one without the Polyakov loop
suppression, and for $\Tc(\Phi\simeq 1)<T$, quark excitations destroy
the chiral condensate and chiral symmetry can be restored.  In this
picture, therefore, the Polyakov loop expectation value, $\Phi$, is
a control parameter that governs the behavior of
$\langle\bar{q}q\rangle$.  Thus, rising $\Phi$ triggers the sizable
change in $\langle\bar{q}q\rangle$ as sketched in the left panel of
Fig.~\ref{fig:schematic}.

Roughly speaking, one may well describe the situation as follows;  in
the PNJL model, chiral symmetry restoration does not occur
as long as confinement, i.e.\ $\Phi\simeq 0$, holds.
If $T_\chi(m=0)$ without the Polyakov loop
effect is smaller than $T_D(m=\infty)$, the chiral restoration
temperature is pushed up by $\Phi$.  This is a quite robust argument
to explain $T_\chi(m)\simeq T_D(m)$ without parameter tuning.  Indeed,
field-theoretical and phenomenological arguments suggest that the
confined phase must break chiral symmetry.  If such a relation between
confinement and chiral symmetry persists to finite temperature,
$\langle\bar{q}q\rangle$ cannot go to zero as long as $\Phi$ is
vanishingly small, as is exactly the situation in the PNJL model.

It is worth mentioning that there is no reason why chiral restoration
\textit{should} happen in the deconfined phase.  More specifically,
if $T_\chi(m=0)$
without the Polyakov loop effect were greater than $T_D(m=\infty)$,
there might have been a new phase where quarks are deconfined but
chiral symmetry is still spontaneously broken as shown in the right
panel of Fig.~\ref{fig:schematic}.  This is not a fictitious
imagination, but one can shift the chiral restoration temperature by
imposing external magnetic fields that primarily couple to quarks
rather than gluons.  The lattice-QCD simulation under strong magnetic
fields~\cite{Bali:2011qj,Bali:2012zg}, however, have revealed that
two crossovers occur like not the right but the left panel of
Fig.~\ref{fig:schematic}.  This important lattice-QCD observation tells us that
deconfinement and chiral restoration are locked together more tightly
than realized in the mean-field PNJL model, presumably through the
back-reaction of the quark polarization in the Polyakov loop
potential.  Actually, the back-reaction is always non-negligible for
phenomenological applications of the PNJL model.

\begin{figure}
  \centering
  \includegraphics[width=0.65\textwidth]{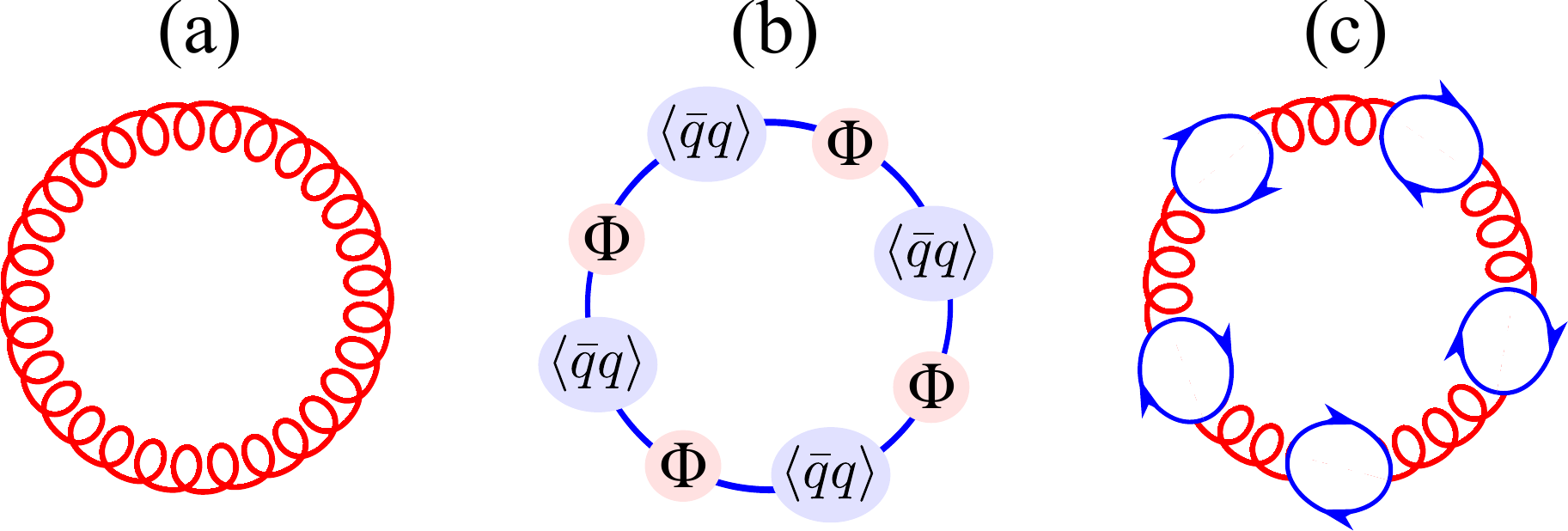}
  \caption{Diagrammatic representation for the building blocks in the
    PNJL model;  (a) Polyakov loop potential from the gluon loop.  (b)
    Medium part with coupling between the Polyakov loop and the chiral
    condensate.  (c) Missing contribution from the back-reaction of the
    quark polarization.}
  \label{fig:loops}
\end{figure}

Here, we give a brief discussion on the back-reaction.  From the
diagrammatic point of view the first term in the model
ingredients~\eqref{eq:VPNJL} represents gluonic loop contributions as shown in
Fig.~\ref{fig:loops}~(a).  The second and the third terms represent
quark loop contributions, and the last term of the medium effect
arises from the quark propagation on top of the Polyakov loop
background;  see Fig.~\ref{fig:loops}~(b).  What is missing in a plain
version of the PNJL model is the back-reaction or the polarization
effects from the processes as in Fig.~\ref{fig:loops}~(c).  Because such
screening diagrams involving quark loops are suppressed in the
large-$\Nc$ limit, one may say that the standard PNJL model implicitly assumes
partial decoupling between the gluonic and the quark sectors in such a
way similar to the large-$\Nc$ limit.

Once the polarization diagrams are considered for the computation of
the Polyakov loop potential, it would promote the strong coupling
constant to the running one with the gluon momentum.  As a result of
the gluon momentum integration, a typical energy scale in the running
coupling constant would be picked up, which turns into the scale
characterizing $T_0$ in the Polyakov loop potential.  If there are
many quark flavors, the strong coupling becomes smaller for a fixed
gluon momentum.  This means that the energy scale in the running
coupling should be smaller to keep the strong coupling unchanged for a
fixed gluon momentum.  The change in the energy scale should be
specified according to the QCD $\beta$-function, which is translated
into $T_0$.  In this way, the diagram (c) can be incorporated in part
into the phenomenological Polyakov loop potential as an
$\Nf$-dependent $T_0$, as first argued in
Ref.~\cite{Schaefer:2007pw}.  Using the QCD $\beta$-function
coefficient in the leading order, $b=(11\Nc-2\Nf)/6\pi$, the Polyakov
loop potential scale is given as
\begin{equation}
  T_0(\Nf) = T_\tau\, e^{-1/(\alpha_0 b)}\;,
\end{equation}
where $\alpha_0=0.304$ is the strong coupling at the ultraviolet scale
and $T_\tau=1.770\;\text{GeV}$.  This parametrization describes the
pure gluonic scale $T_0(0)=270\;\text{MeV}$ that is screened down to
$T_0(\Nf=3)=178\;\text{MeV}$.  This shift is very important for the
PNJL and similar models to quantify physically interested temperatures
on the level to compare with the lattice-QCD results.

Before closing this subsection, let us make a remark on confinement in
the PNJL model.  One might be tempted to identify the last term of
Eq.~\eqref{eq:Vquark} and Eq.~\eqref{eq:medium} as a color-singlet excitation of baryon.  This
interpretation is, however, misleading.  We can understand the problem
by performing a variable change, apart from the flavor sum, as
\begin{equation}
  \int\frac{d^3 p}{(2\pi)^3} \,f(\Nc\sqrt{p^2+m^2}-\Nc\mu)
 = \frac{1}{\Nc^3}\int\frac{d^3 p'}{(2\pi)^3}\,f(\sqrt{p'^2+M^2}-\mu_B)\;,
\label{eq:baryon}
\end{equation}
for an arbitrary function $f$,
where $p'=\Nc p$ is the new integration variable, $M=\Nc m$
approximates the baryon mass, and $\mu_B=\Nc\mu$ is the baryon
chemical potential.  The precise form of $f$
is irrelevant in discussions below.  Now, as is clear from the
above variable change, $V_{\rm medium}[\Phi=0]$ can be surely
translated into a form of the baryonic partition function, but it is
multiplied by a suppression factor $1/\Nc^3$.  So, the PNJL model
cannot reproduce correct thermodynamics of baryonic degrees of
freedom but significantly underestimate it.
This failure of underestimating baryonic contributions by $1/\Nc^3$ is
attributed to incomplete confinement in the model.  For $\Phi=0$,
genuine confinement demands that three quarks be localized within a baryon
radius, while the PNJL model only describes the situation that three
quarks have zero triality (that is a discrete charge associated with
center symmetry) overall and these quarks can travel far from each other.
In this sense, confinement implemented in the PNJL model is sometimes
called the \textit{statistical confinement}.

The lack of local confinement is not a flaw of the model itself, but it
simply results from the mean-field treatment of the Polyakov loop.  If
the Polyakov loop at each spatial point is treated including the
spatial correlations, as is the case in the strong coupling
expansion~\cite{Gocksch:1984yk,Fukushima:2002ew}, the model should, in
principle, be capable of grasping more realistic natures of
confinement, giving rise to mesons and baryons dynamically.

To extend the validity region of the PNJL model toward the hadronic
phase, it would be essential to think of the problem of how to include
the mesonic and the baryonic degrees of freedom.  As long as we treat
the Polyakov loop using $\Phi$ and $\bar{\Phi}$ in the mean-field
approximation, the statistical confinement has nothing to do with
hadrons.  If we go into quark two loops, we can incorporate hadrons as
collectively resonating states.  Such a program to improve the PNJL
model is technically involving.  For the pions the higher-loop
calculations have been done~\cite{Yamazaki:2012ux}, but an extension
to the baryons along the same lines as Ref.~\cite{Yamazaki:2012ux}
would be too complicated.  One might then think that the hadrons could
be introduced by hand as extra degrees of freedom.  However, it is a
physical requirement that such composite states should dissolve at high
enough temperature or at high energy scales in general.
In fact, in proper calculations as in Ref.~\cite{Yamazaki:2012ux},
the pions should disappear due to the momentum
dependent self-energy or the wave-function renormalization.

\subsubsection{PQM model}
\label{sec:PQM}

The Polyakov-loop augmented quark-meson (PQM) model is an efficient
approach to take account of the mesonic loops.  This is a chiral model in
which the pions $\vec{\pi}$ and the sigma meson $\sigma$ are introduced as
point particles.  Instead of four-point fermionic vertices,
fermions interact via interacting meson exchanges.
One may well think that such mesonic fields and
original quark fields overcount physically active degrees of freedom
unless finite-$T$ dissociation is somehow implemented.  The PQM model
is designed in a very nice way to avoid this problem.  

The self-interaction of mesons in the PQM model is given by the
potential that exhibits spontaneous breaking of chiral symmetry.  The
standard choice would be,
\begin{equation}
  V_{\rm QM}[\sigma,\vec{\pi}] = \frac{m_\sigma^2}{8f_\pi^2}(\sigma^2
  +\vec{\pi}^2-f_\pi^2)^2 + f_\pi m_\pi^2\,\sigma\;,
  \label{eq:VQM}
\end{equation}
from which $\sigma$ acquires a non-zero expectation value $\sim f_\pi$
and $\vec{\pi}$'s become light as the NG bosons.  At high temperature,
chiral symmetry is restored, and then the above potential is changed
to a symmetric shape, with which both $\sigma$ and $\vec{\pi}$'s acquire
masses of order $T$ on top of the vacuum contributions.
Because mesons become heavier than quarks in the symmetric phase at high $T$ or high energies
in general, they are effectively decoupled from the dynamics.

Such a procedure to incorporate mesons is systematically formulated in
a framework of the functional renormalization group (FRG) equation~\cite{Schaefer:1999em,Bohr:2000gp}.
In the FRG language, the potential like Eq.~\eqref{eq:VQM} is an
infrared output as a result of the RG flow, and the original input
potential at the ultraviolet scale, $\Lambda$, should be symmetric and
$\sigma$ and $\vec{\pi}$'s are as massive as $\sim\Lambda$.  This should
be so, because the relevant degrees of freedom must be quarks and
gluons only in the ultraviolet sector.  Solving the FRG equations with QCD$+$mesons can
establish a firm theoretical connection between the PQM model and the
QCD-based calculations~\cite{Braun:2009gm,Herbst:2013ail}.

We comment a little more on the necessity of the FRG treatment of the
(P)QM model.  It used to be a long standing problem how to avoid an
artificial first-order phase transition in the linear sigma model beyond 
the mean-field approximation~\cite{AmelinoCamelia:1997dd,Chiku:1998kd}.
Higher order loops may cure the
problem, but momentum dependent self-energies generally have
complicated analytical structures~\cite{Nishikawa:2003js} and it is
not realistic to tackle this problem diagrammatically.  The basic
equations in the FRG framework is amazingly simple particularly in the
local potential approximation and with Litim's optimized
cutoff~\cite{Litim:2000ci}.  These technical advances have modernized
the (P)QM model.

The PQM model is a more suitable approach to study the QCD phase
diagram than the PNJL model for two reasons.  The first reason is that
the PQM model has mesonic fluctuations and so not only the
thermodynamics in the low-$T$ hadronic phase but also the critical
properties around $\Tc$ can be correctly reproduced.  The latter has a
significant phenomenological impact especially for the size of the QCD
criticality region~\cite{Skokov:2010wb,Skokov:2010uh,Herbst:2010rf}.
Another reason is that the recent
developments in the FRG application to QCD has been clarifying
the mapping between the
PQM model and QCD if the baryon density is low.  The FRG-QCD phase
diagram is found in Ref.~\cite{Braun:2009gm}.  Also see
Refs.~\cite{Fischer:2009gk,Fischer:2014ata}.
for the related
functional approach to the QCD phase diagram based on the
Dyson-Schwinger equations~\footnote{As a side remark, 
	in Dyson-Schwinger approach there is 
	a way to extract the Polyakov loop, see Refs.~\cite{Gattringer:2006ci,Bilgici:2008qy}.}. 
We also make a brief comment that the inclusion of heavier hadrons
and resonances may eventually allow for a dual description in which the Polyakov
loop would be dictated by the hadron spectrum~\cite{Megias:2012kb}.

Now, one might wonder about the treatment of meson dissociation in the
PQM model;  that is, being heavy is not the same as being unstable, is
it?  At high temperature, as already explained,
$\sigma$ and $\vec{\pi}$'s are heavy
enough to decouple, but physically speaking, they must be unstable and
decay into quarks.  These two descriptions seem to be different, but actually, they
are not.  If mesons are not dynamical, the kinetic term is vanishing,
or the wave-function renormalization, $Z_\phi$, should be vanishingly
small.  We can understand this from the NJL model;  the meson fields
can be introduced as auxiliary fields through the Hubbard-Stratonovich
transformation, and at this point, they are not yet dynamical and
simply non-propagating fields to replace the fermionic interactions.
The important point is that the wave-function renormalization and the
mass are related to each other.  If the kinetic term for a meson
$\phi$ behaves like $\tilde{Z}_\phi p^2 \phi^2$ with a dimensionless
$\tilde{Z}_\phi$, and the potential curvature $V'' \phi^2$ is
non-zero, the physical mass (after rescaling
$\phi\to \tilde{\phi}=\tilde{Z}_\phi^{1/2}\phi$) should be
$m^2 = V''/\tilde{Z}_\phi\to\infty$ as $\tilde{Z}_\phi\to 0$.  Thus,
the field is physically very massive if it corresponds to
non-propagating mode.  We can also understand the decay into quarks in
the same way.  If the original Yukawa coupling is $g \bar{q}\phi q$, the
rescaling to $\tilde{\phi}$ gives the physical Yukawa coupling that
diverges as $g/\tilde{Z}_\phi^{1/2} \to \infty$
in the limit of $\tilde{Z}_\phi\to 0$, so that
the decay is enhanced as it should be.

To deal with composite states in general within the framework of the
FRG equation, the wave-function renormalization plays an essential
role, which can be explicitly seen in the argument of the so-called
rebosonization procedure~\cite{Gies:2001nw}.  In principle, such an
idea to treat mesons using the rebosonization procedure can be applied
to diquarks and also baryons.  The major problem in the inclusion of
baryons in the PQM-type models is that they do not necessarily become
massive at high density unlike mesons at high temperature.  As a
cutoff effective description for nuclear matter, as attempted in
Ref.~\cite{Drews:2014wba}, the FRG approach is still useful, but to
establish a connection to QCD in the high-density region, the baryonic
counterpart of the rebosonization (which might be called the
rebaryonization) must be addressed.  Such a program is still under
investigations.

\subsubsection{Critical point with heavy quarks}
\label{sec:critical}

So far, we have seen the coupling between the Polyakov loop and the
dynamical quarks in the light-flavor sector, to discuss how chiral
restoration is affected by the Polyakov loop dynamics.  Now, let us
turn to the heavy-flavor sector and consider how the Polyakov loop
dynamics is influenced perturbatively by massive quarks.

For a pure gluonic theory with $\Nc=3$ and no dynamical quarks, the
deconfinement phase transition is of first order.
By adding (heavy) dynamical quarks we can control the ${\rm Z}(3)$
symmetry breaking and weaken the transition;
in this respect quarks act as a background ${\rm Z}(3)$ breaking field.
As the quark mass decreases, the first-order phase transition turns
into a crossover; hence there is a 
critical value of the quark mass where the first-order boundary
terminates at a deconfining critical point of the second order phase
transition.

The location of this critical point is a sensitive probe of the
Polyakov loop potential;
it reveals the properties of the pure gluonic Polyakov loop potential
beyond the expectation value
or the curvature computed at the expectation value.

\begin{figure}
  \centering
  \includegraphics[width=0.47\textwidth]{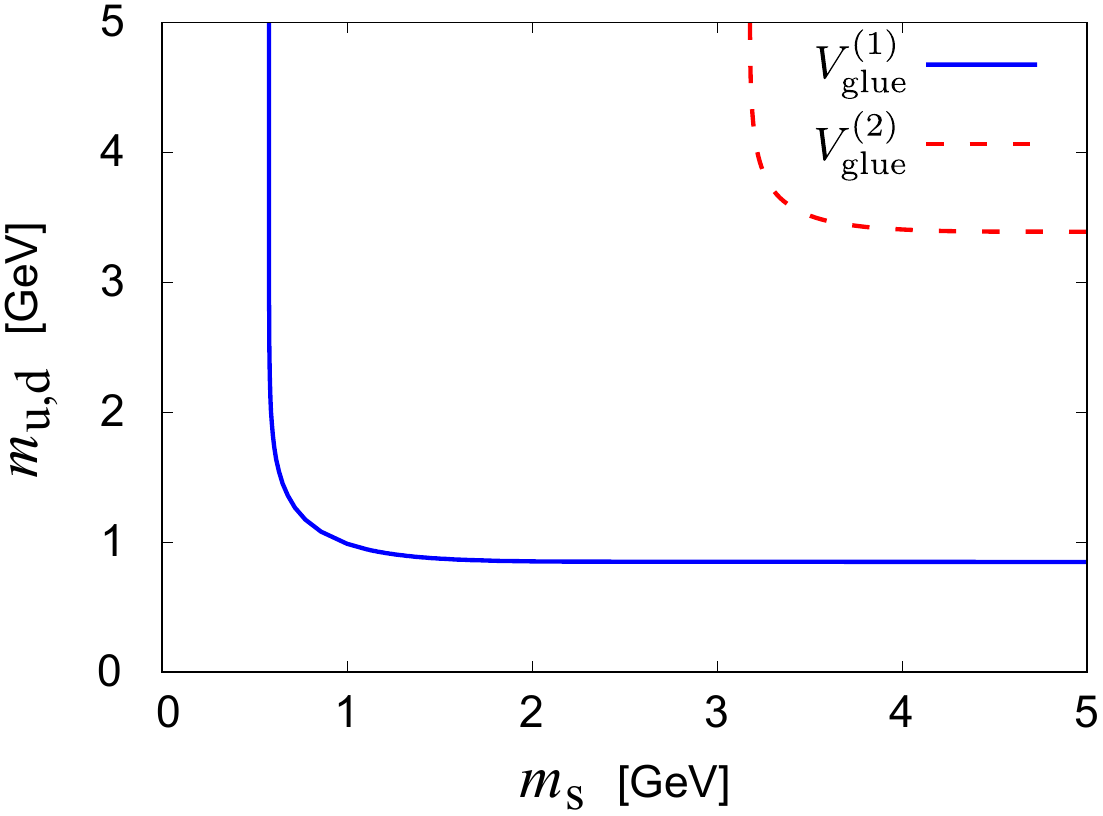}
  \caption{The upper right corner of the Columbia plot.
    The lines correspond to the second order phase transition.
    $V^{(1)}_{\rm glue}$ denotes the model defined by
    Eq.~\eqref{eq:polpotential} and $V^{(2)}_{\rm glue}$ by
    Eq.~\eqref{eq:polpotential2}.}
  \label{fig:Columbia}
\end{figure}

The deconfining critical point is located at rather heavy quark
masses, where we can use the Boltzmann approximation for the quark
thermodynamic potential.
Hence we have to go back to the quark contribution to the effective
potential,  Eq.~\eqref{eq:Vquark}, and
proceed with expansion of each logarithm into a power series (at zero
chemical potential), i.e.
\begin{equation}
	\tr\ln (1 + L e^{-\beta \epsilon_p} ) = 
	\tr L\, e^{-\beta \epsilon_p} - \frac{1}{2} \tr(L^2)
        \, e^{-2 \beta \epsilon_p} + \mathcal{O}(e^{-3 \beta \epsilon_p})\;,
	\label{Eq:BExp}
\end{equation}
where the higher order terms are suppressed exponentially.
We will keep only the first term in the expansion, which corresponds
to the Boltzmann approximation or the leading order in the hopping
parameter expansion.
The momentum integration can be performed analytically to yield,
\begin{equation}
	V_{\rm quark}[q] \approx - \sum_{i} \frac{m_i^2 \, T\, V}{\pi^2}
        K_2\Bigl(\frac{m_i}{T}\Bigr) \bigl( \tr L + \tr L^\dag \bigr)\;,
	\label{Eq:Vheavy}
\end{equation}
where $i$ makes the summation with respect to the flavors.
The advantage of having masses as parameters is that 
by varying the quark masses, one can tilt the Polyakov loop potential 
and probe it at different values of the order parameter. 
Thus it is not surprising that, as was demonstrated in
Ref.~\cite{Kashiwa:2012wa}, 
the critical values of the mass differs significantly in the models
described by Eq.~\eqref{eq:polpotential} and Eq.~\eqref{eq:polpotential2}.   
The calculations at $\Nf=3$ shows that, for the former, the critical
mass is $3.5\;\text{GeV}$;  for the latter, it is about $1\;\text{GeV}$.
A model of Ref.~\cite{Lo:2014vba} constrained by the Polyakov loop
susceptibilities produced yet another value of the 
critical mass, i.e.\ $1.48\;\text{GeV}$.

We demonstrate this striking difference by plotting the most upper
right corner (that is the region where all quark masses are heavy) of
the Columbia plot in Fig.~\ref{fig:Columbia}.
The lines represent the locations of the second-order deconfinement
phase transition, which is found by setting first three derivatives of
the effective potential with respect to the Polyakov loop to zero.
See also Ref.~\cite{Reinosa:2015oua} for other approaches to the
Columbia plot with a different type of confining potential.

The lattice studies for the critical point of the deconfinement phase
transition are available for different number of heavy flavors;
see e.g.\ Ref.~\cite{Saito:2011fs} and references therein. 
They, however, were not extrapolated to the continuum;  this hinders
a direct comparison with the model predictions at present.

\subsection{Systems at Finite Baryon Density}
\label{sec:density}

Once the chemical potential is turned on, the most interested and yet
unsolved problem in QCD is the QCD phase diagram.
There are many theoretical attempts to explore the QCD phase diagram
at finite $T$ and $\mu$ using the PNJL and the PQM models.  Because
baryonic degrees of freedom are underestimated as seen in
Eq.~\eqref{eq:baryon}, it is still a challenging problem to reveal the
correct structures of the dense QCD vacuum even on the qualitative level.
Also, the system has valence quarks for $\mu\neq 0$, and so center
symmetry is more broken by those quarks.  Therefore, the Polyakov loop
would lose its meaning as an approximate order parameter for
deconfinement.  Nevertheless, the Polyakov loop is a theoretically interesting
quantity for the finite-density systems, especially for the analysis of the
sign problem in some algebraic approaches.

\subsubsection{Sign problem}
\label{sec:sign}

We start from a short reminder of the sign problem in QCD.
At finite baryon or quark density the Dirac operator (apart from
the mass term) is no longer anti-Hermitian, and thus the Dirac
determinant may have a complex phase.  This can be seen from the fact
that the Dirac determinant at finite chemical potential is written as
\begin{equation}
  \calM[A;\mu] = \det\bigl( i\feyn{D} - i\gamma_4 \mu - m \bigr)\;.
\end{equation}  
We should recall that in our convention Euclidean $\gamma_\mu$'s are
chosen as anti-Hermitian matrices.  So, $i\feyn{D}$ is an
anti-Hermitian operator having pure imaginary eigenvalues;
$\pm i\lambda$.  It is then clear that $-i\gamma_4\mu$ is Hermitian
and eigenvalues of such a mixed operator, $i\feyn{D}-i\gamma_4\mu$,
are neither pure imaginary nor real.

We can then separate the real part and the imaginary part of the Dirac
determinant as
\begin{equation}
  \calM[A;\mu] = \Re \calM[A;\mu] + i \Im \calM[A;\mu]\;,
\end{equation}
and then we can prove that, under the charge parity transformation;
$A_\mu\to A_\mu^C=-A_\mu$, the real part and the imaginary part,
respectively, change as
\begin{equation}
  \Re\calM[A^C;\mu] = \Re\calM[A;\mu]\;,\qquad
  \Im\calM[A^C;\mu] = -\Im\calM[A;\mu]\;.
\end{equation}
We can understand this from the property of
$\calM^\ast[A;\mu]=\det C^{-1}(i\feyn{D}-i\gamma_4\mu-m)^\ast C
=\calM[A^C;\mu]$, where under the charge parity transformation,
$C^{-1}\gamma_\mu^\ast C\to -\gamma_\mu$.  In the functional
integration with respect to $A_\mu$, any contribution from a certain
$A_\mu$ has a paired contribution from $A_\mu^C$, and
so after taking the functional integration over whole $A_\mu$, the
$C$-odd part is averaged away.  Therefore, the partition function is,
\begin{equation}
  Z = \int \calD A_\mu\, \Re\calM[A;\mu]\,e^{-S_{\rm YM}[A]}\;,
\end{equation}
apart from the gauge fixing.  In the same way for a $C$-even
(real-valued) observable, $\calO_+[A]$, the expectation value is,
\begin{equation}
  \langle \calO_+[A]\rangle
  = \frac{1}{Z} \int \calD A_\mu\, \calO_+[A]\,\Re\calM[A;\mu]\,
  e^{-S_{\rm YM}[A]}\;,
\end{equation}
and for a $C$-odd real-valued observable, $\calO_-[A]$, the
expectation value is,
\begin{equation}
  \langle \calO_-[A]\rangle
  = \frac{i}{Z} \int \calD A_\mu\, \calO_-[A]\,\Im\calM[A;\mu]\,
  e^{-S_{\rm YM}[A]}\;.
\end{equation}
From these arguments we see that the integrand in the functional
integrals is always a real-valued function, and what does really
matter is the sign changes of $\Re\calM[A;\mu]$ and $\Im\calM[A;\mu]$
at finite $\mu$.  This is why we commonly refer to this difficulty as
the ``sign problem'', not the complex problem of the Dirac
determinant.

It is interesting how the sign problem looks like concretely in the
PNJL model.  As we already mentioned near Eq.~\eqref{eq:Vquark}, the
quark contribution, $V_{\rm quark}[q]$, is complex for $\mu\neq 0$ if
$\ell$ is complex.  In the mean-field approximation, $\ell$ and
$\ell^\ast$ are replaced with real-valued expectation values, which are
calculated from
\begin{equation}
  \langle \Re\ell \pm i\Im\ell\rangle
  = \frac{1}{Z}\int\calD A_\mu\, \bigl( \Re\ell\,\Re\calM
  \mp\Im\ell\,\Im\calM \bigr)\,e^{-S_{\rm YM}}\;.
\end{equation}
If the Dirac determinant or the quark potential is expanded as in
Eq.~\eqref{Eq:Vheavy}, $\Im\calM$ is proportional to $\Im\ell$, and
so~\cite{Dumitru:2005ng},
\begin{equation}
  \bar{\Phi} - \Phi = -2i\langle \Im\ell\rangle
  \;\propto\; \langle (\Im\ell)^2 \rangle_{\rm YM}\;.
\end{equation}
For $\mu>0$, we see $\bar{\Phi}>\Phi$, and we can immediately give a
physical explanation for this.  Since $\Phi$ is related to an energy
excess $f_q$ induced by a test static quark, $\bar{\Phi}>\Phi$ means
that a test anti-quark costs less energy than a test quark.  In other
words, we can place an anti-quark more easily than a quark into the
finite-density medium, which should be naturally so due to screening
effects;  see Ref.~\cite{Allton:2002zi} for lattice-QCD results and
physical discussions.

Now, one might think that, once the mean-field approximation is
adopted for the Polyakov loop, one may no longer monitor the sign
problem.  There is no apparent sign problem indeed, but the Polyakov loop
potential has only the saddle points then, which should be regarded as a remnant of the sign
problem.  The saddle point structure was considered in details in
Ref.~\cite{Fukushima:2006uv}.  For the purpose to see the problem, the
simplest lowest order would be sufficient.  The Polyakov loop
potential~\eqref{eq:polpotential} can be approximated in the confined
phase as
\begin{equation}
  T^{-4}V_{\rm glue}(\Phi,\bar{\Phi}) \approx
  -\frac{b_2}{2}\bar{\Phi}\Phi
  = -\frac{b_2}{16}\bigl[ (\bar{\Phi}+\Phi)^2 - (\bar{\Phi}-\Phi)^2 \bigr]\;.
\end{equation}
If we na\"{i}vely take the $\Phi$ derivative and the $\bar{\Phi}$
derivative of $V_{\rm glue}(\Phi,\bar{\Phi})$, we have
$\Phi=\bar{\Phi}=0$ as is reasonable in the confined phase ($b_2>0$).
However, clearly, the above potential is unstable along the direction
of $\bar{\Phi}-\Phi$, and $\Phi=\bar{\Phi}$ is not a potential minimum
but a saddle point.  Surprisingly, this problem occurs not only at
finite density but already at zero density, as soon as we allow for a
possibility of $\bar{\Phi}\neq\Phi$ (which is allowed only at finite
density, so the sign problem matters only at finite density anyway).
In Ref.~\cite{Fukushima:2006uv} it was conjectured that the
saddle-point approximation should work, though it cannot be justified
from the thermodynamic principles.  Recently, the mathematical
justification for the saddle-point prescription
has been proposed with help of the Lefschetz thimble
method~\cite{Tanizaki:2015pua}.  As a final remark it is worth
mentioning that the Polyakov loop potential in terms of $q_i$ (such as the
inverted Weiss potential) has no such problem of the saddle-point
prescription, but as we already confirmed in Fig.~\ref{fig:su3weiss},
the ground state lies in the global minimum of the potential.

\subsubsection{Heavy-dense model}
\label{sec:heavy-dense}

Here, let us introduce a very useful model to consider the sign problem.
We have seen that the hopping parameter expansion picks up a Polyakov
loop in Sec.~\ref{sec:critical}.  There is an interesting limit in
which such an expanded form becomes exact while keeping a non-trivial
phase structure.  A na\"{i}ve heavy mass limit would justify the expansion, but
the coefficient would be too small to cause anything
interesting.  To compensate such smallness of the coefficient, we can
take the high density limit simultaneously.  That is, the Dirac
determinant (on the lattice) simplifies as
\begin{equation}
  \det [i\feyn{D} - i \gamma_4 \mu - m] \;\to\;
  [\det(1+\epsilon L)]^{\Nf/4}\;,
\end{equation}
where the flavor counting is complicated due to the doubler problem.
To avoid unnecessary complication, let us limit our considerations
below to the simplest case of $\Nf=4$.  We should take two limits of
$m\to\infty$ and $\mu\to\infty$ so as to keep their
combination~\cite{Blum:1995cb},
\begin{equation}
  \epsilon \equiv \biggl( \frac{e^{\mu a}}{2m a} \biggr)^{N_\tau}\;,
\end{equation}
to be finite.  We can intuitively understand the above statement in
the following way;  in the heavy limit quarks cannot move and the
static quark and anti-quark propagation at spatial point $\bx$ is written in
terms of $L(\bx)$ and $L^\dag(\bx)$, from which the anti-quark part by
$L^\dag(\bx)$ is negligible in the dense limit.  Thus, the Dirac
determinant should be a function of $L(\bx)$ only.

Interestingly, such a simple expression of $\det(1+\epsilon L)$
captures the important features of the sign problem.  For the color
SU(2) case we immediately see that the Dirac determinant is always
real.  That is, in this case of the color SU(2) group, we can
explicitly take the determinant as
\begin{equation}
  \det (1+\epsilon L) = \prod_{\bx} \bigl[ 1 + \epsilon^2
  + 2\epsilon \ell(\bx) \bigr] \qquad \text{[for SU(2)]}\;,
\end{equation}
which is real because $\ell(\bx)$ is real for the SU(2) case.  We note
that the above determinant is positive definite and this is because we
implicitly assumed $\Nf=4$.

In contrast to the SU(2) case, the sign problem becomes manifested for
the color SU(3) group.  The explicit calculation leads to
\begin{equation}
  \det (1+\epsilon L) = \prod_{\bx} \bigl[ 1 + \epsilon^3
  + 3\epsilon \ell(\bx) + 3\epsilon^2 \ell^\ast(\bx) \bigr]
  \qquad \text{[for SU(3)]}\;.
\end{equation}
Clearly, except for special values of $\epsilon=0$ (i.e.\ zero
density) and $\epsilon=1$ (i.e.\ half-filling), the Dirac determinant
generally has a complex phase.  We note that the determinant is real
for $\epsilon=1$ but it can be negative for $-1<\ell+\ell^\ast<-2/3$.
Because this heavy-dense model setup has a duality, i.e.\ invariance
under the transformation, $\epsilon\leftrightarrow 1/\epsilon$ and
$\ell\leftrightarrow\ell^\ast$, the high density region is described
as a dilute anti-quark (or diquark with anti-triplet color) medium,
which is due to the lattice artifact of saturation.

One can treat the pure gluonic part as it is and simply model it using the
matrix model~\eqref{eq:Zstrong}, or one can completely drop it assuming the
strong coupling limit.  There are many applications of the heavy-dense
model;
for the mean-field study on the sign problem,
see Ref.~\cite{Fukushima:2006uv}.  For the
strong-coupling expansion including the heavy-dense model discussions as a
special example, see Ref.~\cite{Langelage:2014vpa}.  For the test of
the complex Langevin simulation using the heavy-dense model, see
Refs.~\cite{Aarts:2008rr,Aarts:2015yuz}.

\subsubsection{Roberge-Weiss phase transition with imaginary chemical potential}
\label{sec:RW}

There is no sign problem for the imaginary chemical potential, which
is obvious from Eq.~\eqref{eq:Vquark};  $\ell\,e^{\beta\mu}$ and
$\ell^\ast\,e^{-\beta\mu}$ are complex conjugate to each other
because $(\mu)^\ast=-\mu$ then,
and so the quark contribution to the potential is real.  Although
considering the imaginary chemical potential may sound rather
academic, the ground state properties at finite imaginary chemical
potential have quite intriguing aspects.

We can regard the Polyakov loop as a colored imaginary chemical
potential, as we mentioned previously, and so the imaginary chemical
potential for the baryon or quark U(1) symmetry is mixed together with
a center element of the Polyakov loop.  The
dependence on the imaginary chemical potential $\imu$ (where
$\mu=i\imu$) appears only
through a combination of $\ell\,e^{i \beta\imu}$.  The quark effective
potential is thus a function of $\ell\,e^{i\beta\imu}$ and
$\ell^\ast e^{-i\beta\imu}$.  We know that the gluonic potential,
$V_{\rm glue}[\ell,\ell^\ast]$, is center symmetric, and there are
$\Nc$ degenerate minima.  Hence, under the center transformation, the
effective potential changes as
\begin{align}
  V_{\rm glue}[\ell,\ell^\ast]
  + V_{\rm quark}[\ell\,e^{i\beta\imu},\ell^\ast e^{-i\beta\imu}]
  &\to
  V_{\rm glue}[\ell, \ell^\ast]
  + V_{\rm quark}[e^{i2\pi k/\Nc} \ell\,e^{i\beta\imu},
  e^{-i2\pi k/\Nc} \ell^\ast e^{-i\beta\imu}] \notag\\
  &= V_{\rm glue}[\ell,\ell^\ast]
  + V_{\rm quark}[\ell\,e^{i\beta(\imu+2\pi k T/\Nc)},
  \ell^\ast e^{-i\beta(\imu+2\pi k T/\Nc)}]\;.
\end{align}
This in turn means that such a shift in the imaginary chemical
potential by a multiple of $2\pi T/\Nc$ is absorbed by a Z($\Nc$)
transformation on the Polyakov loop.  At low $T$, center symmetry
approximately remains and the Polyakov loop distribution is centered
around $\ell=0$, which implies that such a Z($\Nc$) rotation is
possible without any energy barrier.  At high $T$, on the other hand,
center symmetry is largely broken and the Polyakov loop distribution
is centered around $\Re\ell\gg\Im\ell\simeq 0$.  There, we would
anticipate an energy barrier associated with the Z($\Nc$) rotation.
As a matter of fact, such a structure is evident in the perturbative
potential~\eqref{eq:weiss_quark};  by replacing $\mu$ by $i\imu$, we
find the quark contribution to the Weiss potential to be a function of
$(q_i+1/2+\beta\imu/2\pi)_{\text{mod1}}$.

\begin{figure}
  \begin{center}
  \includegraphics[width=0.47\textwidth]{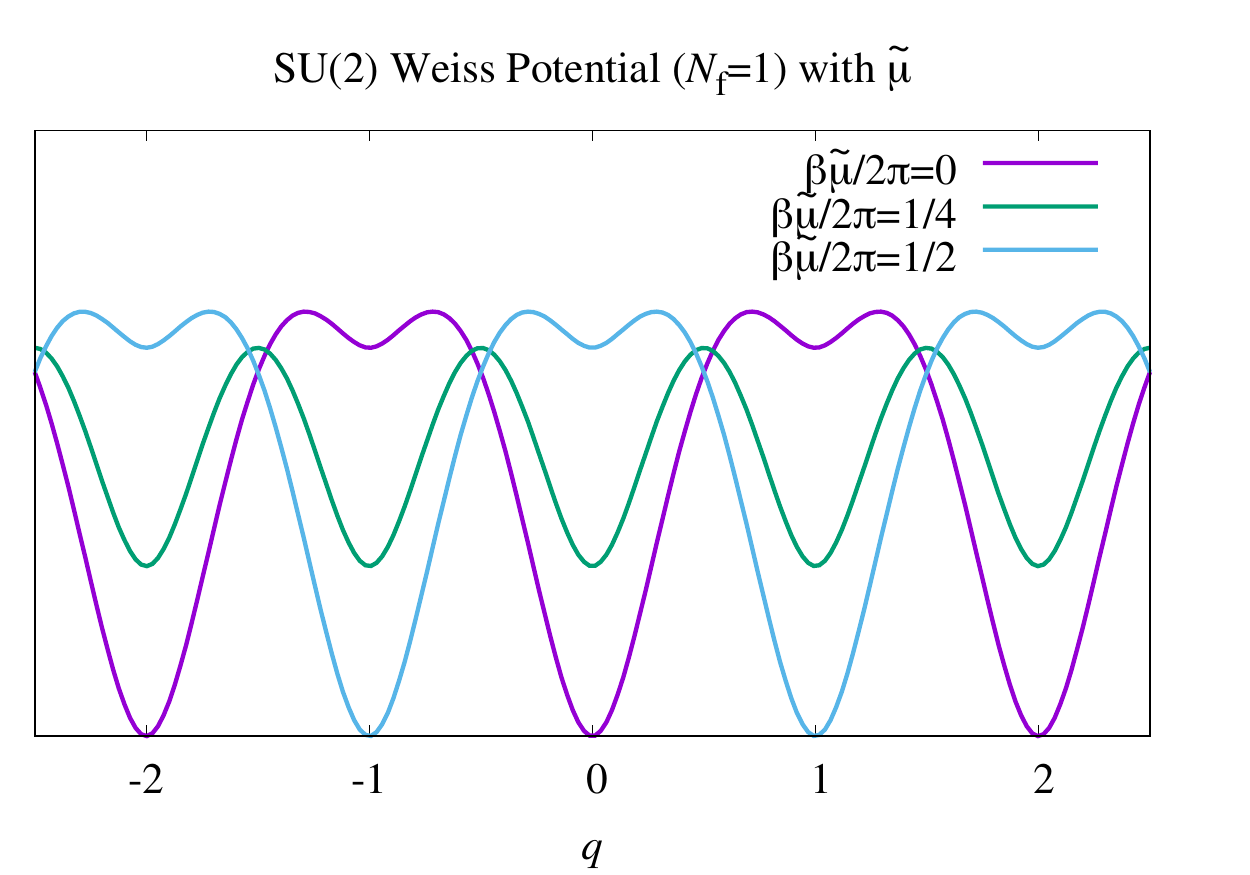}
  \includegraphics[width=0.47\textwidth]{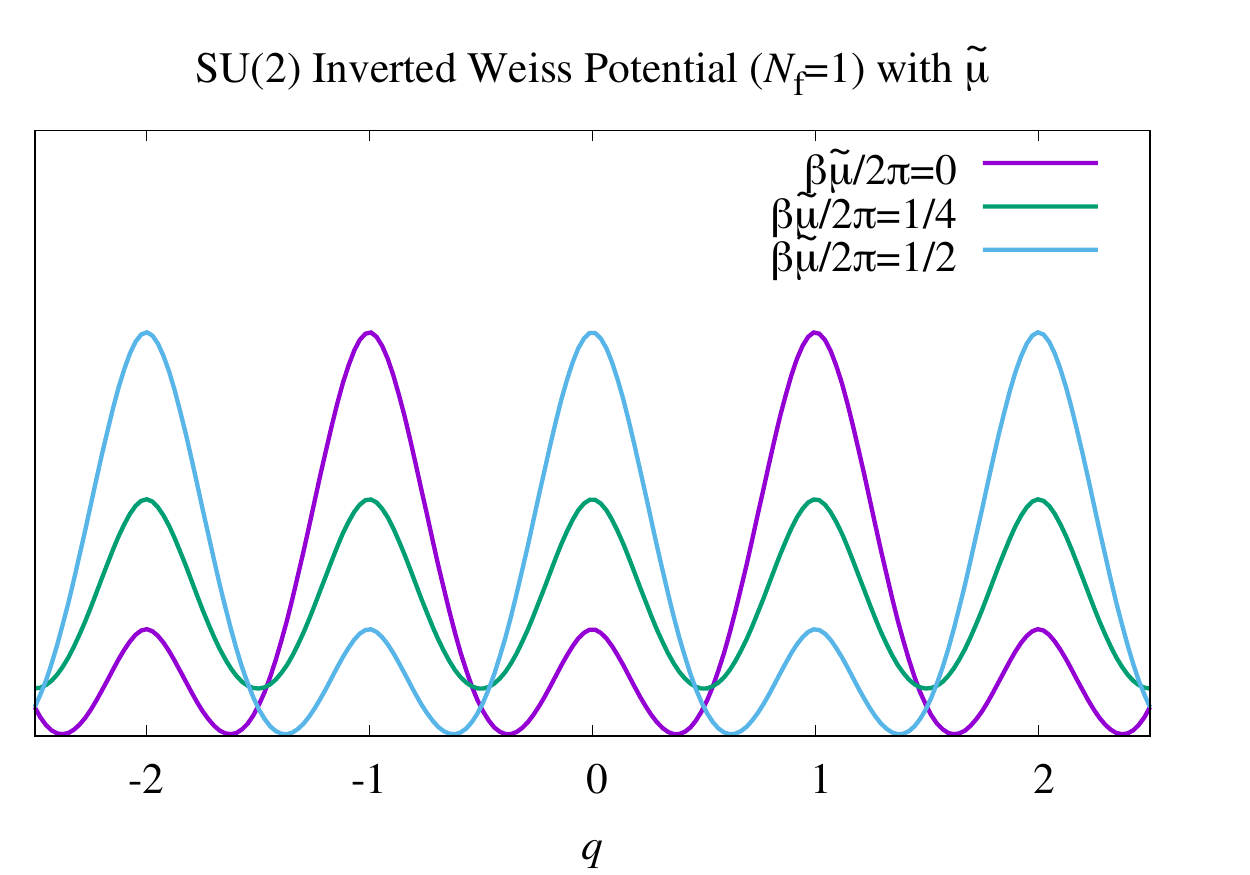}
  \end{center}
  \caption{SU(2) Polyakov loop potential with imaginary chemical
    potential $\imu$ for the perturbative Weiss potential (left) and
    the inverted Weiss potential (right).}
  \label{fig:robertweiss}
\end{figure}

In the deconfined phase the SU(2) perturbative Weiss potential is
modified by non-zero $\imu$ as shown in the left of
Fig.~\ref{fig:robertweiss}.  With increasing $\imu$ the perturbative
vacuum at $q=0$ is pushed up, while the center transformed point at
$q=1$ is pushed down.  Eventually, two minima become degenerated at
$\beta\imu/(2\pi)=1/4$, and then the absolute minimum jumps from $q=0$
to $q=1$, which signifies a first-order phase transition at
$\imu = \pi T/2$.  This first-order phase transition at
$\imu = \pi T/\Nc$ for general $\Nc$ is called the Roberge-Weiss phase
transition~\cite{Roberge:1986mm}.  In the confined phase, in contrast,
there is no such first-order phase transition owing to unbroken
center symmetry.  To see it expressly, we model confinement using the inverted
Weiss potential multiplied by $-1.4$.  Then, as shown in the right of
Fig.~\ref{fig:robertweiss}, the absolute minimum continuously moves
but no jump occurs in this case.  Here we point out an interesting
observation.  The confined phase corresponds to $q=1/2$ (to make
$\Phi=0$) if center symmetry is exact, but the potential minimum
generally differs from $q=1/2$ due to explicit breaking of center
symmetry in the presence of dynamical quarks.  Actually, at $\imu=0$,
the potential minimum is located at $q\simeq 0.377$ leading to
$\Phi\simeq 0.377$.  We can see, however, that the potential minimum
goes to $q=1/2$ exactly for $\beta\imu/(2\pi)=1/4$, which means that
center symmetry is unbroken for this special value of $\imu$.  We note
that with this special value of $\imu$ the quark distribution function
looks like a bosonic one.

It is possible to investigate the phase diagram at finite $\imu$ using
the PNJL model.  The location of the Roberge-Weiss point is robust
regardless of the model choice, but it would be model dependent at
which temperature the Roberge-Weiss phase boundary of first order ends
and at which temperature the crossovers of deconfinement and chiral
restoration occur for general $\imu$ away from the Roberge-Weiss point.  For
(non-local) PNJL analyses, see
Refs.~\cite{Pagura:2011rt,Kashiwa:2011td,Sakai:2008py,Sasaki:2011wu,%
  Morita:2011eu,Scheffler:2011te} for concrete evaluations.

Here, let us point out a suggestive similarity between an imaginary
chemical potential and the strong $\theta$ angle in the light quark
sector.  We must be careful that the $\theta$ angle dependence in the
pure gluonic sector shows resembling behavior to the light quark sector,
but the underlying physics is different~\cite{Witten:1998uka} (and
nevertheless, the light and the heavy sectors could be continuously
linked;  see Ref.~\cite{Mameda:2014cxa}).  If the theory has massless
fermion, the $\theta$ dependence is gauged away with help of the
${\rm U}(1)_{\rm A}$ anomaly.  For massive quarks, the $\theta$ angle
dependence is sensitive to whether chiral symmetry is broken or not.
If chiral symmetry is broken, the phase of the chiral condensate
changes with $\theta$, and then a first-order phase transition occurs
at $\theta=\pi$~\cite{Creutz:2009kx}.  It should be noted that, after
an appropriate ${\rm U}(1)_{\rm A}$ rotation, the $\theta$ dependence
appears only in the quark mass term as
$\bar{q}\,m\,e^{-i\theta\gamma_5}q$, which favors the chiral
condensate $\langle\bar{q}q\rangle$ in a $\theta$-tilted direction.
One may say that, in analogy to the Roberge-Weiss phase transition,
the relationship between center symmetry and the Polyakov loop at finite $\imu$
is comparable to the relationship between chiral symmetry and the chiral condensate at
finite $\theta$.  For further phase diagram research with finite $\theta$,
interested readers can see Refs.~\cite{Boer:2008ct,Boomsma:2009eh} for
the NJL model studies,
Refs.~\cite{Sakai:2011gs,Sasaki:2011cj} for the PNJL model studies, and
Refs.~\cite{DElia:2012pvq,DElia:2013uaf} for the lattice simulation.
Finally we make a note that, even though there are apparent
similarities between $\imu$ and $\theta$ in the phase structures, the
theory at finite $\imu$ is free from the sign problem, while a finite
$\theta$ term is always a complex phase suffering from the sign
problem.

\subsection{Deformed QCD with Center Symmetry}
\label{sec:qcdlike}

We have understood that center symmetry is spontaneously broken in the
deconfined phase.  From this fact, it has been speculated that
non-perturbation information on confinement and chiral symmetry
breaking could be extracted even \textit{perturbatively} by enforcing center
symmetry on QCD.\ \ We will introduce several ideas here, and finally we
will pay our special attention to deformed QCD in $R^3\times S^1$ as a rigorous
and successful framework.

\subsubsection{Canonical ensemble}

The first example of deformation that makes QCD preserve center
symmetry is the canonical ensemble.
There have been controversial arguments about the physical
interpretation of the center symmetry breaking at high temperature
(see Ref.~\cite{Smilga:1996cm} for critical discussions against
regarding the Polyakov loop as a physical quantity;  see also
Ref.~\cite{Hansson:1994ep} for a related work using solvable models).
For field-theoretical objects responsible for the triality screening,
see Ref.~\cite{Polonyi:1988yp}.

Interestingly, it is possible to reformulate hot QCD in such a way
that center symmetry is forced to be exact.  To understand the point
in the easiest manner, let us take a concrete expression of the
symmetry breaking term as Eq.~\eqref{Eq:Vheavy} or,
$-h(\ell + \ell^\ast)$, simply.  Then, the partition function is,
\begin{align}
  Z_{\rm QCD} &\sim \int\calD A_\mu\, e^{-S_{\rm glue}[A] -h \int d^3 x\,
    (\ell+\ell^\ast)} \notag\\
  &= \int\calD A_\mu\, e^{-S_{\rm glue}[A]}
  \frac{1}{\Nc} \Bigl( e^{-h\int d^3 x\,(\ell +\ell^\ast)}
  + e^{-h\int d^3 x\,(\ell + \ell^\ast)} + \cdots \Bigr) \notag\\
  &= \int\calD A_\mu\, e^{-S_{\rm glue}[A]}
  \frac{1}{\Nc}\sum_{k=0}^{\Nc-1} e^{-h\int d^3 x\,(z_k \ell + z_k^\ast \ell^\ast)}\;.
\label{eq:canonical}
\end{align}
Here, from the second to the third line, we used that the integration
measure is center invariant.  This superposition of seemingly
different but equivalent $\Nc$ integrals is nothing but a projection
operator  to the zero $\Nc$-ality (that is a charge associated with
${\rm Z}(\Nc)$ center group) sector.  In other words, the last
line of Eq.~\eqref{eq:canonical} represents the QCD partition function
in the canonical ensemble with respect to the $\Nc$-ality charge.

This is a surprisingly simple idea and was originally proposed to
circumvent the absence of the strict confinement order parameter (see
Refs.~\cite{Fukushima:2002bk,Detar:1982wp} and references therein).
Thanks to the apparently restored center symmetry in the partition
function, the na\"{i}ve expectation was that the Polyakov loop should
be no longer an approximate but an exact order parameter.  The above
way to restore chiral symmetry is somehow an approach opposite to
restricting the functional integration to one fundamental modular
domain~\cite{Zwanziger:1993dh,Lenz:1997su,Lenz:1998qk}, which was also
proposed to account for confinement.

Let us explain the idea a bit more using the physics language.  In
principle, the Dirac determinant can be expanded in terms of gauge
invariant gluonic operators, namely, the Wilson loops.  The most
important trajectories of the Wilson loops in our consideration at
finite $T$ are the ones wrapping around the temporal direction (which is
nothing but the Polyakov loop if the trajectory is straight along the temporal
axis), which counts the excited quark number in the thermal
bath~\cite{Green:1983sd}.  If the trajectory has one winding, its operator
would change non-trivially under the center transformation, but such
an expectation value
should be just vanishing after the functional integration of
entire gluonic configurations.  That is, in the example of
Eq.~\eqref{eq:canonical}, if the symmetry breaking term
is expanded, the first order term is,
\begin{equation}
  \int\calD A_\mu\, e^{-S_{\rm glue}[A]} (-h)\int d^3 x\, \ell(x)
  = 0\;,
\end{equation}
from Elitzur's
theorem~\cite{Elitzur:1975im}, while the second order term contains,
\begin{equation}
  \int\calD A_\mu\, e^{-S_{\rm glue}[A]} \frac{(-h)^2}{2!}
  \int d^3 x\,d^3 y\,\ell(x)\,\ell^\ast(y) \neq 0\;,
\end{equation}
which physically represents a pair of quark at $x$ and anti-quark at
$y$.  One might want to associate this type of pair excitation with
mesons, but just like the statistical confinement in the PNJL model,
$x$ and $y$ can be independently separated.  Also, the $\Nc$-th order
term has a baryonic combination, i.e.
\begin{equation}
  \int\calD A_\mu\, e^{-S_{\rm glue}[A]} \frac{(-h)^{\Nc}}{\Nc!}
  \int d^3 x_1\dots d^3 x_{\Nc}\,\ell(x_1)\dots \ell(x_{\Nc}) \neq 0\;,
\end{equation}
but again, $\Nc$ quarks may be distributed non-locally.  We should
stress that, even without rewriting the partition function as in
Eq.~\eqref{eq:canonical}, center symmetry seems to be unbroken superficially,
and Eq.~\eqref{eq:canonical} is simply a way to make it clear.

This argument about unbroken center symmetry is mathematically correct
as long as the volume is finite.  We should note that the canonical
ensemble could be then not equivalent to the grand canonical ensemble, and
in this sense, nothing is surprising about such unbroken center
symmetry in the canonical ensemble if the volume is finite.  The
question is the infinite volume limit or the thermodynamic limit, in
which the canonical ensemble should become equivalent to the grand
canonical ensemble.  This happens in a very singular way, and actually
the center symmetric vacuum in the canonical ensemble turned out to be
thermodynamically unstable in the large volume limit, as first pointed
out in Ref.~\cite{Hagedorn:1984uy}.  Roughly speaking, the instability
occurs due to abnormally long-ranged interactions as seen in Eq.~\eqref{eq:canonical}.
If the final form in Eq.~\eqref{eq:canonical} is exponentiated, the
interaction looks like,
\begin{equation}
  S_{\rm int} = -\ln\Biggl[ \frac{1}{\Nc}\sum_{k=0}^{\Nc-1}
    e^{-h\int d^3 x\,(z_k \ell + z_k^\ast \ell^\ast)} \Biggr]\;,
\end{equation}
for which the volume is not simply factored out.  Then, in the
infinite volume limit, center symmetry in the canonical ensemble is
always \textit{spontaneously} broken down into the explicitly broken
state in the grand canonical ensemble.  Therefore, unfortunately, the
canonical ensemble approach to restore center symmetry does not work
as na\"{i}vely expected.  Moreover, one can learn an important lesson;
when the finite-density phase transition is investigated in a
small-size canonical ensemble as discussed in
Refs.~\cite{Alexandru:2005ix,Li:2010qf}, it is difficult to make a
reliable conclusion until the thermodynamic limit is taken.

\subsubsection{Center twisted flavors}
\label{sec:twisted}

We can avoid the instability of the canonical ensemble once we change the interaction so that
the volume factor is correctly factored out.  The simplest remedy is
to replace the projection sum in Eq.~\eqref{eq:canonical} with the
projection product~\cite{Kouno:2012zz,Kouno:2013mma,Kouno:2015sja}.
Then, the theory is no longer QCD and the partition function in this
``QCD-like theory'' reads,
\begin{equation}
  Z_{\rm QCD-like} \sim \int \calD A_\mu\, e^{-S_{\rm glue}[A]}
  \prod_{k=0}^{\Nc-1} e^{-h\int d^3 x\,(z_k\ell + z_k^\ast \ell^\ast)}\;.
\label{eq:twisted}
\end{equation}
In the leading order of the expansion, the above is just a trivial
example and the matter parts are averaged away.  It should be obvious how this idea can be generalized
to unexpanded form of the Dirac operator once we realize that
multiplying $z_k$ on $\ell$ is equivalent to adding an imaginary
chemical potential by $i2\pi T k/\Nc$.  That is, this center symmetric
QCD-like theory may be defined as
\begin{equation}
  Z_{\rm QCD-like} = \int \calD A_\mu\,e^{-S_{\rm glue}[A]}
  \prod_{k=0}^{\Nc-1} \det \calM[A;\mu + i2\pi T k/\Nc]\;.
\end{equation}
Such a deformed QCD as defined above has interesting properties.  Because
$\Nc$ quarks as a combination do not break center symmetry explicitly,
the first-order deconfinement phase transition in the pure gluonic
sector survives with a shifted critical point to a lower
temperature, just like in QCD-like theories with adjoint fermions.
What is quite non-trivial is that chiral symmetry breaking persists at
temperatures far above deconfinement (as shown in the right panel of
Fig.~\ref{fig:schematic}), which is because single quark
excitations are projected out as seen in Eq.~\eqref{eq:twisted}.
Moreover, flavor-dependent imaginary chemical potential explicitly
breaks chiral symmetry as
${\rm SU}(\Nc)_{\rm L}\times {\rm SU}(\Nc)_{\rm R}
\to {\rm U}(1)_{\rm V}^{\Nc-1}\times {\rm U}(1)_{\rm A}^{\Nc-1}$.
Although the introduced deformation changes the theory from QCD, this
center symmetric formulation has been attracting interest in
connection to the instanton-dyon model of
confinement~\cite{Larsen:2016fvs}, especially to understand the
interplay between confinement and chiral symmetry breaking.

\subsubsection{Center-stabilized QCD}

A more sophisticated idea to impose center symmetry on QCD was
proposed in Ref.~\cite{Unsal:2008ch} for the pure gluonic sector on
$R^3\times S^1$ in the large-$\Nc$ limit, and later extended to
address continuous chiral symmetry including the matter
sector~\cite{Shifman:2009tp} (see Ref.~\cite{Sannino:2005sk,DelDebbio:2008ur} for earlier
exact results at large $\Nc$ using orientifold field theories, and also
Ref.~\cite{Vuorinen:2006nz} for
related discussions near the transition temperature).
In the original argument in
Ref.~\cite{Unsal:2008ch}, a toroidal compactification is also
discussed, but here, for simplicity, we look over the one dimensional
case only and change the notation of the compactification size from
$L$ to $\beta=1/T$.  The point of the whole idea is that the
perturbation theory inevitably breaks center symmetry as we already
saw in Sec.~\ref{sec:perturbative}, but we can deform the pure gluonic
theory by adding terms that prevent the theory from breaking center
symmetry spontaneously even in the perturbative regime.

For the deformation to stabilize center symmetry, quadratic
interactions such as $|\ell|^2$ and more generally $|\tr(L^n)|^2$ are
added.  Then, for high temperature, the perturbation theory should
work due to small running coupling, but center symmetry is still
unbroken.  The key observation in such center-stabilized QCD is that,
as long as there is no center symmetry breaking, such deformed QCD at
any temperature is smoothly connected to the pure gluonic theory at
zero temperature in the large-$\Nc$ limit.  This opens a possibility
that confinement can be perturbatively investigated if the temperature
is large enough, i.e.\ $T\gg \Nc\LQCD$.

One can intuitively understand the essence of the correspondences from
the following arguments.  We know that in the ordinary pure gluonic
theory the large-$\Nc$ limit makes the low-$T$ and the high-$T$ phases
clearly distinct.  Below the deconfinement transition temperature, all
physical quantities are insensitive to the temperature because the
glueball excitations are $1/\Nc^2$ suppressed as compared to the gluon
excitations that are prohibited by center symmetry.  In the
center-stabilized QCD, there is no phase transition, and so the
temperature dependence is always suppressed for any temperature.
Thus, with imposed center symmetry, the high-$T$ state is equivalent
to the low-$T$ state and the latter is naturally connected to the pure
gluonic theory only with smooth deformation.

In this way, in Ref.~\cite{Unsal:2008ch}, the mass gap and the string
tension have been parametrically estimated.  For the description of
finite mass gap of gauge bosons, the vacuum state is first identified
from the condition to minimize the perturbatively obtained energy, which
favors a special form of the Polyakov loop matrix as
$L=\diag(1,z_1,z_2,\cdots,z_{\Nc-1})$ where $\tr L=0$ is indeed
satisfied.  It should be noted that, if each eigenvalue of $L$ is
associated with different quark flavor, a deformed QCD-like theory in
Sec.~\ref{sec:twisted} emerges.  This background configuration
spontaneously breaks the gauge symmetry from ${\rm SU}(\Nc)$ down to
${\rm U}(1)^{\Nc-1}$, i.e.\ Abelianized, giving a mass gap $\sim T$
for off-diagonal gauge bosons.  Then, including non-perturbative
contributions to the diagonal ``photons'', the mass gap is found to be
$\sim \LQCD(\Nc\LQCD/T)^{5/6}|\ln \Nc\LQCD/T|^{9/11}$.  Also, the
string tension turns out to be
$\sim \LQCD^2(\Nc\LQCD/T)^{-1/6}|\ln \Nc\LQCD/T|^{-3/11}$.
As long as the flavor number is $\calO(1)\ll \Nc$, the above mentioned
argument holds with quarks in the fundamental representation
included.  Then, because chiral symmetry is not broken perturbatively,
the deformed QCD at high enough temperature $\gg \Nc\LQCD$ should
accommodate a state with confinement but no chiral symmetry breaking,
which suggests a conjecture to relate spontaneous breaking of chiral
symmetry to not Abelian but non-Abelian confinement whose critical
temperature is of order of $\sim \Nc\LQCD$~\cite{Shifman:2009tp}.

We shall make a comment on related developments on $R^3\times S^1$.
In Ref.~\cite{Poppitz:2012nz} a twisted partition function for the
gluonic theory with one adjoint Weyl fermion (whose mass is $m$) was
considered on $R^3\times S^1$.  In the $m=0$ limit perturbative
confinement realizes, while the ordinary pure gluonic theory is
recovered in the $m\to\infty$ limit.  It is claimed in
Ref.~\cite{Poppitz:2012nz} that the confinement-deconfinement
transition is continuous as $m$ changes, which means that
non-perturbative confinement may be investigated perturbatively with
$m$ as an interpolating variable.  An interpretation in terms of
monopoles was also given, and then the chiral condensate was
calculated analytically in Ref.~\cite{Cherman:2016hcd}, which shows an
interesting connection to the idea of center twisted flavors as we saw
in the previous subsection.

\section{Phenomenological Implications}
\label{sec:miscellaneous}

In this section we will make a brief overview of two selected topics on
phenomenological applications of the Polyakov loop physics.  There are
many relevant topics especially in the context of heavy-ion
collisions, but we specifically choose the following:  One is the
fluctuation measurement which is expected to have good sensitivity to probe a
phase transition.  The other is an effective description of the
so-called strongly-correlated QGP (sQGP) by means of the Polyakov
loop, or an alternative of sQGP, which is known as the semi-QGP
regime.  For other phenomenological applications such as the chiral
fluid dynamics~\cite{Herold:2013bi}, the domain-wall
dynamics~\cite{Asakawa:2012yv}, and so on, readers can consult recent
proceedings volumes of Quark Matter conference
series~\cite{Braun-Munzinger:2014pya,Proceedings:2016arp}.

\subsection{Higher Cumulants and Polyakov Loop Fluctuations}
\label{sec:cumulants}

The idea is close to the one discussed in Sec.~\ref{sec:critical}, and
the Polyakov loop potential could be probed from the light quark sector.
In Sec.~\ref{sec:PNJL}, 
we discussed two limits $\Phi \to 1$ and   $\Phi \to 0$. 
In the former limit, we demonstrated that the quark thermodynamics 
is dominated by free quarks. In the latter limit,  we showed that
quark and diquark-type excitations are suppressed and only color-singlet 
combinations of three quarks survive.
This radical change in the effective degrees of freedom 
impacts many observables measurable in the heavy-ion collisions.
In particular,
the cumulants of the net-baryon number can be very sensitive  
probes of this change.
It is probably worth mentioning that the very first successful example
of the PNJL model calculations is the quantitative agreement of the
quark number susceptibility with the lattice
result~\cite{Ratti:2005jh}.

From the mathematical point of view, the thermodynamic pressure 
is a cumulant generating functional for the net-baryon number fluctuations. 
Thus, we find the cumulants by differentiating the pressure with respect
to the baryon chemical potential,
\begin{equation}
	\chi^B_n  = \frac{\partial^n (p/T^4)}{\partial (\mu_B/T)^n}\;.
	\label{Eq:chi_n}
\end{equation}
In the same way one can define the cumulants with respect to other
quantum numbers such as the strangeness, the electric charge, etc., as
QGP probes~\cite{Asakawa:2000wh,Jeon:2000wg}.
We here list first few non-trivial coefficients expressed through the moments as 
\begin{align}
	\chi^B_2 &= \frac1{VT^3} \langle (\delta N_B)^2 \rangle\;, \\  
	\chi^B_3 &= \frac1{VT^3} \langle (\delta N_B)^3 \rangle\;, \\  
	\chi^B_4 &= \frac1{VT^3} \left( \langle (\delta N_B)^4 \rangle - 3 \langle (\delta N_B)^2 \rangle^2 \right)\;,
	\label{Eq:chi4}
\end{align}
where $\delta N_B \equiv N_B - \langle N_B\rangle $ and $N_B$ is the net-baryon number.  

The so-called kurtosis defined by the ratio, $\chi^B_4/\chi^B_2$,
is known as a good probe of the deconfinement transition;
as we demonstrate below, at zero chemical potential, 
it reflects the quark content of physical degrees of freedom that
carry the baryon number~\cite{Ejiri:2005wq,Karsch:2010ck}.

Let us first consider \textit{small temperatures} where $\Phi \to 0$ and 
the spontaneous chiral symmetry breaking results in a dynamical quark
mass as $m \gg T$. 
In this limit we can perform the expansion similar to what we saw in
the heavy-quark limit; see Eq.~\eqref{Eq:BExp}.
For simplicity we assume that all flavors are degenerate, and then we
find the quark pressure as
\begin{equation}
  \begin{split}
	p = -\frac{V_{\rm quark}}{V}  =   -2 T \Nf \int \frac{d^3p}{(2\pi)^3} 
		\Bigl\{
		e^{-\beta \varepsilon_p} \bigl( \tr L e^{\beta \mu} +  \tr L^\dagger e^{-\beta \mu}  \big)
		- \frac{1}{2} e^{-2 \beta \varepsilon_p} \bigl[ \tr L^2 e^{2 \beta \mu} +  \tr (L^\dagger)^2 e^{-2 \beta \mu}  \bigr]
		 \\  + \frac{1}{3} e^{-3 \beta \varepsilon_p} \bigl[ \tr L^3 e^{3 \beta \mu} + \tr (L^\dagger)^3 e^{-3 \beta \mu} \bigr] + \dots 
		\Bigr\}\;, 
  \end{split}
  \end{equation}
where the higher order corrections  are suppressed exponentially. 
Differentiating this expression  with respect to the baryon chemical potential ($\mu_B = 3\mu$), 
setting it to zero, and equating $\Phi=\bar{\Phi}$, we arrive in the
mean-field approximation at
\begin{equation}
	\lim_{T\to 0} \frac{\chi^B_4}{\chi^B_2} \approx 
	\frac{\displaystyle \int \frac{d^3p}{(2\pi)^3}  \biggl[ \frac{1}{81} e^{-\beta \varepsilon_p} \Phi - \frac{8}{81} e^{-2\beta \varepsilon_p} (3 \Phi -2) \Phi
	+ e^{-3\beta \varepsilon_p} (27 \Phi^3 - 27\Phi^2 +3)
\biggr] }
{\displaystyle
	\int \frac{d^3p}{(2\pi)^3}  \biggl[ \frac{1}{9} e^{-\beta \varepsilon_p} \Phi - \frac{4}{9} e^{-2\beta \varepsilon_p} (3\Phi -2) \Phi 
	+  e^{-3\beta \varepsilon_p} (27\Phi^3 - 27\Phi^2 +3)
	\biggr]
}\,.
\end{equation}
Now in the limit of vanishing $\Phi$ we see that only the last terms
(containing color-singlet parts) proportional to
$e^{-3\beta \varepsilon_p}$ survive in the numerator and the denominator;
therefore, the ratio $\chi^B_4/\chi^B_2$ goes to the unity, as we expect from the hadron resonance gas model.
We can also check that, if we neglect the Polyakov loop coupling of
quarks (that is, if we set $\Phi\to1$ in the above),
the leading term with $e^{-\beta \varepsilon_p}$ dominates at small temperature and 
we get a smaller result of $\chi^B_4/\chi^B_2 \to 1/9$.

At \textit{high temperatures}, on the other hand, chiral symmetry is
supposed to be restored; hence, we can neglect the quark mass and recover Eq.~\eqref{eq:weiss_quark}. 
By differentiating the pressure obtained from
Eq.~\eqref{eq:weiss_quark}, at zero chemical potential and in the
perturbative limit ($q\to0$), we obtain, 
\begin{equation}
	\lim_{T\to\infty} \frac{\chi^B_4}{\chi^B_2}  \to \frac{2}{3\pi^2}\;. 
\end{equation}
This result is slightly different from $1/9$ as considered in
Ref.~\cite{Ejiri:2005wq}.  In fact, $1/9$ appears in the Boltzmann
approximation that leaves only the leading order of the expansion in
$e^{-\beta\varepsilon_p}$, but such a treatment is no longer valid once
$T\gg m$ owing to the  chiral symmetry restoration at high temperature. 

\begin{figure}
  \centering
\includegraphics[width=0.47\linewidth]{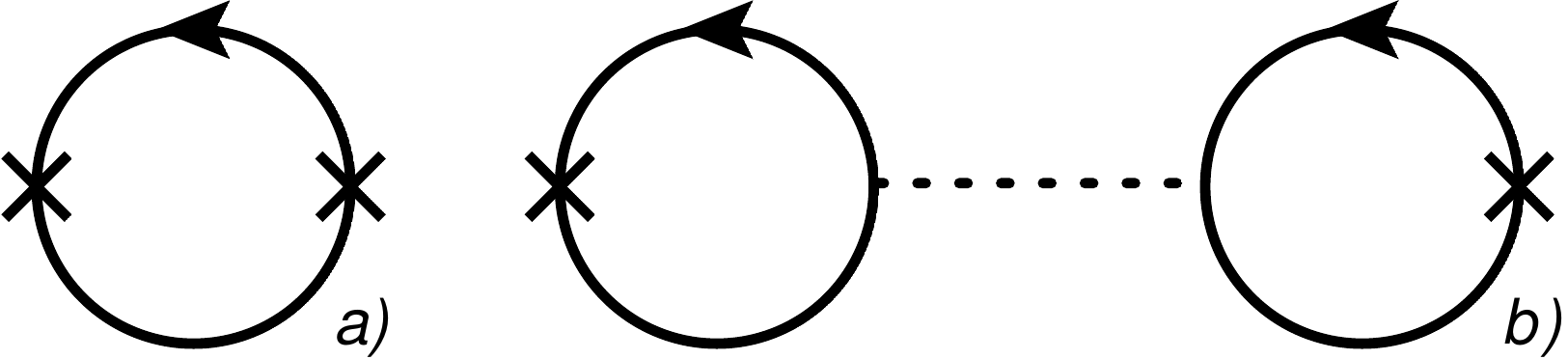}
\caption{Diagrammatic representation of different contributions to
  $\chi^B_2$ in  Eq.~\eqref{Eq:chi2fin}: (a) the one-particle
  irreducible part and
  (b) the one-particle reducible part.  The dashed line corresponds to
  the mean-field exchange defined by the inverse of the mass matrix, namely, the susceptibility $\chi_{ij}$.   
}
\label{fig:c2_diag}
\end{figure}

On top of the above discussions on deconfinement,
the cumulants are also useful probes to study the chiral dynamics and
the effect of the Polyakov loop fluctuations.  
In order to demonstrate this we have to refine the discussions presented above and 
consider a mean-field description with the self-consistently defined
fields.
We here introduce a collective notation, $\phi \equiv (\Phi, \bar\Phi,
\sigma)$, to simplify expressions.
The first derivative of the pressure with respect to the chemical potential is 
\begin{equation}
  \frac{d p}{d \mu_B}  = \frac{\partial p}{\partial \mu_B}
  + 
  \frac{\partial p}{\partial \phi_i}
  \frac{\partial \phi_i}{\partial \mu_B}
  = 
	\frac{\partial p}{\partial \mu_B} \;,
	\label{Eq:c1}
\end{equation}
where the second equality follows from the equations of motion,
$\partial p/\partial \phi_i = 0$.
Continuing to take the second derivative, we find,   
\begin{equation}
	\frac{d^2 p}{d \mu_B^2}  = \frac{\partial^2 p}{\partial \mu_B^2} + \frac{\partial^2 p}{\partial \mu_B \partial \phi_i} \frac{\partial \phi_i}{\partial \mu_B}
	= \frac{\partial^2 p}{\partial \mu_B^2} + \frac{\partial^2 p}{\partial \mu_B \partial \phi_i}  
	\chi_{ij}
	\frac{\partial^2 p}{\partial \mu_B \partial \phi_j}\;,  
	\label{Eq:c2}
\end{equation}
where we defined the susceptibilities, $\chi_{ij}$, as the inverse of
the mass matrix, i.e.
\begin{equation}
	  \chi_{ij}   \biggl( - \frac{\partial^2 p}{\partial \phi_j \partial \phi_k} \biggr)  = \delta_{ik}\;.
	\label{Eq:chi_defin}
\end{equation}
In Eqs.~\eqref{Eq:c2} and \eqref{Eq:chi_defin}, we used the equation
of motion differentiated with respect to the chemical potential (that is justified by the fact that 
the equation of motion holds for any $\mu$),
\begin{equation}
  \frac{d}{d\mu_B} \frac{\partial p}{\partial \phi_i} =  \frac{\partial^2 p}{\partial \mu_B \partial \phi_i} + 
	\frac{\partial^2 p}{\partial \phi_i \partial \phi_j} \frac{\partial \phi_j}{\partial \mu_B} = 0 \;.
\end{equation}
This relation allows us to eliminate $\partial \phi_i/\partial \mu_B$ from Eq.~\eqref{Eq:c2}.
Finally, we see that the second cumulant is related to the field susceptibilities $\chi_{ij}$ as
\begin{equation}
	\chi^B_2 = \frac{\partial^2 (p_{\rm quark}/T^4)} {\partial (\mu_B/T)^2} 
	+ T^2  \frac{\partial^2 (p_{\rm quark}/T^4)}{\partial (\mu_B/T) \partial \phi_i}  
	\chi_{ij}
	\frac{\partial^2 (p_{\rm quark}/T^4)}{\partial (\mu_B/T) \partial \phi_j}\;.
	\label{Eq:chi2fin}
\end{equation}
Here we replaced the total pressure $p$ with its quark part $p_{\rm quark}$ 
in the differentiations with respect to the chemical potential.
This equation has an intuitive diagrammatic representation as shown in
Fig.~\ref{fig:c2_diag}.
Clearly, the baryon number fluctuation is affected by the Polyakov
loop fluctuations in $\chi_{\Phi\Phi}$, $\chi_{\Phi\bar{\Phi}}$, and
$\chi_{\bar{\Phi}\bar{\Phi}}$ as well as the chiral susceptibility.

For even higher-order cumulants the number of contributing diagrams
increases and the full expression becomes tedious;
at the same time, it would not bring any new theoretical insights
beyond those that can already be illustrated in
Eq.~\eqref{Eq:chi2fin}.  Here, we point two useful features out below.

\begin{figure}
  \centering
  \includegraphics[width=0.3\linewidth]{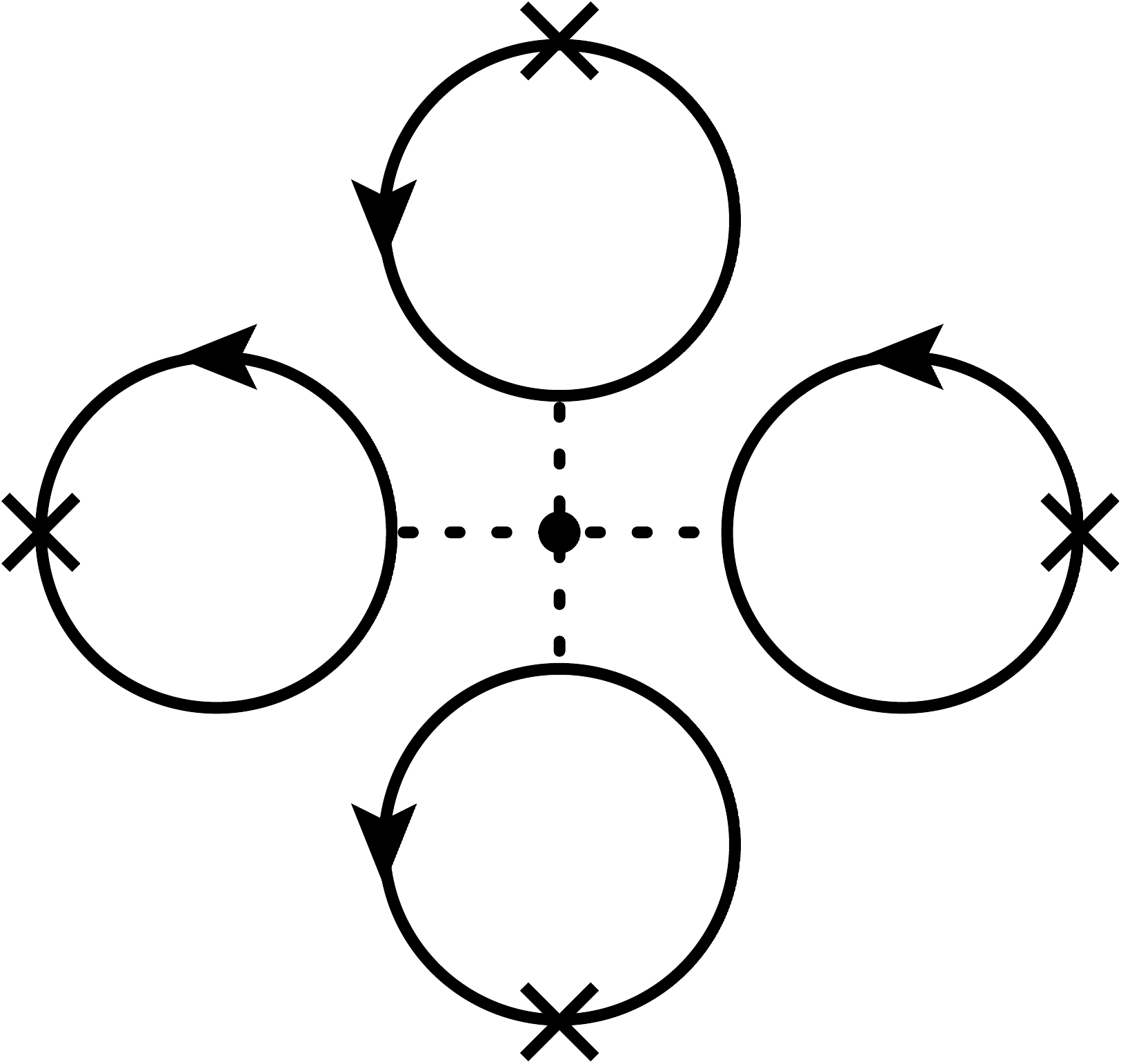}
\caption{Diagrammatic representation of the divergent contribution to $\chi_4$ at chiral critical point.
	The dashed line corresponds to the $\sigma$ propagator. 
}
\label{fig:c4_diag}
\end{figure}

\textit{First}, at the chiral critical point, the $\sigma$
susceptibility diverges and this will be reflected in
the second order cumulant via the one-particle reducible contribution
[see Fig.~\ref{fig:c2_diag}~(b)].
Actually, based on this argument, it is easy to see why higher-order
cumulants exhibit stronger divergence;  they include contributions
with larger number of the $\sigma$ propagator, as illustrated in
Fig.~\ref{fig:c4_diag} for an example of the fourth-order contribution.  
A detailed analysis of the singular contribution to the cumulants shows that
the $n$-th order cumulant of the net-baryon number fluctuations diverges with the correlation length, $\xi$, as~\cite{Stephanov:2008qz}
\begin{equation}
	\chi^{\rm sing}_n \propto \xi^{\frac{n\beta \delta}{\nu} -3 }\,,
\end{equation}
where 
$\beta$, $\delta$ and $\nu$ are the standard notations for critical exponents. 
The QCD critical point belongs to the Z(2) universality class, and thus 
$\beta\approx 0.31$, $\delta \approx 5$, $\nu\approx 0.64$.
Hence the fourth-order cumulant, $\chi^B_4$, is proportional to
$\approx \xi^7$ as predicted in Ref.~\cite{Stephanov:2008qz}.
The sensitivity of the fourth-order cumulant to the correlation length 
has motivated an experimental search for the QCD critical point.
This is an exciting research field and interested readers are guided
to see, e.g.\ Ref.~\cite{Luo:2017faz} for an overview of the
state-of-the-art experimental status.

\textit{Second}, through the one-particle reducible part, the second
order cumulant as well as higher orders are all sensitive to the
Polyakov loop susceptibilities.
This becomes especially important at high temperatures (not
necessarily in the vicinity of the critical point),
where $\sigma$ decouples and the details of the Polyakov loop
potential significantly affect the whole cumulants.
The Polyakov loop potential including fluctuation effects was proposed
in Ref.~\cite{Lo:2013hla}, and recently, the FRG-QCD
calculations~\cite{Fu:2016tey} imply that the experimental data for
the kurtosis $\chi^B_4/\chi^B_2$ can be best reproduced with this
proposed Polyakov loop potential.  This is the first experimental
(other than the numerical experiment of the lattice QCD) indication
to provide us with very concrete information on the Polyakov loop
physics.

\subsection{Semi-QGP Regime}
\label{sec:semiQGP}

In the previous subsection we discussed the effect of the Polyakov loop on the equilibrium static observables.
Here we also consider how the Polyakov loop may change the out-of-equilibrium 
observables such as the shear viscosity, the collisional energy loss,
and the production of direct photons and dileptons.

Experimentally the heavy-ion collisions at ultra-relativistic energies
probe QCD at temperatures above and close to the transition temperature during most of the collision evolution. 
Collective properties confirmed by experimental measurements (e.g.\ elliptic flow) 
demonstrated that the matter created in these collisions can be
described well by 
hydrodynamics with a very small value of the ratio of the shear viscosity, $\eta$, 
to the entropy. At least na\"{i}vely, this fact implies that the QCD coupling constant 
is large near the transition temperature as the extracted 
value of the shear viscosity ratio is explained by a strongly coupled
$\mathcal{N}=4$ supersymmetric SU($\Nc$) theory~\cite{Kovtun:2004de}.

However, strong correlation does not necessarily mean strong coupling,
and moreover, EQCD (that is a three-dimensional effective theory of
hot QCD) demonstrated that 
at the transition temperature the coupling constant stays quite
moderate, i.e.\ $\alpha_s^{\rm EQCD}\approx 0.3$~\cite{Laine:2005ai}.
Motivated by this and constrained by the fact that
such a transitional thermodynamic region cannot be treated reliably neither by a hadron 
resonance gas nor by (resummed) QCD perturbation theory, 
the authors of Ref.~\cite{Hidaka:2009ma} 
explored an alternative explanation for the small value of the shear viscosity 
even with the moderate/weak coupling constant.  
To describe the transition, they considered a new regime with a partial ionization of color, called the ``semi''-Quark-Gluon Plasma (semi-QGP). 
This region of partial ionization/deconfinement can naturally be modeled by
a non-trivial Polyakov loop.
On a small sphere as we discussed in Sec.~\ref{sec:sphere},
one can show that the semi-QGP description is  manifestly an appropriate effective theory. 

Most of the results concluded from the semi-QGP calculations can be
intuitively understood as follows. 
In Minkowski spacetime, the diagrammatic calculations are formulated as
an ordinary 
perturbation theory, except that the background field $A_4$ acts 
like an imaginary chemical potential for color.  For a quark with 
a given color $a$ and for a gluon with a given color pair $a$ and $b$,
the statistical distribution functions are, respectively,
$n_a = [\exp(-\beta \varepsilon + 2\pi i q_a)+1]^{-1}$ and
$n_{ab}=[\exp(-\beta \varepsilon + 2\pi i q_{ab})-1]^{-1}$.
In the Boltzmann approximation, the color-traced distribution function for a 
single quark (and anti-quark) is suppressed by 
the Polyakov loop, that is,
$\Nc^{-1}\sum_a \exp(-\beta \varepsilon + 2\pi i q_a) \propto 
\exp(-\beta \varepsilon)\, \ell$.
For gluons, the suppression is more prominent as
$\exp(-\beta \varepsilon)\, \ell^2$.
Thus, in this way, any physical observable sensitive to the abundance of color charge in the system
will be suppressed in the semi-QGP regime by one or even larger powers
of the Polyakov loop. 

Let us consider the shear viscosity first.
Here we will give a somewhat qualitative account by skipping concrete
calculations.  For technical details, see the original calculations in
Ref.~\cite{Hidaka:2009ma} and also see
Refs.~\cite{Arnold:2000dr,Arnold:2003zc} for the conventional
calculations of the transport coefficients in perturbative QCD.

The shear viscosity of a gas is in general proportional to 
$\eta \propto n \bar{p} \lambda$, where $n$ is the number density, 
$\bar{p}$ is the average momentum, and $\lambda$ is the mean free path. 
We first consider a conventional plasma of fully-ionized gluons to extract 
a parametric dependence of the 
viscosity on the temperature and the coupling constant.
The mean average momentum in this case is proportional to 
the temperature, the mean free path is of the order of
$1/(n\sigma)$ where $\sigma$ is the cross section.
In QCD, the transport cross section is 
$\sigma_{\rm pQCD} \sim g^4 T^{-2} \ln(T/m_{\rm Debye})$ with the
Debye mass of of the order of $gT$, where
the logarithm arises due to an infrared singularity in the forward scattering.
Combining everything together, we conclude that the shear viscosity is
parametrically given by 
$\eta_{\rm pQCD} \sim T^3/(g^4 \ln 1/g)$. 

Now, we shall consider the semi-QGP regime and show that in this case the counting is somewhat different.
The mean free path should be changed due to the color flow.  It 
 is proportional to the average number of the color sources 
divided by the effective collisional area or the collisional cross
section, which leads to
\begin{equation}
  \lambda_{\rm semi-QGP} \;\propto\;
  \frac{\displaystyle \sum_{a,b} n_{ab} }
       {\displaystyle \sum_{a,b,c,d} n_{ab}\, n_{cd}\; \sigma_{ab,cd} }\;.
\end{equation}
The gluon density is proportional to $\ell^2 T^3$ as we discussed above. 
The denominator in $\lambda_{\rm semi-QGP}$, describing a hard
$2\leftrightarrow 2$ scattering 
with a single soft exchange in the $t$-channel, is dominated by the planar 
graph with two of the color indices being equal.  This color index contraction leads to 
a partial cancellation of, say, $q_b$ in $n_{ab}$ and $q_c$ in
$n_{cd}$ in the color sum 
and results in the parametric result 
$\lambda_{\rm semi-QGP} \propto (\ell^2 T^3) /(\ell^2 T^6 \sigma) = (T^3 \sigma)^{-1}$, 
which is approximately $\ell$ independent. 

The shear viscosity of semi-QGP is then
$\eta_{\rm semi-QGP} \sim  T \sum_{a,b} n_{ab} /  \lambda_{\rm semi-QGP} \propto
\ell^2 \,\eta_{\rm pQCD}$, so that the shear viscosity as compared to
the perturbative result is suppressed by the Polyakov loop squared.
In the vicinity of the phase transition in the pure gluonic theory,
the Polyakov loop is roughly $\sim 1/2$, and thus the viscosity of
semi-QGP is smaller than the perturbative estimate by a factor $\sim 4$.
Interestingly, taking account of quark degrees of freedom does not modify this parametric suppression of the 
shear viscosity by the Polyakov loop.

Before moving on to the next example of physical observables, we
mention on an exceptional case, for which only color singlet states
make a contribution. 
Let us consider the dilepton production 
and show that it is not 
sensitive to the Polyakov loop at least at the leading order.
This is to be expected since the dilepton rate arises from the
annihilation process of a quark  and an
anti-quark in a color singlet states going into a virtual photon,
which can eventually decay
into a dilepton pair.

For the sake of simple arguments, we restrict ourselves to the
back-to-back dilepton pair, i.e.\ the momentum of the intermediate 
photon is $p=0$.
In the kinetic theory, the dilepton production rate is given by the product of
quark statistical distribution functions and a squared amplitude.
The energy conservation requires that each quark carries a
half of the dilepton energy, $E$, and the expression is parametrized as
\begin{equation}
	\frac{d\Gamma}{d^4 p} \;\propto\; 
	e^2 \sum_{a=1}^{\Nc} e^{-\beta E/2  + i 2 \pi  q_a}\, e^{-\beta E/2  - i 2 \pi q_a} |M|^2\;,
\end{equation}
where $e$ is the QED coupling and we assumed $E \gg T$ to justify the Boltzmann approximation for quarks.
The eigenvalue, $q_a$, enters in the quark/anti-quark distribution functions
as a colored imaginary chemical potential, and they naturally
contribute with opposite signs.  Then, two $q_a$'s cancel in the
production rate!
The probability for a hard virtual photon, $E\gg T$, to be produced
from a quark/anti-quark annihilation process is, therefore, independent of the Polyakov loop. 
This makes a sharp contrast to the statistical distribution 
function for individual quarks and anti-quarks suppressed by the Polyakov loop.

Interestingly enough, explicit calculations in Refs.~\cite{Lee:1998nz,Gale:2014dfa,Hidaka:2015ima,Satow:2015oha} show that the dilepton production rate
for $\ell < 1$ is not suppressed but always more enhanced than
the $\ell = 1$ case, i.e.
\begin{equation}
	\frac{d \Gamma}{d E d^3p} \biggr|_{\ell}
	= 
	\frac{f(\ell)}{f(\ell=1)}
        \cdot \frac{d \Gamma}{d E d^3p} \biggr|_{\ell=1}
\end{equation}
with
\begin{equation}
	f(\ell) \equiv 1 - \frac{2 T}{3p} 
	\ln\biggl[ 
	\frac{1+3\ell e^{-\beta (E-p)/2} + 3\ell e^{- 2 \beta (E-p)/2}   +e^{-3 \beta (E-p)/2}  } 
	{1+3\ell e^{-\beta (E+p)/2} + 3\ell e^{- 2 \beta (E+p)/2}   +
          e^{-3 \beta (E+p)/2}  }
        \biggr]\;.
\end{equation}
The enhancement due to $\ell < 1$ is very modest; about 10-20 percent
at most.
It is therefore unlikely that the enhancement plays any important role
in the heavy-ion phenomenology.

Let us now turn our attention to a more Polyakov loop sensitive
observable, that is, the real photon production. 
Now, we consider a real photon with a large momentum, $E = p \gg T$.
The leading order processes for the photon production consist of the Compton scattering of a quark or an anti-quark and the annihilation of a quark/anti-quark pair.
Both processes are proportional to $e^2 g^2$ and of the same order.
There is an additional contribution; a quark scatters
with an arbitrary number of soft gluons with energy of order $gT$, and
emits collinear photons.
Such multiple scattering processes are of the same order of $e^2g^2$
because of the Bose-Einstein enhancement for the soft gluon, as seen
as $n(gT) \sim 1/g$ if $\ell=1$.
In the semi-QGP regime, $\ell < 1$ by definition, and this
Bose-Einstein enhancement is diminished for off-diagonal gluons (see Ref.~\cite{Hidaka:2015ima} for calculations of the collinear 
rate in the semi-QGP regime).  
At small energies ($E\ll T$) the gluon distribution function is proportional to 
$1/(e^{-i 2\pi q_{ab}} - 1)$ and is of order the unity if
$q_{ab} = q_a - q_b \sim 1$ for different $a$ and $b$.
The contribution of the diagonal ($a=b$) gluons is 
suppressed by $1/\Nc$. Hence, up to $1/\Nc$ corrections, the production 
of real photons is dominated by $2 \leftrightarrow 2$ processes in the
semi-QGP regime. 
To illustrate the effect of the non-trivial Polyakov loop on the photon 
production, let us consider the case of quark/anti-quark annihilation
into a gluon and a photon; 
the conclusion we will draw is also valid  for the Compton scattering.   
In the Boltzmann approximation, the rate of the 
real photon production is proportional to 
\begin{equation}
	E \frac{d\Gamma_\gamma}{d^3p} \;\propto\; 
	e^2 g^2 \sum_{a,b} 
	e^{-\beta \varepsilon_1 - i 2 \pi q_a} 
	e^{-\beta \varepsilon_2+ i 2 \pi q_b} 
	|M^{ab}|\;, 
	\label{eq:Gg}
\end{equation}
where $\varepsilon_1$ and $\varepsilon_2$ are the energies of the
color $a$ quark and the color $b$ anti-quark, respectively.  The
matrix element of the process is denoted by $M^{ab}$.
In the perturbative limit, $q_a, q_b \to 0$, and the rate is simply
proportional to $e^2 g^2 \Nc^2$.  In states near the confined phase,
the sum $\sum_a e^{-i  2 \pi q_a}$ goes to zero, as it is nothing but
the traced Polyakov loop, so that only $a=b$ terms in
Eq.~\eqref{eq:Gg} make finite
contributions.  This immediately 
results in a suppression factor of $1/\Nc$.
Moreover, the 
matrix element $M^{ab}$ involves a quark-gluon vertex giving 
an additional suppression factor of $1/\Nc$ for $a=b$.
This amounts to an overall $1/\Nc^2$ suppression factor in the confined phase!
Such a large suppression of the real photon production in the semi-QGP regime 
can have a significant impact on the phenomenological
observables;  this mechanism is considered to be potentially 
responsible for resolving the puzzle of the photon azimuthal
anisotropy, as argued in Ref.~\cite{Gale:2014dfa}.  

Similarly, the calculations of the collisional energy loss
show that, at the leading order, 
the energy loss can be also suppressed by powers of the Polyakov loop. 
For small values of the Polyakov loop, this suppression behavior is linear for
the light-quark scattering, and is quadratic for the gluon scattering
or the Compton scattering.
This results can be directly 
attributed to the number of the colored object in the semi-QGP regime, as we 
discussed above.  For a concrete phenomenological setup, for example,
see Ref.~\cite{Xu:2015bbz} in which a semi quark gluon monopole plasma
is assumed.

There are many other transport coefficients calculable in the semi-QGP
regime, such as the heat conductivity, the electric conductivity, the
bulk viscosity, etc.  However, they have not yet been considered in
the literature and remain to be explored in the future.

\section{Summary}
\label{sec:summary}

In writing this review, we followed a rather orthodox approach;
we gave a minimal description about the quantization procedure for QCD
at finite temperature, which is indispensable for understanding the
physical meaning of the Polyakov loop and its relation to the
realization of center symmetry.  Then, we introduced monumental
examples of the Polyakov loop calculations that were milestones in the
history of the Polyakov loop physics.  We made a quick review of
Polyakov's original arguments and calculations in the strong coupling
limit of the pure gluonic theory.  We also covered more extensive
strong coupling analysis to discuss the explicit forms of the Polyakov
loop potentials in the confined phase.

An asymptotic limit of weak coupling realizing at high 
temperatures, where quarks and gluons are deconfined and the perturbation theory can be applied, 
serves as an important example of the Polyakov loop calculation. 
In this case, the calculation
self-consistently leads to center symmetry breaking and we confirmed
that the center broken vacuum is stabilized by a finite Debye
screening mass.  
An interesting and non-trivial question is how these two asymptotic 
phases of confined and deconfined matter transition from one to
the other as the temperature approaches the critical value.  Similarly
to the formation of magnetic domains in spin systems, the
Z($\Nc$) domain walls may be identified as the physical
interfaces between the two phases. A 
domain-wall classical configuration can be easily found from the equation of motion 
obtained from  the perturbative potential.  In QCD, however, the vacuum
can accommodate other, more non-trivial configurations called instantons; 
the center elements are disturbed by instantons and
anti-instantons as well as by the Z($\Nc$) domain walls.  We shortly
reviewed a special class of finite-$T$ instantons with a non-trivial
holonomy (a non-unity Polyakov loop).  
In this context, for the advance of the confinement physics, 
it was important to recognize  that the 
finite-$T$ instantons are the bound states of dyons. 
Up to this point all discussions were based on first-principle QCD 
calculations.

Even if it is impenetrable to extract information on the Polyakov loop
directly from first-principle theory, we can still make use of the Polyakov loop as a
representative variable to characterize the hot QCD system.  In this
way, full thermodynamic behavior in the pure gluonic theory could be
well dictated by a Polyakov loop potential parametrized by a few
coefficients.
Such an ad-hoc  parametrization procedure not only results in successful fitting
of lattice results 
but also suggests the validity of an underlying physical picture that the Haar
measure or the ghost singularity is responsible for confinement.
To improve upon this ad-hoc approach, one can pursue 
a more ambitious goal of deriving the Polyakov loop potential directly from 
QCD supplemented by some approximations; thus, a goal of extending our understanding to the 
new, more fundamental, and QCD-based level. 
We selected and described two
methods along these lines.  The first one, the inverted Weiss
potential, is a very simple but powerful idea providing a qualitatively
concrete description of the confining vacuum.  The
ingredients necessary for the inverted Weiss potential are
the one-loop perturbative expression from QCD calculations, in which the
propagators are slightly modified to implement a clear observation of
ghost dominance seen in the lattice QCD or in the strong coupling
expansion. This method can be regarded as a hybrid
approach to confinement, for it is based on the perturbative technique 
supplemented by the non-perturbative input.  The second method we
discussed is the Polyakov loop matrix model.  This effective
description is derived from QCD by the strong coupling
expansion with an assumption that higher-order Polyakov loop
interactions can be neglected. The theory of the SU($\Nc$) matrix model has
interesting properties on its own;  for instance, we can formulate the
mean-field approximation in such a way  as to manifest the group integration
nature, which is useful to guarantee gauge invariance.

The most profound application of the Polyakov loop matrix model is
found in the large $\Nc$ limit.  Although some analytical studies are
possible for large-$\Nc$ QCD, the nature of the phase transition is
not yet fully revealed.  We showed that rich phase structures of
first, second, and third order phase boundaries emerge as coefficients
in the Polyakov loop potential vary.  Such a phase structure is
relevant also to QCD in a different setup, namely, QCD on a small 
sphere.  The running coupling constant is small on a
small 3-sphere just like in the  high temperature environment justifying 
the perturbation theory approach.  Perturbative
calculations result in an effective theory of the Polyakov loop matrix
model.

Coupling to quarks turns  the Polyakov loop physics into  a quite fertile
research field.  The biggest unsolved puzzle is the structure of
the QCD phase diagram at finite temperature and density.  For this
purpose various chiral models have predicted various phase boundaries
associated with chiral symmetry restoration.  These models can be
augmented with the Polyakov loop potential and the coupling to the quarks; such augmented models provide  handy
theoretical tools for a  discussion of  the interplay between chiral restoration
and deconfinement.  Instead of presenting model-dependent examples of the
phase diagrams, we focused on model-independent features of the
Polyakov loop coupling to quarks.  For further quantifying the model
studies, it is crucial to establish a precise determination of the
Polyakov loop potential.  We introduced an idea to constrain the
Polyakov loop potential from the heavy quark sector.

Even with sufficient knowledge on the Polyakov loop potential, the
sign problem prevents us from making any reliable prediction at finite
density.  In particular, the Polyakov loop and the anti-Polyakov loop
are sensitive to  quark and  anti-quark excitations in the medium,
so that the Polyakov loop model can be a minimal setup that shares the
same sign problem with original QCD.\ \ For one example
we gave a brief explanation of one of the simplest and thus most
frequently used models called the heavy-dense model.  To circumvent the
sign problem, it is a common strategy to change the underlying theory 
such that to evade the issue. This can be done is several ways, and one of the most 
well-understood is to analytically continue 
the chemical potential to imaginary values. QCD with imaginary chemical potential 
shows a different pattern
of the phase structure from the finite-density QCD, and we exemplified it 
by the perturbative Weiss potential, which exhibits the Roberge-Weiss
phase transition, and by the inverted Weiss potential.

Not only the external parameter such as the chemical potential but also the
definition of the theory itself could be deformed in such a way as to make it 
convenient for theoretical considerations.  The deformed QCD, although related to the original theory, 
does not describe physical QCD and should be regarded as a QCD-like model; nevertheless,  it  offers a useful insight to
deepen our understanding on the non-perturbative nature of QCD.\ \
In
QCD, chiral symmetry breaking and confinement are entangled in a
complicated way, which is of course the most interesting aspect of 
QCD; a disentanglement  of these transitions opens a possibility 
to investigate the confinement physics {\it exclusively}.  
In order to achieve this, one has to deform the theory or introduce the 
matter fields in such a way as to keep the center symmetry   
unbroken and, consequently, as to preserve the status of the  Polyakov loop  
as the exact order parameter.  In this review, we selected three approaches out of many 
existing ones in the literature.  We made this specific
choice based on  relevance to the other parts of this review
and based on their future prospects.

The Polyakov loop is gauge invariant and thus physically observable, but 
since the Polyakov loop is defined in Euclidean
spacetime, one may conclude that it hardly has any relevance for experimentally
measurable quantities. This is not correct;  the
Polyakov loop is not an academic phantom as it does affect
experimental observables measured in the relativistic heavy-ion collisions.  We took
a quick look at two examples;  one is the baryon number fluctuations
and the other is the so-called semi-QGP regime.  The baryon number
fluctuations play an important role of a signature for deconfinement, which is
naturally sensitive to the Polyakov loop.  Moreover, the baryon number
fluctuations provide a possible measure to detect and locate  the QCD critical point; 
experimental data at various collision energies are already
available (to be precise, not the baryon number but rather the
proton number is measurable).  Because the Polyakov loop has a direct
coupling to  quark, the baryon number fluctuations  receive a
contribution from the Polyakov loop fluctuations or the curvature of
the Polyakov loop potential.  One should now notice that not only the
Polyakov loop fluctuations but the expectation value of the Polyakov
loop can change the real-time physical observables including  the
transport coefficients.  This is the idea of the semi-QGP regime,
which is applicable more generally near the QCD crossover.  We know
that around the QCD phase transition temperature the Polyakov loop is
still small, and this means, that the quark and the anti-quark excitations
are significantly screened by the effect of the Polyakov loop.  Since
gluons are colored particles, small Polyakov loops also suppress
gluonic excitations, which could be an alternative description of the
strongly correlated QGP with small shear viscosity to the entropy
density ratio.

In this review we stressed  that the Polyakov loop
is not just another observable measured in the lattice-QCD simulation; as in 
theoretical studies it is sometimes useful to regard the
Polyakov loop as a control parameter to access the confining
sector of QCD.\ \ It is even possible to treat the Polyakov loop
as a theoretical device to establish an intuitive understanding of hot gluonic
medium, similar to the concept of a constituent quark in hadronic physics. 
We chose the
topics covered in this review so that our discussions bear 
some degree of generality and could be applied not only to QCD but also to other systems,
e.g.\ condensed matter systems with emergent gauge symmetry.
This is the reason why we did not cover important but rather
specific subjects including higher-order perturbative
calculations of the Polyakov loop expectation
value~\cite{Brambilla:2010xn,Berwein:2015ayt}, subtleties in the
renormalization
prescriptions~\cite{Kaczmarek:2002mc,Dumitru:2003hp,Petreczky:2015yta}
(see Ref.~\cite{Bazavov:2016uvm} for latest results with the $Q\bar{Q}$
renormalization procedure),
Casimir scaling~\cite{Gupta:2007ax,Megias:2013xaa}, magnetic field coupling to the
Polyakov loop~\cite{Bruckmann:2013aza,Ozaki:2015yja}, etc.

We close this review by reiterating that there is a common consensus
that the Polyakov loop is an approximate order
parameter for the deconfinement phase transition in QCD.\ \
The traditional way of treating the Polyakov loop was to
use it to characterize a thermal system away from confinement
rather than coping with the low-$T$ confinement problem itself.  
However, in view of the recent developments, the
Polyakov loop is becoming an increasingly more important  player in the arena of
confinement physics. We hope that this review makes a strong case for
the paramount significance of the Polyakov loop as the quantity defining the
properties of pure gluonic theory and QCD near zero as well as
near the transition temperature.

\section*{Acknowledgments}
The authors thank Jens~Braun,  Yoshi~Hatta, Yoshimasa~Hidaka,
Hiroaki~Kouno, Larry~D.~McLerran, Hiromichi~Nishimura,
Jan~M.~Pawlowski, Rob~Pisarski,
Janos~Polonyi, Bernd-Jochen~Schaefer, Edward~Shuryak, Yuya~Tanizaki,
Mithat~\"{U}nsal, and Wolfram~Weise for discussions.
K.~F.\ was partially supported by JSPS KAKENHI Grant No.\ 15H03652 and
15K13479.


\bibliographystyle{utphys.bst}
\bibliography{polyakov,polyakov_vl,polyakov_spires}

\end{document}